\documentclass[aps,prx,twocolumn,10pt,amsmath,amssymb,floatfix,superscriptaddress]{revtex4-2}

\usepackage{graphicx} 
\usepackage{bm} 
\usepackage{hyperref} 
\usepackage{braket,bbold,color}
\usepackage{hyperref}
\usepackage{multirow}
\usepackage[table]{xcolor}
\begin{document}

\title{Exciton Berryology}

\author{Henry Davenport}
\affiliation{Blackett Laboratory, Imperial College London, London SW7 2AZ, United Kingdom}
\author{Johannes Knolle}
\affiliation{Technical University of Munich, TUM School of Natural Sciences, Physics Department, 85748 Garching, Germany}
\affiliation{Munich Center for Quantum Science and Technology (MCQST), Schellingstr. 4, 80799 M\"unchen, Germany}
\affiliation{Blackett Laboratory, Imperial College London, London SW7 2AZ, United Kingdom}
\author{Frank Schindler}
\affiliation{Blackett Laboratory, Imperial College London, London SW7 2AZ, United Kingdom}

\begin{abstract}
In translationally invariant semiconductors that host exciton bound states, one can define an infinite number of possible  exciton Berry connections. These correspond to the different ways in which a many-body exciton state, at fixed total momentum, can be decomposed into free electron and hole Bloch states that are entangled by an exciton envelope wave function. Inspired by the modern theory of polarization, we define an exciton projected position operator whose eigenvalues single out two unique choices of exciton Berry phase and associated Berry connection -- one for electrons, and one for holes. We clarify the physical meaning of these exciton Berry phases and provide a discrete Wilson loop formulation that allows for their numerical calculation without a smooth gauge. As a corollary, we obtain a gauge-invariant expression for the \emph{exciton polarisation} at a given total momentum, \emph{i.e.} the mean separation of the electron and hole within the exciton wave function. In the presence of crystalline inversion symmetry, the electron and hole exciton Berry phases are quantized to the same value and we derive how this value can be expressed in terms of inversion eigenvalues of the many-body exciton state. We then consider $C_2 \mathcal{T}$ symmetry, for which no symmetry eigenvalues are available as it is anti-unitary, and confirm that the exciton Berry phase remains quantized and still diagnoses topologically distinct exciton bands. The notion of shift excitons, whose exciton Wannier states are displaced from those of the non-interacting bands by a quantized amount, can therefore be generalised beyond symmetry indicators.
\end{abstract}

\maketitle
\section{Introduction} 

\begin{figure}[t]
    \centering
    \includegraphics[width=0.98\linewidth]{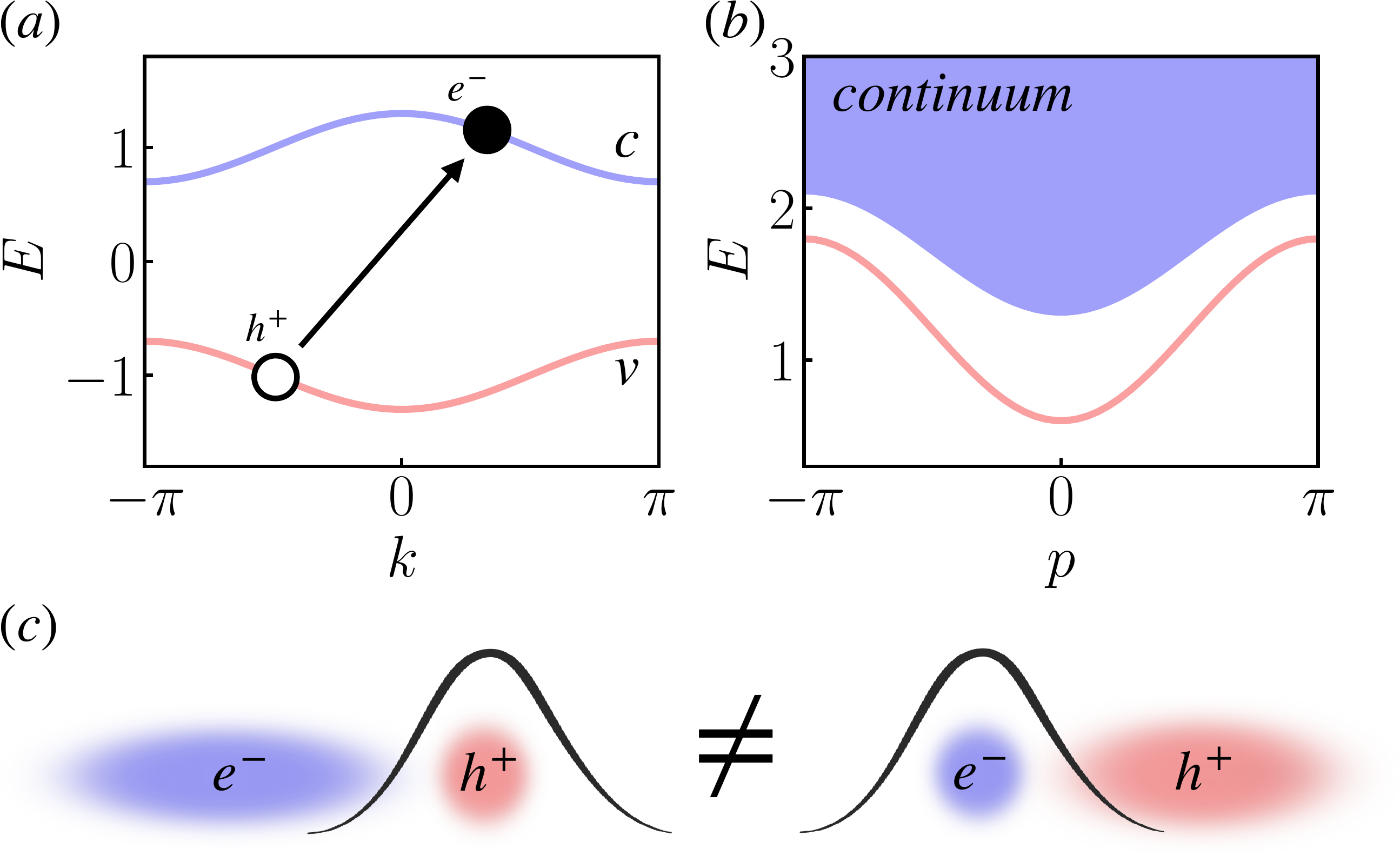}
    \caption{
    Exciton Wannier states. $(a)$ An illustrative semiconductor electronic band structure. Energy $E$ is plotted against momentum $k$. At half filling the valence band ($v$) is occupied and the conduction band ($c$) is empty. An electron $e^-$ can be excited from the conduction band leaving a hole in the valence band $h^+$. $(b)$ The $e^-$ and $h^+$ can become bound due to the Coulomb interaction. The electron-hole excitation energy ($E$) plotted against total momentum ($p$) shows an exciton band (red) gapped from the continuum of unbound electron-hole states (purple). $(c)$ Exciton Wannier functions can maximally localise the hole (red) or the electron (purple). These are not always equal and can have different centres of mass. The different exciton Wannier centres are associated with the different exciton Berry phases we introduce.}
    \label{fig:Fig1Intro}
\end{figure}
Topology has proven essential for the classification and prediction of novel quantum phases in condensed matter systems~\cite{HasanKaneRMP10, QSHEffectGrapheneMele2005, QSHEffectBernevig2006}. In particular, the topological classification of ground states in non-interacting electron systems—including under crystalline and local symmetries—is increasingly well understood~\cite{FuTCI11, topocrysInsulatorsRobertJan2013, Ryu_2010, Ashvin230SINatComm17, AndreiTQC17}. Given this success, there is increasing interest in extending these ideas to systems where electron-electron interactions play a crucial role. One natural direction is to focus on the topology of interacting ground states~\cite{MartinaInteractingTQC23, ShiozakiSPTClassification23, NarenManjunathTCIClassification24, JonahInteractingTQCNatComm24, WenZooTopPhases, Rachel_2018}.
However, interactions can also significantly alter the excitation spectrum, motivating an alternative approach: the topological classification of interaction-induced excitations. Since the simplest excitations in insulators and semiconductors are excitons, the study of exciton topology has emerged as a rapidly developing field~\cite{ExcitonTopologyDavenport, ExcitontopologySlager, ExcitonTopologyPaperKwan, ExcitonTopologyPaperQian, ExcitonTopologyPaperTitus, ExcitonTopologyPaperUchoa, thompson2025topologicallyenhancedexcitontransport, paiva2024shiftpolarizationexcitonsquantum, topExcitonsMoire, FuChernExcitons, QuantumGeometricDipole}. 

In insulators and semiconductors, the ground state is gapped. In the absence of electron-electron interactions the ground state consists of all bands below the Fermi energy occupied - these bands are known as the valence bands. The bands above the Fermi energy are known as the conduction bands and are unoccupied in the ground state. As shown in Fig.~\ref{fig:Fig1Intro}a, an electron can be excited from a valence band to a conduction band. This leaves a (positively charged) hole in the valence band which, due to the Coulomb interaction, can become bound to the excited electron. The spectrum of electron-hole excitations as a function of total momentum (Fig.~\ref{fig:Fig1Intro}b) typically consists of a continuum - \emph{i.e.} states where the electron and hole are effectively free - as well as bound states which form in the many-body gap. These bound states, known as excitons, are of technological significance since they dominate the optical properties of semiconductors. They are therefore of vital importance in various fields from photovoltaics~\cite{nelson2003physics} to light emitting diodes~\cite{forrest2020organic}. 

The study of exciton topology has initially focussed on excitons formed within topological electronic bands~\cite{ExcitonTopologyPaperKwan, ExcitonTopologyPaperQian, ExcitonTopologyPaperUchoa} but more recently, attention has expanded to study the \emph{crystalline} topology of excitons even in trivial electronic bands~\cite{ExcitonTopologyDavenport, ExcitontopologySlager}. In this context, in an earlier work, we introduced the concept of shift excitons in inversion-symmetric systems~\cite{ExcitonTopologyDavenport}. Shift excitons have maximally localised exciton Wannier centres~\cite{NeatonExcitonMLWFsPRB23} that are shifted by a quantised amount with respect to the electronic Wannier states of the underlying conduction and valence bands~\cite{ExcitonTopologyDavenport}. Therefore, even in systems where the underlying electronic bands are both trivial (\emph{i.e.} with electronic Wannier centres at the centre of the unit cell), the interactions can bind excitons that are non-trivial (\emph{i.e.} with the exciton Wannier centres shifted to the unit cell edge). This leads to the formation of exciton edge states even in systems where the underlying electronic bands do not exhibit edge states. Shift excitons are an example of interaction-induced topology beyond the topological classification of single electron band topology. One particular outstanding research area is to fully understand, and disentangle, the various contributions to the topology of excitons, because nontrivial exciton topological invariants can both be \emph{inherited} from the non-interacting bands as well as \emph{induced} by the interactions~\cite{ExcitonTopologyDavenport}.

In this paper we address foundational questions concerning the physical meaning of the exciton Berry phase and the exciton Wannier functions. We begin by showing that there are infinitely many equally valid definitions for the exciton Berry phase, each of which may evaluate to a distinct value. We then show that these Berry phases contain different \emph{physical} information about the exciton band. We do this by defining two projected position operators for excitons—one corresponding to the electron and the other to the hole position—whose eigenvalues yield two unique, gauge-invariant Wilson loops. These Wilson loops give the shift of the exciton Wannier centre when either the electron or hole is maximally localised within the exciton, in general, the two are not equal (see Fig.~\ref{fig:Fig1Intro}c). 
We further examine how these shifts transform under inversion and $C_2 \mathcal{T}$ symmetry in systems with one spatial dimensional (1D). These are the only symmetries that quantise exciton Berry phases in physically reasonable 1D systems. We end by describing the general case where these symmetries are broken.

\section{A multitude of Exciton Berry Connections}\label{sec:FirstExcitonBerryPhaseSec}
We begin by demonstrating that there are infinitely many valid exciton Berry connections. Our approach will be to determine what constraints the gauge freedoms of both the exciton and electronic wave functions place on the exciton Berry connection. We will show that there is not one unique Berry connection but an infinite family of Berry connections that obey these constraints. 

We consider translationally symmetric systems in 1D with periodic boundary conditions (PBC). We label unit cells by $R\in \{0, 1, \dots, L-1\}$ (for system size $L$). Without loss of generality, we will exclusively consider excitons formed from two electronic bands where $c_{k, c}$ ($c_{k, v}$) annihilates an electron in the conduction (valence) band at crystal momentum $k \in [-\pi,\pi)$ (we set the lattice parameter $a=1$). The allowed momenta are spaced by $\Delta = 2\pi/L$. The creation operators expressed in the real space basis are $c^\dagger_{k, c/v} = \sum_{R, i} \psi_{R, i}^{k, c/v} c^\dagger_{R, i}$ using $\psi_{R, i}^{k, c/v} = e^{\mathrm{i}kR}\braket{i|u_{k, c/v}}$ in terms of the Bloch functions $\ket{u_{k, c/v}}$ and atomic orbitals $\ket{i}$. Note that we assume the atomic sites are in the centre of the unit cell throughout this paper. In the ground state the valence band is filled $\ket{\mathrm{GS}} = \prod_{k} c^\dagger_{k, v} \ket{0}$, where $\ket{0}$ is the fermionic vacuum. Due to translational symmetry, exciton eigenstates can be labelled by total momentum $p$. There are many ways of writing equivalent exciton wave functions at total momentum $p$, we can parametrise these by the real (and rational) parameter $\alpha$,
\begin{equation} 
\ket{p_{\mathrm{exc}}} = \sum_k \phi^{p}_{k - \alpha p} c^\dagger_{(1-\alpha) p+k, c} c_{-\alpha p + k, v} \ket{\mathrm{GS}}.
\label{eq:excitonEigenstate}
\end{equation}
The exciton wave function $\phi^{p}_{k - \alpha p}$ depends on $\alpha$ in such a way that the total $\ket{p_{\mathrm{exc}}}$ remains equal regardless of $\alpha$. We note that all $\alpha$ are equally valid and therefore $\alpha$ is \emph{not} related to the electron and hole masses. The total momentum of $\ket{p_{\mathrm{exc}}}$ is $p$ (again regardless of $\alpha$) since $(1-\alpha) p + k + \alpha p - k = p$. The exciton state is normalised such that $\braket{p_\mathrm{exc}|p_{\mathrm{exc}}} = 1$, this implies that the exciton wave function is normalised to \mbox{$\sum_k |\phi^p_{k-\alpha p}|^2 = 1$}.  When we take the thermodynamic limit ($L\rightarrow\infty$) we replace $\sum_{k}\rightarrow \int_0^{2\pi} \frac{\mathrm{d}k}{\Delta}$. Hence the normalisation condition becomes $\int_0^{2\pi} \frac{\mathrm{d}k}{\Delta}|\phi^p_k|^2 = 1$.

We will now show that we can use $\alpha$ to tune between an infinite number of valid exciton Berry connections. We begin by attempting to construct an exciton Berry connection $A_{\mathrm{exc}}^\alpha(p)$ informally using the requirement that, under a global gauge transformation of the exciton state ($\ket{p_{\mathrm{exc}}} \rightarrow e^{\mathrm{i}\theta(p)} \ket{p_{\mathrm{exc}}}$), the Berry connection transforms covariantly [\emph{i.e.} $A_{\mathrm{exc}}^\alpha(p) \rightarrow A_{\mathrm{exc}}^\alpha(p) - \nabla \theta(p)$]. Berry connections are only defined in the thermodynamic limit since they require momentum to be a continuous quantity. By analogy with the \emph{electronic} Berry connection we might first consider an \emph{exciton} Berry connection which is simply $\int_{0}^{2\pi} \frac{\mathrm{d}k}{\Delta} \;\bar \phi^{p}_{k - \alpha p} \mathrm{i}\partial_p \phi^{p}_{k - \alpha p}$ where the bar denotes complex conjugation. This expression is gauge covariant under the transformation above. However it has a different gauge problem - it is not gauge \emph{invariant} under gauge transformations of the underlying \emph{electronic} bands. Under gauge transformations of the \emph{electronic} bands ($c^\dagger_{p, v}\rightarrow e^{\mathrm{i}\eta(p)}c^\dagger_{p, v}$ and $c^\dagger_{p, c}\rightarrow e^{\mathrm{i}\xi(p)}c^\dagger_{p, c}$) the total exciton state $\ket{p_{\mathrm{exc}}}$ should be invariant. To account for the change in the electronic gauge the exciton wave function transforms oppositely, $\phi^{p}_{k - \alpha p} \rightarrow e^{\mathrm{i}\eta [-\alpha p + k]} e^{-\mathrm{i} \xi[(1-\alpha) p + k]} \phi^{p}_{k - \alpha p}$. Hence, under gauge transformations of the electronic bands, the naive \emph{exciton} Berry connection defined above acquires additional terms proportional to the derivatives of the functions $\eta(p)$ and $\xi(p)$ (see Appendix~\ref{sec:ApxFirstExcitonBerryPhaseSec}). This would mean that the exciton Berry \emph{phase} would change if we changed the gauges of the electronic bands. This is undesirable since we want the exciton Berry phase to contain physical information about the excitons. We can, however, construct a gauge-invariant exciton Berry connection by including correction terms which depend on the  \emph{electronic} Berry connections of both the valence and conduction bands,
\begin{widetext}
\begin{equation}
\begin{aligned}
A^\alpha_{\mathrm{exc}}(p) =  \int_{0}^{2\pi}\frac{\mathrm{d}k}{\Delta}\: \left(\bar \phi^{p}_{k - \alpha p} \mathrm{i} \partial_p \phi^{p}_{k - \alpha p}  +|\phi^{p}_{k - \alpha p}|^2 \left\{\alpha A_{\mathrm{elec}, v} \left(k-\alpha p\right)+(1-\alpha) A_{\mathrm{elec}, c}\left[(1-\alpha) p+k\right]\right\} \right),
\label{eq:naiveBerryConnection}
\end{aligned}
\end{equation}
\end{widetext}
where the \emph{electronic} Berry connections are $A_{\mathrm{elec}, c/v}(k) = \bra{u_{q, c/v}} \mathrm{i}\partial_q \ket{u_{q, c/v}}|_{q = k}$.  
As shown explicitly in Appendix~\ref{sec:ApxFirstExcitonBerryPhaseSec}, under electronic gauge transformations these additional terms transform in the opposite way to the first and so the Berry connection (for each $\alpha$) is both gauge \emph{covariant} under global transformations and gauge \emph{invariant} under transformations of the electronic bands. In Appendix~\ref{apd:CellPeriodicWavefunctions} we present an alternative derivation of the family of exciton Berry connections in terms of the different possible \emph{cell-periodic} exciton wave functions that can be extracted from the many body state $\ket{p_\mathrm{exc}}$.

We now have infinitely many exciton Berry connections, one for each $\alpha$, and each can be integrated over the Brillouin zone (BZ) to give a seemingly different exciton Berry phase. This raises several questions. Do these Berry connections really give different values for the Berry phase? What physical information do the Berry phases contain? And how many of these Berry phases give unique information? In the next section, we derive the two unique exciton Wilson loops—and their corresponding Berry connections—in a manner that reveals the physical interpretation of these different exciton Berry connections.

\section{Exciton projected position operator}
We now derive the exciton Berry phase in a way that clarifies its physical meaning. In non-interacting electronic band topology, the \emph{electronic} Berry phase is found to be related to the electronic Wannier centres. This can be demonstrated using the electronic position operator projected into the electronic band of interest. We first recap this connection before generalising the method to excitons.

\subsection{Recap of the \emph{electronic} projected position operator}\label{sec:ElectronicProjPosition}
We begin by revisiting the relationship between the electronic Berry phase, electronic Wannier states and the projected position operator in non-interacting systems. Excitons are excitations above some ground state and so we consider an analogous scenario in the non-interacting case -- a single electron in the conduction band and a filled valence band. Our presentation has the novelty that we use a genuinely many-body operator to find the Berry phase of the excitations. This means that our results here are readily generalisable to the exciton case and  we will see that this elucidates the physical meanings of the various exciton Berry phases. We consider excitations of a single electron (with momentum $p$) above the occupied valence band,
\begin{equation}
\ket{p} = c^\dagger_{p, c} \ket{\mathrm{GS}},~\label{eq:singlelectronExcitation}
\end{equation}
where $c_{p, c/v}$ annihilates an electron in the conduction/valence band and the ground state is $\ket{\mathrm{GS}} = \prod_{q} c^\dagger_{q, v} \ket{0}$. The many-body Wannier states are the linear combinations of $\ket{p}$,
\begin{equation}
\ket{\mathcal{W}^R} = \sum_{p} e^{-\mathrm{i} p R} e^{\mathrm{i}\phi(p)} \ket{p},\label{eq:generalWannierStateEq}
\end{equation}
where $e^{\mathrm{i}\phi(p)}$ captures the freedom in the choice of gauge for the states $\ket{p}$. Modifying this phase factor affects the localisation of the Wannier states. One desirable way to determine the phase factors is to choose them such that the Wannier functions are maximally localised~\cite{MarzariVanderbiltSmoothGaugePRB97}. For excitations of the form Eq.~\eqref{eq:singlelectronExcitation}, this means finding the gauge where the Wannier states maximally localise the conduction band electrons in real space. This requires an operator which gives the density of the electrons in the conduction band; we use the band projected number operator. This is constructed by expressing the full number operator in the band basis and only retaining the terms which are quadratic in the creation and annihilation operators for the band of interest. For the conduction band this gives $\hat n_{R, i}^c = \sum_{k, k'} \bar\psi^{k, c}_{R, i} \psi^{k', c}_{R, i} \, c^\dagger_{k, c} c_{k', c}$. To maximally localise the excitation Wannier states we minimise the localisation functional,~\cite{MarzariVanderbiltSmoothGaugePRB97}
\begin{equation}
\Omega = \bra{\mathcal{W}^{0}} \hat Y_c \ket{\mathcal{W}^0} - \left[ \bra{\mathcal{W}^{0}}\hat X_c\ket{\mathcal{W}^0} \right]^2.
\end{equation}
where $\hat X_c = \sum_{R, i} R \,\hat n^c_{R, i}$ is the band-projected position operator and $\hat Y_c = \sum_{R, i} R^2 \,\hat n^c_{R, i}$. This is equivalent to minimising the spatial variance of the Wannier state. 

In 1D the Wannier states which minimise this functional are the eigenfunctions of the projected position operator $\hat{\mathcal{X}} =\hat P_{\mathrm{elec}}\hat X_c \hat P_{\mathrm{elec}}$, where $\hat P_{\mathrm{elec}}$ is the projector onto the excitation band $\hat P_{\mathrm{elec}} = \sum_p \ket{p}\bra{p}$~\cite{MarzariVanderbiltSmoothGaugePRB97}. This operator has an awkward property however -- it is not periodic in real space.
This means that only in the thermodynamic limit ($L\rightarrow \infty$) are the eigenstates (\emph{i.e.} the Wannier states) related by translational symmetry and of the form of Eq.~\eqref{eq:generalWannierStateEq}~\cite{Neupert_2018}.  
To show that the eigenstates of $\hat{\mathcal{X}}$ are the Wannier states we must therefore go immediately to the thermodynamic limit form of the position operator and replace $\hat X_c\rightarrow \mathrm{i}\partial_k$~\cite{Neupert_2018}. The physical meaning of this substitution is, perhaps, still clear for the electronic excitations we are considering here, but this substitution does obscure the physical meaning for  excitons -- would this replacement constitute the position operator for the electron, the hole or the centre of mass of the exciton? To anticipate and ameliorate this difficulty we introduce a genuinely periodic \emph{many-body} operator from which we can derive the excitation Wannier states even for finite sized systems. We switch to a new operator ($\hat Z_c$) that encodes position in a periodic system and project it into the excitation band to give $\hat{\mathcal{Z}}$,
\begin{equation} \label{eq:Zoperator}
\hat Z_c = \sum_{R} e^{\mathrm{i} \Delta R} \hat n^c_{R, i}, \quad
\hat{\mathcal{Z}} = \hat P_{\mathrm{elec}} \hat Z_c \hat P_{\mathrm{elec}},
\end{equation}
where, as before, $\Delta = 2\pi/L$. A direct calculation of the eigenstates of $\hat{\mathcal{Z}}$ (shown in Appendix~\ref{Sec:ApxElectronicProjPosition}) gives the standard expression for the maximally localised Wannier states, each labelled by an integer $R \in \{0,1, \dots,  L-1\}$ and normalised by $\mathcal{N}$,
\begin{equation}
\ket{\mathcal{W}^R} = \mathcal{N}\sum_{p} e^{-\mathrm{i} pR} (W_c)^{-\frac{p}{2\pi}} \prod_{k = 0}^{p-\Delta} t_{k}\ket{p},\label{eq:electronicWannierStates}
\end{equation}
where $t_k = \braket{u_{k+\Delta, c}|u_{k, c}}$ and $W_c$ is the Wilson loop for the conduction band,
\begin{equation}
\prod_{k = 0}^{2\pi - \Delta} t_k = \prod_{k = 0}^{2\pi - \Delta} \braket{u_{k+\Delta, c}|u_{k, c}} = W_{c}.
\end{equation}
In the thermodynamic limit the Wilson loop has modulus one and can be written as $W_c \rightarrow e^{\mathrm{i} \gamma_c}$ where $\gamma_c$ is the conduction band electronic Berry phase~\cite{Neupert_2018}. We can see how Eq.~\eqref{eq:electronicWannierStates} modifies  the (arbitrary) gauge of $\ket{p}$ in order to maximally localise the Wannier functions. Firstly, the product of Bloch function overlaps ($t_k$) maximally aligns the phases of the Bloch functions ($\ket{u_{p, c}}$) at adjacent momenta. The Wilson loop term [$(W_c)^{-\frac{p}{2\pi}} \rightarrow e^{-\mathrm{i }\frac{\gamma_c p}{2\pi}}$] then equally distributes the total Berry phase across the BZ. This results in a smooth and continuous gauge across the whole BZ (known as the \emph{twisted parallel-transport gauge}), this gauge choice maximally localises the Wannier functions in 1D~\cite{MarzariVanderbiltSmoothGaugePRB97, Neupert_2018}.

Given that the eigenstates of $\hat{\mathcal{Z}}$ are the maximally localised Wannier functions ($\ket{\mathcal{W}^R}$) we expect the corresponding eigenvalues $\lambda_R$ to give the centres of the Wannier states. In the thermodynamic limit, the eigenvalues of $\hat{\mathcal{Z}}$ are,
\begin{equation}
\lambda_R = e^{\mathrm{i}\Delta(\frac{\gamma_c}{2\pi} + R)},
\end{equation}
as shown in Appendix~\ref{Sec:ApxElectronicProjPosition}. This is the expected form for the eigenvalues since we exponentiated the positions $R$ to make the operator $\hat{\mathcal{Z}}$ periodic. The real space centres of the excitation Wannier states ($s_{R}$) can therefore be extracted by undoing this and taking the logarithm of the eigenvalues $\lambda_R$. We find that the excitation Wannier centres are,
\begin{equation}
s_{R} = -\mathrm{i} \log(\lambda_R)/\Delta = \frac{\gamma_c}{2\pi} + R,\label{eq:excitationWannierCentres}
\end{equation}
which is the standard expression for the electronic Wannier centre positions in terms of the Berry phase~\cite{Neupert_2018}. We see that the Wannier centres are equally spaced in real space (as a result of translational symmetry) and the shift of the Wannier states from the centre of the unit cell is proportional to the Berry phase. 

To summarise, we have introduced a new way to derive the electronic Wannier states and the Berry phase in non-interacting systems. Crucially our method avoids taking the thermodynamic limit and is performed using a genuine many-body operator $\hat{\mathcal{Z}}$. We note that this is different from the many-body operator, $e^{\mathrm{i} \Delta\hat X}$,  introduced by Resta in Ref.~\onlinecite{RestaQMPositionOperator} (where the position operator is $\hat X = \sum_R R\: \hat n_{R, i}$). The Berry phase/polarisation of the \emph{ground state} can be found by calculating the expectation value of this operator in the \emph{ground state}~\cite{RestaQMPositionOperator}. However, we are interested in Berry phases for excited states so this operator cannot be used. The alternative many-body operator ($\hat{\mathcal{Z}}$) that we introduced above we see \emph{does} allow us to extract the Berry phase for excitations.
In addition our method allows us to avoid having to write down a first quantised wave function for the excitations, instead we can directly work with the many-body wave function -- this will again be essential once we study excitons. We showed that the eigenvectors of the projected periodic ``position" operator $\hat{\mathcal{Z}}$ are equal to the expressions obtained from the standard treatment using the naive projected position operator $\hat{\mathcal{X}}$ in the thermodynamic limit. Furthermore, the eigenvalues were shown to give the Wannier centres. In Appendix~\ref{Sec:ApxElectronicProjPosition} we discuss why $\hat{\mathcal{Z}}$ has such a close relationship with $\hat{\mathcal{X}}$ in the thermodynamic limit. In the next section we generalise the method introduced here so that we can calculate exciton Wannier states, understand the infinite family of exciton Berry phases found in Sec.~\ref{sec:FirstExcitonBerryPhaseSec}, and obtain gauge invariant Wilson loops to calculate the exciton Berry phases in finite systems and without a smooth gauge.

\subsection{The \emph{exciton} projected position operators}
In Sec.~\ref{sec:FirstExcitonBerryPhaseSec} we constructed exciton Berry connections in a somewhat arbitrary way by considering how the connection changed under certain gauge transformations. The different possible Berry connections were seen to be linked to the convention we had used for the exciton wave function. However, we would like to have a \emph{physical} interpretation for these distinct exciton Berry connections since the quantities of physical importance will be agnostic as to the convention in which the exciton wave function is written. Hence these different exciton Berry connections must contain information about different physical properties of the excitons. We take inspiration from the approach introduced in Sec.~\ref{sec:ElectronicProjPosition} and now derive the exciton Berry phases using a many-body operator. 

The major difference for excitons, however, is that, unlike for the single electron excitations described in Sec.~\ref{sec:ElectronicProjPosition}, there are different -- and equally valid -- position operators that we could use for excitons. This is because excitons are \emph{two} particle bound states and so we can consider the position operator for the electron (in the conduction band) as well as that for the hole (in the valence band). To calculate the electron/hole densities within the exciton, we require operators that yield the spatial electron and hole densities within the two bands. We use the band projected number operators introduced in Sec.~\ref{sec:ElectronicProjPosition}. The conduction band projected electron number operator is $\hat n_{R, i}^c = \sum_{k, k'} \bar\psi^{k, c}_{R, i} \psi^{k', c}_{R, i} \, c^\dagger_{k, c} c_{k', c}$ as introduced earlier, but in addition we use the hole number operator projected into the valence band $\hat n_{R, i}^v = \sum_{k, k'} \psi^{k, v}_{R, i} \bar \psi^{k', v}_{R, i} \, c_{k, v} c^\dagger_{k', v}$. These band projected number operators are a key quantity in studying interactions in flat bands and so it is natural that they also are of use here where we consider interactions between (two) particles in different bands~\cite{TBGIII, TBGIV, TrionsTBG}. This leads us to use one of two possible periodic position operators ($\hat Z_{c/v}$) and  we consider these projected into the exciton band ($\hat{\mathcal{Z}}_{c/v}$),
\begin{equation}
\hat Z_{c/v} = \sum_{R, i} e^{\mathrm{i} \Delta R} \hat n^{c/v}_{R, i},\quad 
\hat{\mathcal{Z}}_{c/v} = \hat P_{\mathrm{exc}}  \hat Z_{c/v} \hat P_{\mathrm{exc}},
\end{equation}
where once again $\Delta = 2\pi/L$ for system size $L$ and the exciton projector is $\hat P_{\mathrm{exc}} = \sum_p \ket{p_{\mathrm{exc}}} \bra{p_{\mathrm{exc}}}$. We expect the eigenvectors of $\hat{\mathcal{Z}}_{c/v}$ to be the \emph{exciton} Wannier functions obtained from maximally localising either the electron ($\ket{\mathcal{W}^R_{\mathrm{exc}, c}}$) or the hole ($\ket{\mathcal{W}^R_{\mathrm{exc}, v}}$) within the exciton. A direct calculation of the eigenvectors is shown in Appendix~\ref{Sec:ApxExcitonProjPosition}. The eigenvectors (normalised by $\mathcal{N}$) are,
\begin{equation}
\ket{\mathcal{W}^R_{c/v}} = \mathcal{N}\sum_{p} e^{-\mathrm{i} p R} (W_{\mathrm{exc}, c/v})^{-\frac{p}{2\pi}} \prod_{k = 0}^{p-\Delta} t^{\mathrm{exc}, c/v}_{k}\ket{p_{\mathrm{exc}}},\label{eq:ExcitonWannierStates}
\end{equation}
where the two sets of phase factors 
\begin{align}
t^{\mathrm{exc}, c}_{k} &=  \sum_q  \bar \phi^{k+\Delta}_{q- k} \phi^{k}_{q- k} \braket{u_{q+\Delta, c}| u_{q, c}},\\
t^{\mathrm{exc}, v}_{k} &= \sum_q \bar \phi^{k+\Delta}_{q} \phi^{k}_{q+\Delta} \braket{u_{q+\Delta, v}| u_{q, v}},
\end{align}
maximally localise either the electron or the hole component of the exciton Wannier states. The exciton Wilson loops are, 
\begin{align}
W_{\mathrm{exc}, c} &= \prod_p t^{\mathrm{exc}, c}_{p}=\prod_p \left[\sum_k\bar \phi^{p+\Delta}_{k- p} \phi^{p}_{k- p} \braket{u_{k+\Delta, c}| u_{k, c}}\right]\label{eq:WilsonloopElec},\\
W_{\mathrm{exc}, v} &= \prod_p t^{\mathrm{exc}, v}_{p}=\prod_p \left[\sum_k \bar \phi^{p+\Delta}_{k} \phi^{p}_{k+\Delta} \braket{u_{k+\Delta, v}| u_{k, v}}\right]\label{eq:WilsonloopHole},
\end{align}
where $p$ goes over the incontractible loop through the BZ $p \in \{0, \Delta, \dots 2\pi -\Delta\}$ and $k$ sums over all momenta. Just like for the \emph{electronic} Wilson loop, the \emph{exciton} Wilson loops $W_{\mathrm{exc}, c/v}$ both have modulus 1 in the thermodynamic limit. Therefore, they can be rewritten as $e^{\mathrm{i}\gamma_{\mathrm{exc}, c/v}}$ where $\gamma_{\mathrm{exc}, c/v}$ is an exciton Berry phase. It follows that the exciton Wannier states ($\ket{\mathcal{W}^R_{\mathrm{exc},c/v}}$) take a very similar form to the \emph{electronic} Wannier states in Eq.~\eqref{eq:electronicWannierStates}. The key difference however is that we  have two \emph{different} expressions for the exciton Wannier states when we either maximally localise the electrons ($\ket{\mathcal{W}^R_{\mathrm{exc},c}}$) or the holes ($\ket{\mathcal{W}^R_{\mathrm{exc},v}}$) in the exciton.

The eigenvalues of $\hat{\mathcal{Z}}_{c/v}$ give the exciton Wannier centres. In the thermodynamic limit, the eigenvalues are $\lambda_{R, c/v} = \exp\left[{\mathrm{i} \Delta\left(\frac{\gamma_{\mathrm{exc}, c/v}}{2\pi} + R\right)}\right]$ (see Appendix~\ref{Sec:ApxExcitonProjPosition}). The exciton Wannier  centres are obtained by taking the logarithm of $\lambda_{R, c/v}$ [following  Eq.~\eqref{eq:excitationWannierCentres}]. The exciton Wannier centres for $\ket{\mathcal{W}^R_{\mathrm{exc}, c/v}}$ are therefore $\frac{\gamma_{\mathrm{exc}, c/v}}{2\pi} + R$. What do these two sets of exciton Wannier centres physically mean? They are the mean electron (hole) position within the exciton Wannier states that maximally localise the electron (hole) component of the exciton $\ket{\mathcal{W}^R_{\mathrm{exc}, c}}$ ($\ket{\mathcal{W}^R_{\mathrm{exc}, v}}$)

Our previous discussion (Sec.\ref{sec:FirstExcitonBerryPhaseSec}) found an infinite family of possible exciton Berry connections. We would like to understand how the exciton Wilson loops we have derived are related to this family. By taking the thermodynamic limit 
($L\rightarrow\infty$) of the two exciton Wilson loops we can find expressions for the associated exciton Berry phases (since $\lim_{L\rightarrow \infty}W_{\mathrm{exc}, c/v}=e^{\mathrm{i} \gamma_{\mathrm{exc}, c/v}}$). The two exciton Berry phases $\gamma_{\mathrm{exc}, c/v}$ are integrals over the two Berry connections (derived in Appendix~\ref{sec:ApxExcitonBerryConnections}),
\begin{align}
A_{\mathrm{exc}, c}(p) &= \int_{0}^{2\pi}\frac{\mathrm{d}k}{\Delta}\:\left[ \bar \phi^{p}_{k} \mathrm{i} \partial_p \phi^{p}_{k} +|\phi^{p}_{k}|^2 A_{\mathrm{elec}, c}(p+k)\right],\label{eq:BerryConnectionElectron}\\
A_{\mathrm{exc}, v}(p) &= \int_{0}^{2\pi}\frac{\mathrm{d}k}{\Delta}\:\left[ \bar \phi^{p}_{k - p} \mathrm{i} \partial_p \phi^{p}_{k - p} +|\phi^{p}_{k}|^2 A_{\mathrm{elec}, v}(k)\right].\label{eq:BerryConnectionHole}
\end{align}
Comparing with Eq.~\eqref{eq:naiveBerryConnection}, we can see immediately that $A_{\mathrm{exc}, c}(p) = A^{\alpha = 0}_{\mathrm{exc}}(p) $ and $A_{\mathrm{exc}, v}(p) = A^{\alpha = 1}_{\mathrm{exc}}(p)$. Therefore, at least the $\alpha = 0$ and $\alpha = 1$ Berry connections introduced in Eq.~\eqref{eq:naiveBerryConnection} \emph{do} have different physical interpretations. The Berry phases calculated using these Berry connections give the centres of the two sets of possible Wannier states: they give the electron (hole) position within the set of exciton Wannier states that maximally localise the electron (hole) component of the exciton $\ket{\mathcal{W}^R_{\mathrm{exc}, c}}$ ($\ket{\mathcal{W}^R_{\mathrm{exc}, v}}$).

We can extend these results to find the physical meaning of the Berry connections in Eq.~\eqref{eq:naiveBerryConnection} beyond $\alpha = 0, 1$. First we note that any $A^{\alpha}_{\mathrm{exc}}(p)$ can be written as a weighted sum of the two Berry connections $A^{\alpha = 0, 1}_{\mathrm{exc}}(p)$ as, 
\begin{equation}
\begin{aligned}
A^{\alpha}_\mathrm{exc}(p) =& \alpha A_{\mathrm{exc}, v}(p) + (1-\alpha)  A_{\mathrm{exc}, c}(p).\label{eq:decompInTermsofAlpha}
\end{aligned}
\end{equation} 
Hence the Berry phase calculated from $A^{\alpha}_\mathrm{exc}(p)$ corresponds to the shift of some weighted sum of the electron and hole positions $R_\alpha = \alpha r_{\mathrm{hole}}+ (1-\alpha) r_{\mathrm{elec}}$ with respect to the lattice. Of particular interest for example is the centre of mass of the exciton \emph{i.e.} $\alpha = 1/2$. This gives exactly the Berry connection that has been chiefly studied in previous works~\cite{ExcitontopologySlager, ExcitonTopologyPaperKwan} which relates to the average shift of the electron and the hole with respect to the unit cell origin. 

We see, therefore, that the entire family of exciton Berry phases can be expressed in terms of the two connections we derived $A_{\mathrm{exc}, c/v}(p)$. But when are these Berry connections different and does their difference have a physical interpretation? The difference between the Berry connections is
\begin{align}
&\mathcal{F}(p) = A_{\mathrm{exc}, c}(p) - A_{\mathrm{exc}, v}(p)~\label{eq:Fofp}\\
&= \int_0^{2\pi} \frac{\mathrm{d}k}{\Delta}\:\left\{\bar\phi^p_k \mathrm{i} \partial_k  \phi^p_k + |\phi^p_k|^2 \left[A_{\mathrm{elec},c}(p+k)  - A_{\mathrm{elec},v}(k)\right]\right\}\nonumber.
\end{align}
This is a gauge \emph{invariant} quantity under gauge transformations of both the electronic and exciton wave functions. It must therefore have a physical interpretation at each momentum $p$. Indeed, the difference $\mathcal{F}(p)$ is equal to the expectation value of the electron-hole separation, that is, the \emph{electric polarisation} for the exciton at momentum $p$ (see Appendix~\ref{sec:ApxElectronicPolarisationExcitons})~\cite{paiva2024shiftpolarizationexcitonsquantum, QuantumGeometricDipole}. In appendix~\ref{sec:ApxElectronicPolarisationExcitons} we also show that can rewrite $\mathcal{F}(p)$ so that it can be calculated in the absence of a smooth gauge for the electronic Bloch vectors,
\begin{equation}
e^{\mathrm{i}\Delta \mathcal{F}(p)} =\sum_k \bar \phi^p_{k+\Delta} \phi^p_{k}\braket{u_{p+k+\Delta,c}| u_{p+k, c}} \braket{u_{k,v}| u_{k+\Delta, v}}\label{eq:discreteVersionFofP}.
\end{equation} 
We show higher dimensional generalisations of these expressions in Appendix~\ref{apx:PolarisationHigherDim}. 

Using this physical understanding of $\mathcal{F}(p)$ we can immediately understand when the two sets of exciton Wannier states in Eq.~\eqref{eq:ExcitonWannierStates} are identical. If the electron and the hole remain the same distance apart at all momenta then it is intuitive that maximally localising the hole is equivalent to maximally localising the electron and hence the two sets of exciton Wannier states are equal. In Appendix~\ref{sec:ApxExcitonWannierStates} we show that this condition [$\mathcal{F}(p) = \mathrm{const}.$] is the necessary and sufficient condition for the exciton Wannier states to be equal. In general however the electron and hole distance \emph{does} vary as a function of momentum and hence the two sets of exciton Wannier states differ (see Fig~\ref{fig:Fig1Intro}c).

Although the above expressions have been derived for 1D systems, the results readily extend to higher dimensions. See Appendices~\ref{Sec:ApxExcitonProjPosition},~\ref{sec:ApxExcitonBerryConnections}, and ~\ref{sec:ApxElectronicPolarisationExcitons} for explicit expressions in higher dimensions. Higher dimensional invariants, including the Chern number, mirror Chern number, $\mathbb{Z}_2$ invariant protected by TRS as well as Euler/Stiefel Whitney invariants, can be expressed in terms of 1D Wilson loops. For example, the winding of the 1D Wilson loop spectrum gives the Chern number~\cite{Neupert_2018}. Using our formalism, we can immediately determine whether the exciton Chern number is well defined \emph{i.e.} is it equal regardless of which of the two exciton Berry connections we use? In 2D we can define a family of (now vector-valued) Berry connections $\boldsymbol{A}^{\alpha}_{\mathrm{exc}}(\boldsymbol{p})$ (see Appendix~\ref{sec:ApxExcitonBerryConnections}). The Chern number can be calculated by integrating the curl of the Berry connection for a given $\alpha$, $C_{\mathrm{exc}, \alpha} = \frac{1}{2\pi}\int_{BZ} \boldsymbol{dS}\cdot \nabla_{\boldsymbol{p}} \times \boldsymbol{A}^{\alpha}_{\mathrm{exc}}(\boldsymbol{p})$. Since the decomposition in Eq.~\eqref{eq:decompInTermsofAlpha} also holds in higher dimensions (see Appendix~\ref{sec:ApxExcitonBerryConnections}), it follows that the Chern number can be written as,
\begin{equation}
C_{\mathrm{exc}, \alpha} = \alpha C_{\mathrm{exc}, v} +(1-\alpha) C_{\mathrm{exc}, c},
\end{equation}
where $C_{\mathrm{exc}, c/v} = \frac{1}{2\pi}\int_{BZ} \boldsymbol{dS}\cdot \nabla_{\boldsymbol{p}} \times \boldsymbol{A}_{\mathrm{exc},  c/v}(\boldsymbol{p})$. Since all the Berry connections labelled by $\alpha$ are valid Berry connections therefore $C_{\mathrm{exc}, \alpha}$ must be an integer for all $\alpha$, similarly $C_{\mathrm{exc}, c/v}$ are also integers. We can smoothly change $\alpha$ and yet $C_{\mathrm{exc}, \alpha}$ must remain an integer. This is only true if  the two Chern numbers $C_{\mathrm{exc}, c/v}$ are equal. Therefore all the Chern numbers ($C_{\mathrm{exc}, \alpha}$) calculated using any of the connections must be equal. We note that all previous papers have used the $\boldsymbol{A}^{\alpha = 1/2}_{\mathrm{exc}}(\boldsymbol{p}) = \frac{1}{2} \left[\boldsymbol{A}_{\mathrm{exc}, c}(\boldsymbol{p}) + \boldsymbol{A}_{\mathrm{exc}, v}(\boldsymbol{p})\right]$ Berry connection to define the exciton Chern number (\emph{e.g.} Refs.~\cite{ExcitonTopologyPaperKwan, ExcitonTopologyPaperTitus, topExcitonsMoire}). We have  demonstrated however that all other $\alpha$ choices would also give the same Chern number although the Berry curvature distribution is in general not the same.


We now have a clear physical understanding of the various exciton Berry connections that can be defined. Since all of them are weighted combinations of the electron and hole contributions [\emph{i.e.} the $A_{\mathrm{exc}, c}(p)$, $A_{\mathrm{exc}, v}(p)$] we can restrict our analysis to these two connections, which together describe the whole family of Berry connections. In the remainder of this paper, we examine the exciton Berry phases and Wannier states in the presence of inversion symmetry and $C_2 \mathcal{T}$ symmetry. These exhaust the physically realisable symmetries that quantise exciton Berry phases in 1D. Lastly we consider the physical information contained in the exciton Berry phases when there are no crystalline symmetries.

\section{Crystalline Symmetries}
In the tenfold way classification of topological insulators, the only symmetries that give rise to nontrivial topological phases in 1D are particle-hole and chiral symmetry~\cite{Ryu_2010, Kitaev_2009}. Both these symmetries invert the energy eigenvalue $E \rightarrow -E$ of the state they act on. While this condition may be physical in mean-field superconductors and certain insulators with sublattice symmetry, it cannot be realised on excitonic spectra: the exciton dispersion relation is bounded from below by the energy of the many-body ground state $\ket{\mathrm{GS}}$, and from above by the electron-hole continuum; it is therefore inherently asymmetric. To stabilise nontrivial exciton topology in 1D, and in particular a nonzero and quantised exciton Berry phase, we need to involve crystalline symmetries. There are two inequivalent choices that both reflect a spatial coordinate $x \rightarrow -x$ about the centre of the unit cell: inversion symmetry $\mathcal{I}$, which is unitary, and spinless $C_2\mathcal{T}$ symmetry, which amounts to an inversion operation followed by time-reversal and is therefore anti-unitary. (To conform with previous literature, we use spinless $C_2\mathcal{T}$ here instead of $\mathcal{I}\mathcal{T}$, these are mathematically equivalent in 1D.)

\subsection{Inversion symmetry}
Inversion symmetry ($\mathcal{I}$) quantises the Berry phase of non-interacting \emph{electronic} gapped bands to $0, \pi$~\cite{InversionWilsonLoop}. Correspondingly, the electronic Wannier centres are either at the origin (0) or at the edge of the unit cells (1/2), assuming a convention where all atomic orbitals are located at the unit cell origin. For electronic bands we refer to the latter case as an \emph{obstructed} atomic insulator, since the electronic Wannier states are obstructed from sitting on the atomic sites at the centre of the unit cells~\cite{AndreiTQC17}. Bands with distinct Berry phases can be differentiated using the symmetry indicators of $\mathcal{I}$ symmetry~\cite{AndreiTQC17, Ashvin230SINatComm17, RobertJanPRX17, InversionWilsonLoop}. Since inversion maps $p \rightarrow -p$, the Bloch states at the inversion symmetric momenta $p = 0, \pi$ are eigenstates of inversion. The Berry phase of the band can be diagnosed by the inversion eigenvalues. In this section we show that this remains true for exciton bands: the inversion eigenvalues of the many-body exciton eigenstates [Eq.~\eqref{eq:excitonEigenstate}] at the high-symmetry points of the BZ can be used to diagnose the exciton Berry phases. 

It is first important to determine whether the two exciton Berry connections we derived above give the same (quantised) exciton Berry phase in the presence of inversion symmetry. It can be intuitively shown that this is true. Consider an exciton with momentum $p$, under inversion the momentum changes sign $p\rightarrow -p$ and the exciton polarisation $\mathcal{F}(p)$ also changes sign such that inversion symmetry implies $\mathcal{F}(p) = -\mathcal{F}(-p)$. We saw before that the difference between the two exciton Berry phases is equal to the integral of $\mathcal{F}(p)$ across the BZ~[Eq.~\eqref{eq:Fofp}]. But $\mathcal{F}(p) = -\mathcal{F}(-p)$ means that this integral vanishes. Therefore, in inversion symmetric systems, the two exciton Berry phases must be equal. It follows that, although the exciton Wannier \emph{states} that maximally localise the electron or the hole \emph{do not have to be equal}, they must have the same Wannier centres since these are determined by the exciton Berry phases.

We can relate this \emph{unique}, and quantised, exciton Berry phase to the inversion eigenvalues of the many-body exciton eigenstate [$\lambda_I^{\mathrm{exc}}(\bar p)$] at the high-symmetry momenta $\bar p \in \{0, \pi\}$. 
In terms of the exciton Wilson loops defined in Eqs.~\eqref{eq:WilsonloopElec}, \eqref{eq:WilsonloopHole} and in the thermodynamic limit $L \rightarrow \infty$ we find (Appendix~\ref{sec:ApxSymmetries}), 
\begin{align}
W_{\mathrm{exc}, c/v} &= e^{\mathrm{i} \int_0^{2\pi} A_{\mathrm{exc, c/v}}(p) \mathrm{d} p} = \lambda_I^{\mathrm{exc}}(0)\lambda_I^{\mathrm{exc}}(\pi).
\end{align}
We tabulate all possibilities in Tab.~\ref{tab:inversioneigenvalues}. 
\begin{table}[t]
\centering
\caption{The dependence of the exciton Wilson loops ($W_{\mathrm{exc}, c/v}$) on the non-interacting band Wannier state centres $x_{c/v}$ and the product of the exciton inversion eigenvalues at the high-symmetry points $\lambda_I^{\mathrm{exc}}(\bar p)$. The $s_{\mathrm{exc}, c/v}$ denote the shift of the electron/hole in the exciton Wannier state from the corresponding \emph{non-interacting} electron/hole Wannier centre. The grey highlighted rows indicate the \emph{shift} excitons of Ref.~\onlinecite{ExcitonTopologyDavenport}.}
\vspace{5pt}
\label{tab:inversioneigenvalues}
\begin{tabular}{cc| c |cc |cc}
\hline\hline
&&&&&\\[-1.6ex]
$\;\;x_{c}\;\;$ & $\;\;x_{v}\;\;$ & $\;\;\prod_{\bar p}\lambda_I^{\mathrm{exc}}(\bar p)\;\;$ & $\;W_{\mathrm{exc}, c/v}\;$ & $\;\gamma_{\mathrm{exc}, c}/2\pi\;$ & $\;s_{\mathrm{exc}, c}\;$ & $\;s_{\mathrm{exc}, v}\;$\\[1ex]
\hline
\multirow{2}{*}{0} & \multirow{2}{*}{0} & 1 & 1 & 0 & 0 & 0  \\
 &  & \cellcolor{gray!20}\textbf{-1} & \cellcolor{gray!20}\textbf{-1} & \cellcolor{gray!20}$\mathbf{1/2}$ & \cellcolor{gray!20}$\mathbf{1/2}$ & \cellcolor{gray!20}$\mathbf{1/2}$\\
\hline
\multirow{2}{*}{$1/2$} & \multirow{2}{*}{0} & 1 & 1 & 0 & $1/2$ & 0 \\
 &  & -1 & -1 & $1/2$ & 0 & $1/2$ \\
\hline
\multirow{2}{*}{0} & \multirow{2}{*}{$1/2$} & 1 & 1 & 0 & 0 & $1/2$ \\
 &  & -1 & -1 & $1/2$ & $1/2$ & 0 \\
\hline
\multirow{2}{*}{$1/2$} & \multirow{2}{*}{$1/2$} &\cellcolor{gray!20}\textbf{1} & \cellcolor{gray!20}\textbf{1} & \cellcolor{gray!20}\textbf{0} & \cellcolor{gray!20}$\mathbf{1/2}$ & \cellcolor{gray!20}$\mathbf{1/2}$\\
& & -1 & -1 & $1/2$ & 0 & 0 \\
\hline
\hline
\end{tabular}
\end{table}
We show all possible combinations of the \emph{electronic} Wannier centres for the conduction and valence bands ($x_{c/v}$) and exciton band symmetry indicators. We give the Wilson loop values and their relation to the shift of the \emph{exciton} Wannier centres from the centre of the unit cell  ($\gamma_{\mathrm{exc}, c/v}/2\pi$). In particular, we are interested in how interactions can induce exciton topology \emph{beyond} that of the non-interacting bands. Therefore, we also tabulate the shift of the electrons and holes in the exciton Wannier centres with respect to the corresponding non-interacting electron/hole Wannier centres. This shift we denote as $s_\mathrm{exc, c/v}  = \gamma_{\mathrm{exc}, c/v}/2\pi - x_{c/v}$~\cite{ExcitonTopologyDavenport}. We can see that the exciton symmetry indicators, in conjunction with the electronic band symmetry indicators, can be used to identify shift excitons unambiguously (see highlighted rows in Tab~\ref{tab:inversioneigenvalues}).

Tab.~\ref{tab:inversioneigenvalues} also shows two other interesting scenarios, excitons in electronic bands where $x_c = 1/2, x_v = 0$ or vice versa. We observed that, in inversion symmetric systems, the electron/hole maximally localised exciton Wannier centres are equal. Hence when $x_c \neq x_v$, the shift of the electrons/holes (within the exciton) \emph{with respect to their corresponding electronic Wannier states} ($s_{exc, c/v}$) must differ for the holes and the electrons. These excitons occupy a middle ground between trivial and shift excitons where one, but not both, of the electrons or holes are shifted with respect to their corresponding electronic Wannier centres. This occurs in all inversion-symmetric systems where the electronic Wannier centres of the valence and conduction bands differ, and the exciton band is fully separated from the continuum. Physically this means that the underlying electronic bands can consist of one trivial atomic band and one obstructed atomic band, yet \emph{both} the electron and hole components within the exciton band are obstructed from the centre of the unit cell.

We have seen that inversion symmetry quantises the unique exciton Berry phase and that its value can be directly obtained from the inversion eigenvalues of the many-body exciton eigenstate. However, Berry phases can also be quantized by anti-unitary symmetries which do not allow for symmetry indicators~\cite{hwang2025stablerealspaceinvariantstopology}. Next, we demonstrate how exciton band topology, including shift excitons, can be still be identified in such cases.

\subsection{$C_2\mathcal{T}$ symmetry}
Spinless $C_2 \mathcal{T}$ symmetry combines the unitary symmetry of a spinless $C_2$ rotation about some axis perpendicular to the 1D chain, which squares to $C_2^2 = 1$ (equivalent to inversion in 1D), with the anti-unitary spinless time-reversal symmetry $\mathcal{T}$ that squares to $\mathcal{T}^2 = 1$. While Berry phases are quantised by $C_2\mathcal{T}$ symmetry just like by inversion, anti-unitary symmetries lack an eigenspectrum~\cite{hwang2025stablerealspaceinvariantstopology}. We therefore cannot use symmetry indicators to calculate the Berry phase from $C_2\mathcal{T}$ eigenvalues. 

As with inversion symmetry, we first want to determine whether the two possible exciton Berry phases are equal in the presence of $C_2\mathcal{T}$ symmetry. We saw in Eq.~\eqref{eq:Fofp} that this was true when the integral of $\mathcal{F}(p)$ over the BZ vanished. We will show that this is true by finding how $C_2 \mathcal{T}$ symmetry constrains the exciton polarisation $\mathcal{F}(p)$. We consider how the exciton polarisation/electron-hole separation transforms under $C_2\mathcal{T}$. First we apply time-reversal to an exciton at momentum $p$. This maps the exciton momentum from $p$ to $-p$ but has no effect on the polarisation. Next we apply $C_2$, this transforms the exciton momentum back to the original momentum $p$. In addition, the exciton polarisation $\mathcal{F}(p)$ changes sign since we are inverting the exciton. Therefore, under the combined $C_2 \mathcal{T}$ we have mapped $\mathcal{F}(p)\rightarrow -\mathcal{F}(p)$. In $C_2 \mathcal{T}$ symmetric systems, the exciton wave function must be invariant under this transformation. The electron and hole must therefore have an average separation of 0 [$\mathcal{F}(p) = 0$] for all momenta. It follows that the two exciton Berry phases must be identical. This also has an important consequence for the exciton Wannier states. We saw that the two sets of exciton Wannier states were equal if and only if the electron and hole had the same (average) separation for all momenta. Since $C_2\mathcal{T}$ forces $\mathcal{F}(p) = 0$ for all momenta, the exciton Wannier states that maximally localise the electron vs. the hole must be equal (\emph{i.e.} $\ket{\mathcal{W}^R_{\mathrm{exc}, c}} = \ket{\mathcal{W}^R_{\mathrm{exc}, v}}$).

\begin{figure}
    \centering
    \includegraphics[width=1.0\linewidth]{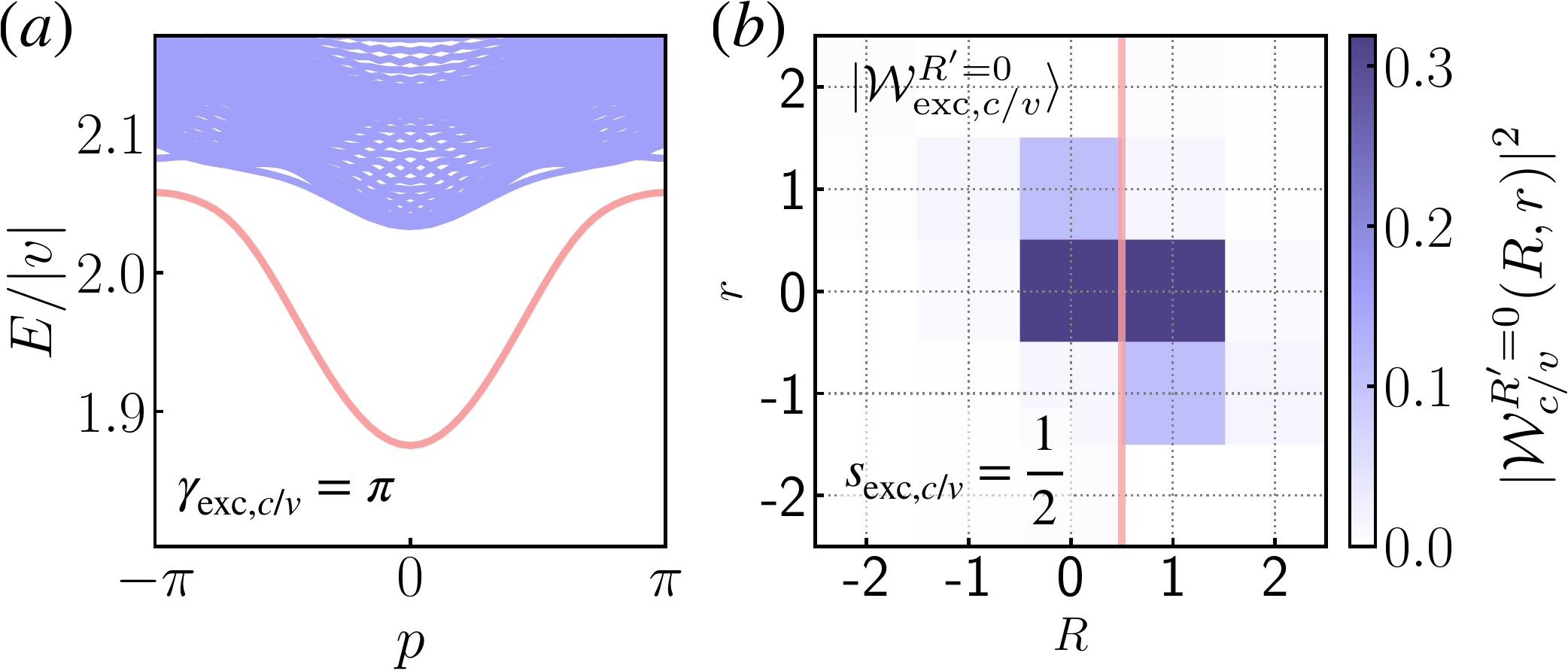}
    \caption{
    Excitons in $C_2\mathcal{T}$ symmetric systems. 
    $(a)$ Exciton dispersion relation for a simple $C_2\mathcal{T}$ symmetric system  (see Hamiltonian in Eq.~\ref{eq:C2THamiltonian}). The shift exciton band is highlighted in red. The hopping parameters are $v = 0.6 + 0.4\,\mathrm{i} $, $w = 0.03$, and $t = 0.01$. The interaction parameters are $U = 0.1, U' = 0.15$. $(c)$ The components of the corresponding \emph{exciton} Wannier function in the basis of the \emph{electronic} Wannier functions \emph{i.e.} $\ket{\mathcal{W}^{R'=0}_{\mathrm{exc}, c}} =\ket{\mathcal{W}^{R'=0}_{\mathrm{exc}, v}}  = \sum_{R, r} \mathcal{W}^{R' = 0}_{c/v}({R, r})\: c^\dagger_{R+r, c} c_{R, v}\ket{\mathrm{GS}}$. The red line marks the exciton Wannier centre ($s_{\mathrm{exc}, c/v} = 1/2$), corresponding to an exciton Berry phase of $\gamma_{\mathrm{exc}, c/v}  = \pi$.}
    \label{fig:C2TSymmetry}
\end{figure}
As a result $C_2\mathcal{T}$ gives one unique exciton Berry phase that is quantised to $0, \pi$. Both the electron- and hole-localising exciton Wannier centres are equal and the quantisation of the exciton Berry phases mean that they must sit at either the edge or the centre of the unit cell. It follows that shift excitons are topologically protected in $C_2 \mathcal{T}$ symmetric systems, even though they are not diagnosable by symmetry indicators~\cite{ExcitontopologySlager, thompson2025topologicallyenhancedexcitontransport}. Fig.~\ref{fig:C2TSymmetry} illustrates this situation in a simple model. The non-interacting component of the Hamiltonian is a variant of the Su-Schrieffer-Heeger (SSH) model that breaks $\mathcal{I}$ symmetry whilst retaining $C_2 \mathcal{T}$. The interactions consist of the intraunit and interunit cell Hubbard interactions between neighbouring sites (with strengths $U$ and $U'$). These are the simplest interaction terms that are allowed in this spinless model. The full model has the Hamiltonian
\begin{align}
\hat H &=\sum_{R} \big( v \,c^\dagger_{R,A} c^{\vphantom{\dagger}}_{R, B} +  v^* \, c^\dagger_{R,B} c^{\vphantom{\dagger}}_{R, A}\big)\label{eq:C2THamiltonian}  \\&\quad+w\sum_{R} \big( \,c^\dagger_{R, B} c^{\vphantom{\dagger}}_{R+1, A} +   \, c^\dagger_{R+1,A} c^{\vphantom{\dagger}}_{R, B}\big)\nonumber\\ &\quad+t\;\sum_{R, i\neq j}\big( c^\dagger_{R, i} c^{\vphantom{\dagger}}_{R+2, j}+c^\dagger_{R+2, j} c^{\vphantom{\dagger}}_{R, i} \big) \nonumber\\ &\quad+ \:U \sum_{R} n_{R, A} n_{R, B} + U'\sum_{R} n_{R, B} n_{R+1, A} \nonumber,
\end{align}
where $i$ sum over the sites $A, B$ in the unit cell. The number operator for unit cell $R$ and site $i$ is $n_{R, i} = c^\dagger_{R, i}c_{R, i}$. 
This model is $\mathcal{I}$ symmetric when $v$ is real but breaks $\mathcal{I}$ symmetry when $v$ is complex. In contrast, for all $v$ it is $C_2 \mathcal{T}$ symmetric.  

Shift excitons have an exciton Berry phase of $\pi$ despite the underlying electronic bands being trivial (with \emph{electronic} Berry phase $0$). In Fig.~\ref{fig:C2TSymmetry}a we show the exciton band structure of the above model. The electronic bands are trivial and yet the lowest energy exciton band has exciton Berry phase $\pi$. Hence the exciton Wannier centres are shifted to the edge of the unit cells (Fig.~\ref{fig:C2TSymmetry}b) despite the electronic Wannier centres being in middle of the unit cell. The exciton Wannier centres can be diagnosed using the two Wilson loops [Eqs.~\eqref{eq:WilsonloopElec} \& \eqref{eq:WilsonloopHole}]. These both evaluate to $W_{\mathrm{exc}, c/v} = -1$ consistent with exciton Wannier centres of $\frac{1}{2} + R$.

This example demonstrates that shift excitons can exist in systems that break inversion symmetry but retain $C_2 \mathcal{T}$ symmetry. As symmetry indicators are not available, shift excitons with $C_2 \mathcal{T}$ symmetry are uniquely characterised by an exciton Berry phase that differs from the non-interacting electronic Berry phase. Despite the absence of inversion symmetry, the spatial shift of the exciton Wannier centres will have the same physical consequences as discussed in Ref.~\onlinecite{ExcitonTopologyDavenport}. In particular, shift excitons can stabilise \emph{exciton} edge states in open boundary conditions (OBC). Therefore, in OBC, excitons can be excited at energies within the bulk gap between the exciton band and the particle-hole continuum. Crucially, these excitations are localised to the system's boundaries —they can be excited at the edges but not in the bulk — and can be detected experimentally using local optical conductivity measurements~\cite{ExcitonTopologyDavenport}.

We have seen that in the presence of $\mathcal{I}$ or $C_2\mathcal{T}$ symmetry, the two exciton Berry phases (localising electrons or holes within the exciton Wannier state, respectively) have to be equal. This raises the question: in 1D, is it ever possible for the two exciton Berry phases, and therefore the two sets of exciton Wannier centres, to be different? We answer this question in the affirmative in the next section. 

\section{No Crystalline Symmetries}
\label{sec:NoSymmetries}
\begin{figure}
    \centering
    \includegraphics[width=1\linewidth]{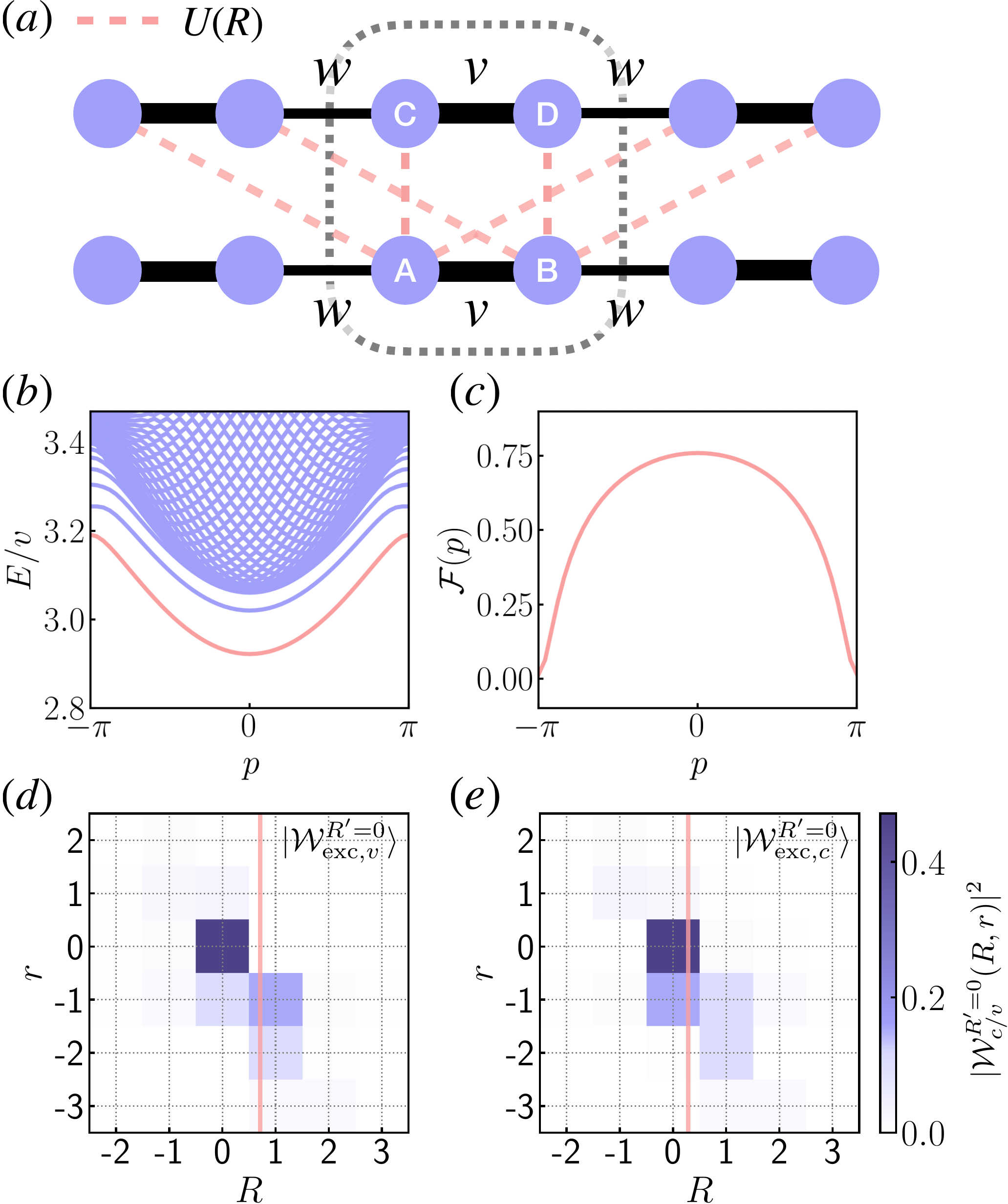}
    \caption{Excitons in a system without $\mathcal{I}$ and $C_2 \mathcal{T}$ symmetry. $(a)$ is the tight-binding model which consists of two stacked SSH models, both with hoppings $v = 1.0, w=0.2$. In addition the SSH model constructed from the sites $C, D$ has a uniform on-site potential $V=4.5$. At the single-particle level, the two SSH models are uncoupled. The Hubbard interaction $U(R)$ between the unit cells separated by distance $R$ couples them. We consider excitons at half filling. See Appendix~\ref{sec:ApxNoSymmetries} for the full model specification. $(b)$ The exciton band structure. The red band highlights the exciton band  studied in the remaining panels. $(c)$ The electronic \emph{polarisation} of the exciton [$\mathcal{F}(p)$]. $(d), (e)$ are the spatial profiles of the maximally localised exciton Wannier functions in the basis of the \emph{electronic} Wannier functions \emph{i.e.} $\ket{\mathcal{W}^{R'=0}_{\mathrm{exc}, c/v}}  = \sum_{R, r} \mathcal{W}^{R' = 0}_{c/v}({R, r})\: c^\dagger_{R+r, c} c_{R, v}\ket{\mathrm{GS}}$. $(d)$ is the hole maximally localised Wannier function $\ket{\mathcal{W}_{\mathrm{exc}, v}^{R'=0}}$, $(e)$ is the electron maximally localised Wannier function $\ket{\mathcal{W}_{\mathrm{exc}, c}^{R'=0}}$ . The Wannier centres extracted from the Wilson loops $W_{\mathrm{exc}, c/v}$ are indicated by a red line.} 
    \label{fig:NoSymmetries}
\end{figure}
For the two exciton Wilson loops in Eqs.~\eqref{eq:WilsonloopElec},~\eqref{eq:WilsonloopHole} to differ, we must break both $\mathcal{I}$ and $C_2\mathcal{T}$ symmetries. We now introduce a simple tight-binding model that illustrates the different physical information contained in the two Wilson loops and exciton Wannier states when these symmetries are absent. We consider the model shown schematically in Fig.~\ref{fig:NoSymmetries}a. This is a model with two identical stacked SSH chains that at the non-interacting level are uncoupled. We add uniform (on-site) potentials to one of the SSH models and study it half filling. We need to break inversion symmetry - we do this at the level of the interactions by adding a Hubbard interaction,
\begin{equation}
\hat H_{\mathrm{int}} = \sum_{\substack{R, R', \\i\in\{A, B\}, j\in\{C, D\}}} U(R - R') n_{R, i} n_{R', j},
\end{equation}
where $U(R - R')$ breaks inversion symmetry \emph{i.e.} $U(R)\neq U(-R)$. We choose an interaction that exponentially decays in $|R|$ but with different rates for $R>0$ compared to $R< 0$. In particular, we choose $U(|R|) > U(-|R|)$ \emph{i.e.} the electron is more attracted to the hole when it is to the right of the hole. [See Appendix~\ref{sec:ApxNoSymmetries} for the full specification of the model]. At half filling we calculate the exciton spectrum shown in Fig.~\ref{fig:NoSymmetries}b and we study the lowest energy exciton band. The electric polarisation of the excitons  $\mathcal{F}(p)$ (see Fig.~\ref{fig:NoSymmetries}c) varies significantly as a function of momentum. The electron is to the right of the hole at all momenta giving a positive polarisation because the interaction $U(R)$ is largest at positive $R$. Therefore the exciton is more tightly bound when the electron is to the right of the hole. However the competing effects of the kinetic energy mean that the electron and the hole localise at roughly the same real space positions around $p=\pi$.

Importantly, unlike for $\mathcal{I}$ and $C_2 \mathcal{T}$ symmetric systems, the function $\mathcal{F}(p)$ now does not integrate to 0 over the BZ. This means that the two exciton Berry phases ($\gamma_{\mathrm{exc}, c/v}$) differ. Since the exciton Berry phases determine the two sets of exciton Wannier centres, this means that these also differ. Furthermore in this general case, since $\mathcal{F}(p)$ is not constant over the BZ, the two sets of exciton Wannier functions ($\ket{\mathcal{W}^R_{\mathrm{exc}, c/v}}$) are not equal; the profile of $\ket{\mathcal{W}^R_{\mathrm{exc}, v}}$ is plotted in Fig.~\ref{fig:NoSymmetries}d and $\ket{\mathcal{W}^R_{\mathrm{exc}, c}}$ in Fig.~\ref{fig:NoSymmetries}e. Both sets of exciton Wannier states have the majority of their weight on on-site excitations \emph{i.e.} excitations between the hole and electron Wannier states in the same unit cell. This is because the interaction decays exponentially with distance and so the lowest energy exciton is tightly bound. The structure of the two sets of exciton Wannier states do differ however as a result of $\mathcal{F}(p)$. 
The associated Wilson loop gives a Berry phase $\gamma_{\mathrm{exc}, c}\sim 1.80 \;\mathrm{rad}$ leading to an exciton Wannier centre of $0.29$ (in units where the lattice constant is $a = 1$). In contrast, the hole-localised Wannier state has its centre shifted within the unit cell with Berry phase $\gamma_{\mathrm{exc}, v}\sim -1.80\;\mathrm{rad}$  and Wannier centre $0.71$. We see therefore that an electron-hole separation that varies with momentum leads to differences in the two sets of exciton Wannier centres.

To conclude, even when $\mathcal{I}$ and $C_2\mathcal{T}$ symmetries are broken and the exciton Wilson loops are not quantised, the exciton Berry phases and Wannier states give important physical information about the nature of the exciton bands. We can calculate both the electron- and hole-localised exciton Wannier centres, as well as the real space distance between the electron and hole, directly from the Wilson loops in Eqs.~\eqref{eq:WilsonloopElec},~\eqref{eq:WilsonloopHole} and the expressions for $\mathcal{F}(p)$ in Eqs.~\eqref{eq:Fofp},~\eqref{eq:discreteVersionFofP}.

\section{Discussion}
We have introduced a family of different Berry phases for excitons as well as their associated gauge invariant Wilson loops. We have elucidated the physical meaning of these Berry phases in terms of two sets of exciton Wannier functions - one set maximally localises the electron and another the hole. We have shown that the two sets of exciton Wannier centres must be equal in $\mathcal{I}$ and $C_2 \mathcal{T}$ symmetric systems. The two exciton Berry connections derived from the Berry phases differ by a gauge invariant quantity $\mathcal{F}(p)$, the electric polarisation of the exciton at a given total momentum. This polarisation is the mean separation of the electron and hole within the exciton wave function. Our work clarifies that, despite the existence of an infinite family of exciton Berry phases, the entire family can be constructed from just two Wilson loops Eqs.~\eqref{eq:WilsonloopElec}, \eqref{eq:WilsonloopHole} or equivalently the two exciton Berry connections Eqs.~\eqref{eq:BerryConnectionElectron}, \eqref{eq:BerryConnectionHole}. 
Furthermore, we showed how both symmetry indicators and explicit calculations of the Wilson loops can be used to diagnose exciton topology in 1D, including the shift excitons introduced in Ref.~\onlinecite{ExcitonTopologyDavenport}. 

We gained an explicit expression for the electronic polarisation of the excitons $\mathcal{F}(p)$. The expression shows that it depends on the exciton wave function and the electronic Bloch functions. A promising direction for future research is to fully explore how symmetries as well as exciton and electronic band topology constrain $\mathcal{F}(p)$. Already in 1D we saw that symmetries constrained $\mathcal{F}(p)$ significantly. Understanding the exciton polarisation is of importance for a variety of applications. For example the electron-hole separation may have important consequences for the shift current in photocells. Shift currents occur when an electron is photo-excited and its spatial position changes. This leads to a current. It is known that, at the non-interacting level, the shift current depends on the electronic band topology~\cite{shiftcurrent}. However, there could be additional excitonic contributions which could be explored through the study of $\mathcal{F}(p)$~\cite{shiftcurrentExcitons1, shiftcurrentExcitons2}.

Although we have explored all the physically reasonable symmetries  that quantise the exciton Berry phase in 1D, there remain foundational questions in higher dimensions where other topological invariants (such as the Chern number and $\mathbb{Z}_2$ index) may be generalisable to excitons. A particularly promising research direction involves understanding how single-electron topology (characterized by quantities like the Chern number) relates to and influences exciton topology. We have started to explore this in a companion paper in which we introduce rhombohedral graphene as a tunable platform for exploring the interplay of electronic and exciton topology~\cite{upcoming}. Understanding the exciton Berry phase therefore paves the way for a full topological classification of exciton bands that would work out the roles of crystalline and local symmetries on the topological classification of exciton bands in 2 and 3 dimensions.

Previous work on exciton topology has exclusively used the $\alpha = \frac{1}{2}$ exciton Berry connection~\cite{ExcitonTopologyPaperKwan, ExcitonTopologyPaperQian, ExcitonTopologyPaperTitus, ExcitontopologySlager, paiva2024shiftpolarizationexcitonsquantum}. This is an equal weighted sum of $A_{\mathrm{exc}, c}(p)$ and $A_{\mathrm{exc}, v}(p)$. We also note that the concept of shift excitons (that we first introduce in Ref.~\cite{ExcitonTopologyDavenport}) is distinct from the shift vector defined in Ref.~\onlinecite{paiva2024shiftpolarizationexcitonsquantum}. We define exciton shift as the difference between the exciton Wannier centres and the electronic Wannier centres. The shift is therefore a property of an exciton \emph{band}, not excitons at specific total momenta $p$ (unlike the shift vector defined in Ref.~\onlinecite{paiva2024shiftpolarizationexcitonsquantum}).  We note that shift excitons have clear physical consequences, for example they exhibit a bulk-boundary correspondence~\cite{ExcitonTopologyDavenport}. From our projected position operators we also derive an expression for the polarisation, this is identical to that derived in both Refs.~\onlinecite{paiva2024shiftpolarizationexcitonsquantum, QuantumGeometricDipole}. Unlike in Ref.~\cite{paiva2024shiftpolarizationexcitonsquantum} we retain the sum over the relative momentum $k$. The sum over $k$ ensures that the electron and hole is bound in real space, the summation is therefore required for Eq.~\eqref{eq:Fofp} to represent the true exciton polarisation.

In electronic systems, the electronic Berry curvature influences the semi-classical transport properties~\cite{BerryPhaseElectronicEffects}. Therefore, we expect a similar relationship between the \emph{exciton} Berry curvature and the exciton transport properties. In Ref.~\onlinecite{ExcitonTopologyPaperQian}, it was found that the time derivative of the exciton centre (\emph{i.e.} the average position of the electron and hole) has an anomalous term related to the exciton Berry curvature. The particular exciton Berry curvature that enters these equations is calculated using $A^{\alpha = 1/2}_{\mathrm{exc}}(p)$. When we have layer polarised excitons however, one might expect some observables to instead depend on $A_{\mathrm{exc},c}(p)$ and $A_{\mathrm{exc},v}(p)$ individually. An interesting future avenue of work is therefore to explore the role that the $\alpha \neq \frac{1}{2}$ exciton Berry curvatures have on the exciton dynamics.

Finally, although our results are formulated for excitons, the methods we present here can be readily extended to other two-body excitations (\emph{e.g.} magnons or Cooper pairs). In these systems, the Berry connection ambiguity we discussed also arises; however, two distinct Wilson loops—analogous to those defined for excitons—can still be constructed. Again the difference between the corresponding Berry connections will give a gauge invariant quantity equal to the expected separation between the 2 bodies in the bound state. Our results are not limited to 2-body bound states; in the general $N$-body case, our results imply that $N$ distinct Wilson loops can be defined.
In addition, there will be $N(N-1)/2$ \emph{independent} gauge invariant quantities, analogous to $\mathcal{F}(p)$, that are equal to the distances between any two of the $N$ bodies. We therefore anticipate that our results will be valuable for interpreting the topological invariants of excitations and few-body states beyond excitons. 
\\

\begin{acknowledgments}
We are very grateful to Yoonseok Hwang for his detailed comments on the manuscript, and insights regarding the exciton Chern number. In addition we would like to thank Jonah B. Haber, Johannes Lischner, and Alessandro Romito for useful discussions. We thank Peru d'Ornellas for suggesting the title of the paper. This work was supported by a UKRI Future Leaders Fellowship MR/Y017331/1. HD acknowledges support from the Engineering and Physical Sciences Research Council (grant number EP/W524323/1).  JK acknowledges support from the Deutsche Forschungsgemeinschaft (DFG, German Research Foundation) under Germany’s Excellence Strategy–EXC– 2111–390814868, DFG grants No. KN1254/1-2, KN1254/2- 1, and TRR 360 - 492547816; as well as the Munich Quantum Valley, which is supported by the Bavarian state government with funds from the Hightech Agenda Bayern Plus. We acknowledge support from the Imperial-TUM flagship partnership.
\end{acknowledgments}

\section*{Data Availability}
The data that support the findings of this article are openly available~\cite{davenport_2025_17722211}.

\bibliography{refs}

\newpage

\onecolumngrid

\appendix
\section{Infinite family of exciton Berry connections}
\label{sec:ApxFirstExcitonBerryPhaseSec}

We begin with the exciton eigenstate introduced in the main text,
\begin{equation} 
\ket{p_{\mathrm{exc}}} = \sum_k \phi^{p}_{k - \alpha p} c^\dagger_{(1-\alpha) p+k, c} c_{-\alpha p + k, v} \ket{\mathrm{GS}}.
\label{eq:ApxexcitonEigenstate}
\end{equation}
The total momentum of the exciton is $p$. We consider 1D systems where the momenta ($p$ and $k$) take values in the range $[-\pi, \pi)$ (assuming the lattice parameter $a = 1$). For a system of $L$ unit cells (in 1D) in periodic boundary conditions, the allowed momenta are spaced by $\Delta = 2\pi/L$. We normalise the exciton eigenstate $\braket{p_{\mathrm{exc}}|p_{\mathrm{exc}}} = 1$. For a finite system, this implies $\sum_{k}|\phi^p_k|^2 = 1$. When we take the thermodynamic limit we replace $\sum_{k}\rightarrow \int_0^{2\pi} \frac{\mathrm{d}k}{\Delta}$. Hence the normalisation condition becomes $\int_0^{2\pi} \frac{\mathrm{d}k}{\Delta}|\phi^p_k|^2 = 1$.

We wish to construct the exciton Berry connection heuristically. We take the thermodynamic limit $L\rightarrow \infty$ so that momentum becomes continuous. We might first consider constructing the exciton Berry connection analogously to the electronic Berry connection [$A_{\mathrm{elec}, c}(p) = \bra{u_{p, c}}\mathrm{i}\partial_p  \ket{u_{p, c}}$], this would be
\begin{equation}
\int_{0}^{2\pi} \frac{\mathrm{d}k}{\Delta} \bar \phi^{p}_{k - \alpha p} \mathrm{i}\partial_p \phi^{p}_{k - \alpha p}.
\label{eq:naiveConnectionNo1}
\end{equation}
Note that we use bar to denote complex conjugation. As desired for a Berry connection, this is gauge covariant under 
a global gauge transformation of the exciton wave functions ($\ket{p_{\mathrm{exc}}} \rightarrow e^{\mathrm{i}\theta(p)} \ket{p_{\mathrm{exc}}}$). Under such a global gauge transformation it transforms as,
\begin{equation}
\int_{0}^{2\pi} \frac{\mathrm{d}k}{\Delta} \bar \phi^{p}_{k - \alpha p} \mathrm{i}\partial_p \phi^{p}_{k - \alpha p} \rightarrow \int_{0}^{2\pi} \frac{\mathrm{d}k}{\Delta} \bar \phi^{p}_{k - \alpha p} \mathrm{i}\partial_p \phi^{p}_{k - \alpha p} - \nabla \theta(p).
\end{equation}
Despite this, Eq.~\eqref{eq:naiveConnectionNo1} is not a valid exciton Berry connection because it is not gauge invariant under gauge transformations of the electronic bands. We can see this by considering how the exciton wave function $\phi^p_k$ transforms under gauge transformations of the single electron bands. Consider the gauge transformations of the single electron bands ($c^\dagger_{p, v}\rightarrow e^{\mathrm{i}\eta(p)}c^\dagger_{p, v}$ and $c^\dagger_{p, c}\rightarrow e^{\mathrm{i}\xi(p)}c^\dagger_{p, c}$). When we make gauge transformations of the electronic bands, then this should still leave the exciton state $\ket{p_{\mathrm{exc}}}$ invariant. Hence the exciton wave function must transform oppositely to the electronic bands, 
\begin{equation}
\phi^{p}_{k - \alpha p} \rightarrow \tilde \phi^p_{k-\alpha p} = e^{\mathrm{i}\eta [-\alpha p + k]} e^{-\mathrm{i} \xi[(1-\alpha) p + k]} \phi^{p}_{k - \alpha p}.
\end{equation} 
Under this transformation the naive Berry connection [Eq.~\eqref{eq:naiveConnectionNo1}] transforms as,
\begin{align}
\int_{0}^{2\pi} \frac{\mathrm{d}k}{\Delta} \bar \phi^{p}_{k - \alpha p} \mathrm{i}\partial_p \phi^{p}_{k - \alpha p} &\rightarrow \int_{0}^{2\pi} \frac{\mathrm{d}k}{\Delta}  (\tilde \phi^{p}_{k - \alpha p})^* \mathrm{i} \partial_p\tilde \phi^{p}_{k - \alpha p} \\
&\rightarrow\int_{0}^{2\pi} \frac{\mathrm{d}k}{\Delta} \bar \phi^{p}_{k - \alpha p} \mathrm{i}\partial_p \phi^{p}_{k - \alpha p}+ |\phi^{p}_{k - \alpha p}|^2 \left\{\partial_p \xi[(1-\alpha) p+k] - \partial_p \eta(-\alpha p + k)\right\}\\
&\rightarrow\int_{0}^{2\pi} \frac{\mathrm{d}k}{\Delta} \bar \phi^{p}_{k - \alpha p} \mathrm{i}\partial_p \phi^{p}_{k - \alpha p}+ |\phi^{p}_{k - \alpha p}|^2 \big[(1-\alpha )\partial_q \xi(q)|_{q =(1-\alpha) p+k} + \alpha \partial_q \eta(q)_{q = -\alpha p + k}\big].\label{eq:ApxNaiveBerryConnectionTransformation}
\end{align} 
Therefore the expression in Eq.~\eqref{eq:naiveConnectionNo1} is not gauge invariant (or even covariant) under gauge transformations of the electronic bands. This means it is not a valid exciton Berry connection since the resulting exciton Berry phase would change under gauge transformations of the electronic bands. The exciton Berry phase therefore cannot physical information. We wish to add correction terms to Eq.~\eqref{eq:naiveConnectionNo1} to make it gauge invariant under gauge transformations of the electronic bands. We will add terms which depend on the \emph{electronic} Berry connections,
\begin{equation}
A_{\mathrm{elec}, c/v}(k) = \bra{u_{q, c/v}} \mathrm{i}\partial_q \ket{u_{q, c/v}}|_{q = k}.
\end{equation}
Under the electronic gauge transforms described above, the conduction band Berry connection transforms as,
\begin{align}
A_{\mathrm{elec}, c}(q) &\rightarrow  \tilde A_{\mathrm{elec}, c}(q)\\
\bra{u_{q, c/v}} \mathrm{i}\partial_q \ket{u_{q, c/v}} &\rightarrow \bra{u_{q, c/v}}e^{-\mathrm{i}\xi(q)} \mathrm{i}\partial_q e^{\mathrm{i}\xi(q )}\ket{u_{q, c/v}}\\
&\rightarrow \bra{u_{q, c/v}} \mathrm{i}\partial_q\ket{u_{q, c/v}} - \partial_q \xi(q)\\
&\rightarrow A_{\mathrm{elec}, c}(q) - \partial_q \xi(q).
\end{align}
We can see that this gives terms depending on $\partial_q \xi(q)$. Likewise, the valence band Berry connection transforms as
\begin{align}
A_{\mathrm{elec}, v}(q) &\rightarrow \tilde A_{\mathrm{elec}, v}(q)\\&\rightarrow A_{\mathrm{elec}, v}(q) - \partial_q \eta(q).
\end{align}
Therefore, we can add terms which depend on $A_{\mathrm{elec}, c/v}(q)$ to cancel out the problematic final two terms in Eq.~\eqref{eq:ApxNaiveBerryConnectionTransformation}. Consider adding the following two additional terms to our first guess for the exciton Berry connection,
\begin{equation}
\begin{aligned}
A^\alpha_{\mathrm{exc}}(p) =  \int_{0}^{2\pi} \frac{\mathrm{d}k}{\Delta} \bar \phi^{p}_{k - \alpha p} \mathrm{i} \partial_p \phi^{p}_{k - \alpha p} +|\phi^{p}_{k - \alpha p}|^2 \big\{ (1-\alpha) A_{\mathrm{elec}, c}\big[(1-\alpha) p+k\big] + \alpha A_{\mathrm{elec}, v}\big[-\alpha p+k\big]\big\}.
\label{eq:ApxnaiveBerryConnection}
\end{aligned}
\end{equation}
We can rewrite this in the different electronic gauge ($c^\dagger_{p, v}\rightarrow e^{\mathrm{i}\eta(p)}c^\dagger_{p, v}$ and $c^\dagger_{p, c}\rightarrow e^{\mathrm{i}\xi(p)}c^\dagger_{p, c}$),
\begin{align}
A^\alpha_{\mathrm{exc}}(p) &=  \int_{0}^{2\pi} \frac{\mathrm{d}k}{\Delta} \bar \phi^{p}_{k - \alpha p} \mathrm{i} \partial_p \phi^{p}_{k - \alpha p} +|\phi^{p}_{k - \alpha p}|^2 \big\{ (1-\alpha) A_{\mathrm{elec}, c}\big[(1-\alpha) p+k\big] +\alpha A_{\mathrm{elec}, v}\big[-\alpha p+k\big]\big\}\\
&=  \int_{0}^{2\pi} \frac{\mathrm{d}k}{\Delta}  (\tilde \phi^{p}_{k - \alpha p})^* \mathrm{i} \partial_p\tilde \phi^{p}_{k - \alpha p} - |\tilde \phi^{p}_{k - \alpha p}|^2 \big[(1-\alpha)\partial_q \xi(q)|_{q =(1-\alpha) p+k} + \alpha \partial_q \eta(q)_{q = -\alpha p + k}\big]\\ &\quad+|\tilde \phi^{p}_{k - \alpha p}|^2  \big\{ (1-\alpha) (\tilde A_{\mathrm{elec}, c}\big[(1-\alpha) p+k\big] + \partial_q \xi(q)|_{q = (1-\alpha) p+k})\nonumber \\&\quad+ \alpha (\tilde A_{\mathrm{elec}, v}\big[-\alpha p+k\big] + \partial_q \eta(q)|_{q = {-\alpha p+ k}})\big\}\nonumber\\
&= \int_{0}^{2\pi} \frac{\mathrm{d}k}{\Delta}  (\tilde \phi^{p}_{k - \alpha p})^* \mathrm{i} \partial_p\tilde \phi^{p}_{k - \alpha p} +|\tilde \phi^{p}_{k - \alpha p}|^2 \big\{ (1-\alpha) \tilde A_{\mathrm{elec}, c}\big[(1-\alpha) p+k\big] + \alpha \tilde A_{\mathrm{elec}, v}\big[-\alpha p+k\big]\big\}\\
&=\tilde A^\alpha_{\mathrm{exc}}(p).
\end{align}
We see therefore that for all $\alpha$ the exciton Berry connection $A^\alpha_{\mathrm{exc}}(p)$ is both gauge invariant under gauge transformations of the electronic bands and gauge covariant under a global gauge transformation of the exciton wave function. For all $\alpha$ this gives a valid Berry connection. But these different exciton Berry connections aren't necessarily equal and in general they can give different values for the exciton Berry phase. In this paper we find the physical interpretation of this family of distinct Berry connections.
\subsection{Cell-periodic exciton wave functions}
\label{apd:CellPeriodicWavefunctions}
An alternative perspective on the above construction can be obtained by looking at the different cell-periodic exciton wave functions which can be extracted from the many-body exciton state $\ket{p_{\mathrm{exc}}}$. First we consider how we extract can the cell-periodic part of the \emph{non-interacting} electronic wave function. If $c^\dagger_{R, i}$ creates an electron at unit cell $R$, atomic site $i$ then we define,
\begin{equation}
c^\dagger_{p, i} = \frac{1}{\sqrt{L}} \sum_R e^{\mathrm{i}pR} c^\dagger_{R, i},
\end{equation}
where $p$ is the momentum. The creation operator for band $\sigma$, momentum $p$ can then be written in terms of the Bloch cell-periodic wave function ($u_{i}^{p, \sigma}$) as,
\begin{equation}
c^\dagger_{p, \sigma} = \sum_{i} u^{p, \sigma} _i c^\dagger_{p, i}.
\end{equation}
We then define the Berry connection as,
\begin{align}
A_{\mathrm{elec}, \sigma}(p) &= \sum_i \bar u^{p, \sigma}_{i} \mathrm{i}\partial_p u^{p, \sigma}_{i}\label{eq:BerryConnectionElectronic}\\
&= \bra{u_{p, \sigma}} \mathrm{i}\partial_p \ket{u_{p, \sigma}}.
\end{align}
We can do a similar thing for the exciton wave function,
\begin{align}
\ket{p_{\mathrm{exc}}} &= \sum_k \phi^p_{k - \alpha p}     c^\dagger_{(1-\alpha)p + k, c} c_{-\alpha p + k, v} \ket{\mathrm{GS}}\\
&= \sum_{k, i, j} \left(\phi^p_{k - \alpha p} \:u^{(1-\alpha)p + k, c}_{i}\: \bar u^{-\alpha p + k, v}_{j} \right)   c^\dagger_{(1-\alpha)p + k, i} c_{-\alpha p + k, j} \ket{\mathrm{GS}}.
\end{align}
We see that the cell-periodic exciton wave function at total momentum $p$ is labelled by both the relative momentum $k$ as well as the two sublattice indices $i, j$. We see in addition that the cell-periodic exciton wave function will have an $\alpha$ dependence. We now generalise the method of Eq.~\eqref{eq:BerryConnectionElectronic} to generate the family of exciton Berry connections $A^\alpha_{\mathrm{exc}}(p)$,
\begin{align}
A^\alpha_{\mathrm{exc}}(p)  &= \sum_{k, i, j} \left(\bar \phi^p_{k - \alpha p} \:\bar u^{(1-\alpha)p + k, c}_{i}\:  u^{-\alpha p + k, v}_{j} \right) \mathrm{i}\partial_p \left( \phi^p_{k - \alpha p} \: u^{(1-\alpha)p + k, c}_{i}\:  \bar u^{-\alpha p + k, v}_{j} \right)\label{eq:FirstQuantisedExcitonBerryConnection} \\
&= \sum_{k, i, j}\bar \phi^p_{k - \alpha p} \mathrm{i}\partial_p \phi^p_{k - \alpha p} + \sum_{k} |\phi^p_{k - \alpha p}|^2 \left(  \sum_j u^{-\alpha p + k, v}_{j} \mathrm{i}\partial_p \bar u^{-\alpha p + k, v}_{j}+\sum_i \bar u^{(1-\alpha)p + k, c}_{i} \mathrm{i}\partial_p u^{(1-\alpha)p + k, c}_{i} \right)\\
&= \sum_{k, i, j}\bar \phi^p_{k - \alpha p} \mathrm{i}\partial_p \phi^p_{k - \alpha p} + \sum_{k} |\phi^p_{k - \alpha p}|^2 \left(-  \sum_j \bar u^{-\alpha p + k, v}_{j} \mathrm{i}\partial_p u^{-\alpha p + k, v}_{j}+\sum_i \bar u^{(1-\alpha)p + k, c}_{i} \mathrm{i}\partial_p u^{(1-\alpha)p + k, c}_{i} \right)\\
&=  \sum_k \left(\bar \phi^{p}_{k - \alpha p} \mathrm{i} \partial_p \phi^{p}_{k - \alpha p}  +|\phi^{p}_{k - \alpha p}|^2 \left\{\alpha A_{\mathrm{elec}, v} \left(k-\alpha p\right)+(1-\alpha) A_{\mathrm{elec}, c}\left[(1-\alpha) p+k\right]\right\} \right).
\end{align}
Therefore our family of different exciton Berry connections can be constructed from the different possible cell-periodic exciton wave functions.
\subsection{Exciton wave functions with arbitrary $\alpha$}
In this section we clarify the definition of the exciton wave function and demonstrate that it can written with arbitrary \emph{rational} $\alpha$. The first subtlety relates to how $k$ transforms as $p\rightarrow p+2\pi$. We consider particle-hole excitations of the following form,
\begin{equation}
\ket{p, k} = c^\dagger_{k_1 = (1-\alpha)p+k, c} c_{k_2 = -\alpha p+k}\ket{\mathrm{GS}},
\end{equation}
labelled by $\alpha$. These are our basis for the exciton eigenstates. We can relate $k_1, k_2$ and $p, k$ using a matrix equation,
\begin{equation}
\begin{pmatrix}
p \\
k 
\end{pmatrix} = \begin{pmatrix}
1 & -1 \\
\alpha & 1-\alpha  
\end{pmatrix}\begin{pmatrix}
k_1\\
k_2,
\end{pmatrix}
\end{equation}
We define $\boldsymbol{k} = (k_1, k_2)^T$. This is invariant under transformations of $k_1$ and $k_2$ by integer multiples of $2\pi$ (independently). Hence $\boldsymbol{k}$ is invariant up to the reciprocal lattice vectors $\boldsymbol{\mathcal{G}}$. They can be expressed in this basis as,
\begin{equation}
\boldsymbol{\mathcal{G}} \in \left\{n \begin{pmatrix}
    2\pi \\0
\end{pmatrix}+m \begin{pmatrix}
    0\\2\pi
\end{pmatrix}\bigg| n, m \in \mathbb{Z}\right\}.
\end{equation}
We can transform these vectors into the $p, k$ basis to see the appropriate reciprocal lattice vectors in the $p, k$ basis, 
\begin{equation}
\boldsymbol{\mathcal{G}}_\alpha \in \left\{n' \begin{pmatrix}
    2\pi \\ 2\pi\alpha
\end{pmatrix}+m' \begin{pmatrix}
    2\pi \alpha \\ 2\pi(1-\alpha)
\end{pmatrix}\bigg| n', m' \in \mathbb{Z}\right\}.
\end{equation}
This tells us that a transformation of $k_{1/2}$ by $2\pi$ translates both $p$ and $k$. However, note that it does not mean that $k$ \emph{alone} is periodic under a transformation $k\rightarrow k+2\pi n'\alpha$. We must both translate $p$ by $p\rightarrow p+2\pi$ and $k \rightarrow k+2\pi \alpha$ for the basis state $\ket{p, k}$ to remain invariant. We have seen therefore that changing $p$ by $2\pi$ in the basis state $\ket{p, k}$ also changes $k$ to an \emph{inequivalent} momentum unless $\alpha = 0, 1$. Despite this, the exciton wave function remains well-defined under the summation sign. Taking $p\rightarrow p+ 2\pi $ in the summation leaves the total exciton state unchanged,
\begin{align}
\ket{(p+2\pi )_{\mathrm{exc}}} &= \sum_k \phi^p_{k - \alpha (p+2\pi )}     c^\dagger_{(1-\alpha)(p+2\pi) + k, c} c_{-\alpha (p+2\pi) + k, v} \ket{\mathrm{GS}}\\
&= \sum_k \phi^p_{(k - 2\pi \alpha ) - \alpha p}     c^\dagger_{(1-\alpha)p + (k - 2\pi \alpha ), c} c_{-\alpha p + (k - 2\pi \alpha ), v} \ket{\mathrm{GS}}.
\end{align}
To show that $\ket{(p+2\pi )_{\mathrm{exc}}}
= \ket{p_{\mathrm{exc}}}$ we need to change the summation variable $k\rightarrow (k - 2\pi \alpha)$. To do this we require that $2\pi \alpha \in \frac{2\pi}{L} \mathbb{Z}$ or equivalently $\alpha \in \frac{n}{L}$ where $n\in \mathbb{Z}$. If this is true then we can change the summation variable and we see that $\ket{(p+2\pi )_{\mathrm{exc}}}
= \ket{p_{\mathrm{exc}}}$.

A related subtlety is that the $k$-point grid depends on $p$. If we assume that $k_1, k_2 \in \frac{2\pi}{L}\mathbb{Z}$, then the grid for the relative momentum $k$ depends on $p$ for all $\alpha \neq 0, 1$.  For example in the symmetric choice $\alpha = 1/2$, when $p$ is an even multiple of the lattice spacing $2\pi/L$ then the relative momentum are $k\in\frac{2\pi}{L}\mathbb{Z}$. But if it is an odd multiple then the allowed $k$ are $k \in \frac{2\pi}{L}(\mathbb{Z}+\frac{1}{2})$. The consistent way of dealing with this is to only calculate the exciton wave function at even multiples of the lattice spacing. However doing this is not computationally efficient as we cannot calculate the Berry connection at all momenta $p \in \frac{2\pi}{L} \mathbb{Z}$. The advantage of the Wilson loops we introduce is that we can calculate the Berry phase using all $p$, not just even multiples of $\frac{2\pi}{L}$. Hence they will require smaller system sizes to converge. 

There is a similar restriction on $p$ which occurs for arbitrary rational $\alpha = \frac{r}{q} $ for $r, q\in \mathbb{Z} \:\backslash \{0\}$. In this case, we only calculate the exciton wave function at $p$ such that $p \frac{r}{q} =\frac{2\pi}{L}\mathbb{Z}$. The allowed momenta are $p = \frac{q}{r} \frac{2\pi}{L} \mathbb{Z}$. Importantly, in the thermodynamic limit in some interval $\delta p$, there are $\frac{r}{2\pi q}L\cdot\delta p$ allowed momenta. This scales with system size $L$ and so as desired $p$ becomes continuous in the thermodynamic limit.

\newpage
\section{Projected position operator for non-interacting electron excitations}\label{Sec:ApxElectronicProjPosition}
Excitons are particle-hole excitations. In order to understand the approach we take for calculating the exciton Berry connection we begin by showing the results for single electron excitations. We will show the relationship between the projected position operator, the electronic Wannier states and the electronic Berry phase.

Consider single-particle excitations for two bands where $c_{k, c}$ ($c_{k, v}$) annihilates an electron in the conduction (valence) band at crystal momentum $k \in [-\pi,\pi)$ and with associated Bloch function $\ket{u_{k, c/v}}$. The creation operators expressed in real space are $c^\dagger_{k, c/v} = \sum_{R, i} \psi_{R, i}^{k, c/v} c^\dagger_{R, i}$ using $\psi_{R, i}^{k, c/v} = e^{\mathrm{i}kR}u_i^{k, c/v}$. The term $u_i^{k, c/v}=\braket{i|u_{k, c/v}}$ is the component of the Bloch function $\ket{u_{k, c/v}}$ on the atomic orbital $\ket{i}$. In the ground state the valence band is filled $\ket{\mathrm{GS}} = \prod_{k} c^\dagger_{k, v} \ket{0}$, where $\ket{0}$ is the fermionic vacuum. The single particle excitations we consider are,
\begin{equation}
\ket{p} = c^\dagger_{p, c} \ket{\mathrm{GS}}.
\end{equation}
For these excitations we want to use the projected \emph{periodic} position operator (introduced in the main text) to recover the single-electron Berry connection,
\begin{equation}
A_{\mathrm{elec}, c}(p) = \bra{u_{p, c}} \mathrm{i}\partial_p \ket{u_{p, c}}.
\label{eq:elecBerryCon}
\end{equation}
The eigenvectors of the projected \emph{periodic} position operator we want to be the Wannier functions which maximally localise (in real space) the electron in the conduction band. We therefore use the number operator projected into the conduction bands. First we rewrite the number operator in the band basis,
\begin{align}
n_{R, i} &= c^\dagger_{R, i} c_{R, i}\\
&= \sum_{p, q, \alpha, \beta} \bar\psi^{p, \alpha}_{R, i} \psi^{q, \beta}_{R, i} c^\dagger_{p, \alpha} c_{q, \beta},
\end{align}
where $c^\dagger_{p, \alpha} = \sum_{R, i} \psi^{p, \alpha}_{R, i} c^\dagger_{R, i}$ for band $\alpha$. The band projection involves restricting the sum to only $\alpha = \beta = c$. This gives the operator,
\begin{align}
n_{R, i}^c &= \sum_{p, q} \bar\psi^{p, c}_{R, i} \psi^{q, c}_{R, i} c^\dagger_{p, c} c_{q, c}\\
n_{R, i}^c &= \sum_{p, q} e^{\mathrm{i}(q - p)R}\bar u^{p, c}_{i} u^{q, c}_{i} c^\dagger_{p, c} c_{q, c}.
\end{align}

Consider the periodic position operator,
\begin{equation}
\hat Z_c = \sum_{R, i} e^{i\Delta R} \hat n^c_{R, i}\label{eq:ApxZcElectronic}
\end{equation}
where $R$ labels the unit cells, $i$ the sites/orbitals and $\Delta = 2\pi/L$ for system size $L$. We project $\hat Z_c$ into the excitation band using the projector,
\begin{equation}
\hat P_{\mathrm{elec}} = \sum_p \ket{p}\bra{p}.
\end{equation}
The projected position operator then becomes,
\begin{equation}
\hat{\mathcal{Z}} = \hat P_{\mathrm{elec}}\hat Z_c \hat P_{\mathrm{elec}}.
\end{equation}
We can calculate the eigenvalues of this operator,
\begin{align}
\hat{\mathcal{Z}} &= \sum_{k, l} \ket{k}\bra{k}\sum_{R, i} e^{i\Delta R} \hat n^c_{R, i} \ket{l}\bra{l}\\
 &=\sum_{k, l, R, i} e^{i\Delta R}\bra{k}  \hat n^c_{R, i} \ket{l}\ket{k}\bra{l}\\
  &=\sum_{k, l} \mathcal{Z}_{k, l} \ket{k}\bra{l}.
\end{align}
We evaluate the matrix elements $\mathcal{Z}_{k, l}$,
\begin{align}
\mathcal{Z}_{k, l} &= \sum_{R, i}e^{\mathrm{i}\Delta R} \bra{k} \hat n_{R, i}^c \ket{l}\\
 &= \frac{1}{L}\sum_{R, i, p, q} e^{\mathrm{i}(q-p+ \Delta)R} \bar u^{p, c}_{i}  u^{q, c}_{i} \bra{\mathrm{GS}} c_{k, c} c^\dagger_{p, c} c_{q, c} c^\dagger_{l, c} \ket{\mathrm{GS}}\\
&= \frac{1}{L}\sum_{R, i, p, q} e^{\mathrm{i}(q-p+ \Delta)R} \bar u^{p, c}_{i}  u^{q, c}_{i} \delta_{k, p} \delta_{q, l}\\
&= \frac{1}{L}\sum_{R, i} e^{\mathrm{i}(l-k+ \Delta)R} \bar u^{k, c}_{i}  u^{l, c}_{i}\label{eq:twoterms}.
\end{align}
We used Wick's theorem to simplify the expectation value. We can further simplify the matrix elements,
\begin{align}
\mathcal{Z}_{k, l} &= \frac{1}{L}\sum_{R, i} e^{\mathrm{i}(l-k+ \Delta)R} \bar u^{k, c}_{i}  u^{l, c}_{i}\\
&=  \delta_{k = l+\Delta}\sum_{i}\bar u^{k, c}_{i}  u^{l, c}_{i}\\
&= \delta_{k = l+\Delta}\sum_{ i}\bar u^{l+\Delta, c}_{i}  u^{l, c}_{i}.
\end{align}
It follows that we can write $\hat{\mathcal{Z}}$ as
\begin{equation}
\hat{\mathcal{Z}} = \sum_q t_q \ket{q+\Delta} \bra{q}
\end{equation}
where,
\begin{align}
t_q &= \sum_{ i}\bar u^{q+\Delta, c}_{i}  u^{q, c}_{i}\\
&=  \braket{u_{q+\Delta, c}|u_{q, c}} .
\end{align}

We now show that the eigenvectors of $\hat{\mathcal{Z}}$ are the maximally localised Wannier states and the eigenvectors give the Wannier centres. We label the $R^{th}$ Wannier state as,
\begin{equation}
\ket{\mathcal{W}^R} = \sum_{p} \mathcal{W}^R({p}) \ket{p},\label{eq:eigenstateElec}
\end{equation}
which has associated eigenvalue $\lambda_R$,
\begin{align}
\hat{\mathcal{Z}} \ket{\mathcal{W}^R} &= \sum_p \mathcal{W}^R(p)  \hat{\mathcal{Z}}\ket{p} \\
&=\sum_p \mathcal{W}^R(p) t_p \ket{p+\Delta}\\
&=\sum_p \mathcal{W}^R({p-\Delta}) t_{p-\Delta} \ket{p}.\label{eq:ZonW}
\end{align}
In addition since $\ket{\mathcal{W}^R}$ is an eigenvector of $\hat{\mathcal{Z}}$,
\begin{align}
\hat{\mathcal{Z}} \ket{\mathcal{W}^R} &= \lambda_R \ket{\mathcal{W}^R}\\
&= \sum_p \mathcal{W}^R(p) \lambda_R \ket{p}.~\label{eq:eigvalueWannierstateeq}
\end{align}
We solve for the components of the eigenvector ($\ket{\mathcal{W}^R}$) by equating  Eq.~\eqref{eq:eigvalueWannierstateeq} and Eq.~\eqref{eq:ZonW}. We first find the relationship between $\mathcal{W}^R({p-\Delta})$ and $\mathcal{W}^R(p)$,
\begin{align}
\mathcal{W}^R(p - \Delta) t_{p-\Delta} &= \mathcal{W}^R(p)  \lambda_R,
\end{align}
therefore,
\begin{align}
\mathcal{W}^R(p+\Delta)  &=\frac{1}{\lambda_R} t_{p}\mathcal{W}^R(p).
\end{align}
Component $\mathcal{W}^R(p+2\Delta)$ is related to $\mathcal{W}^R(p)$ by,
\begin{align}
\mathcal{W}^R(p+2\Delta)  &=\frac{1}{(\lambda_R)^2} t_{p+\Delta}t_{p}\mathcal{W}^R(p).\label{eq:WannierStart}
\end{align}
We can generalise this to give the relationship between arbitrary components $\mathcal{W}^R(q)$ and $\mathcal{W}^R(p)$,
\begin{align}
\mathcal{W}^R(p) &=\frac{1}{(\lambda_R)^{(p-q)/\Delta}} \left(\prod_{k = p-\Delta}^q t_{k}\right)\mathcal{W}^R(q).
\end{align}
Finally, we find an expression for the eigenvalues ($\lambda_R$) by choosing $p = q+2\pi$, 
\begin{align}
\mathcal{W}^R({q+2\pi})  &= \frac{1}{(\lambda_R)^{L}} \left(\prod_{k = q+2\pi - \Delta}^q t_{k}\right)\mathcal{W}^R(q),
\end{align}
and using $\mathcal{W}^R({q+2\pi}) = \mathcal{W}^R(q)$. This means that,
\begin{equation}
 \frac{1}{(\lambda_R)^L}\prod_{k = q+2\pi - \Delta}^q t_{k} = 1.
\end{equation}
It follows that,
\begin{align}
(\lambda_R)^L &=  \prod_{k = q+2\pi - \Delta}^q t_{k}\label{eq:definitionofWilsongeneral}\\
&= \prod_{k = q+2\pi - \Delta}^q \braket{u_{k+\Delta, c}|u_{k, c}}\\
&= \prod_{k = 0}^{2\pi - \Delta} \braket{u_{k+\Delta, c}|u_{k, c}}\\
&= \braket{u_{0, c}|u_{2\pi - \Delta, c}}\bra{u_{2\pi - \Delta, c}}\dots \ket{u_{\Delta, c}} \braket{u_{\Delta, c}| u_{0, c}}\\
&= W_c.
\end{align}
Using $\prod_{k = q+2\pi - \Delta}^q t_{k} = \prod_{k = 0}^{2\pi - \Delta} t_{k}$ by rearranging terms. We see that $(\lambda_R)^L$ is equal to the conduction band Wilson loop ($W_c$)~\cite{Neupert_2018}. In the thermodynamic limit ($L\rightarrow\infty$) we know that this is equal to $e^{\mathrm{i}\gamma_c}$ where $\gamma_c$ is the Berry phase for the conduction band since,
\begin{align}
W_c &= \lim_{L\rightarrow\infty}\prod_{k = 0}^{2\pi - \Delta} \braket{u_{k+\Delta, c}|u_{k, c}}\\
&= \lim_{L\rightarrow\infty}\prod_{k = 0}^{2\pi - \Delta} \left[\bra{u_{k, c}}  +\Delta \partial_k(\bra{u_{k, c}})\right]\ket{u_{k, c}}\\
&= \lim_{L\rightarrow\infty}\prod_{k = 0}^{2\pi - \Delta} (1  +\Delta \partial_k(\bra{u_{k, c}})\ket{u_{k, c}}\\
&= \lim_{L\rightarrow\infty}\prod_{k = 0}^{2\pi - \Delta} (1  -\Delta \bra{u_{k, c}}\partial_k\ket{u_{k, c}})\\
&= \lim_{L\rightarrow\infty}\prod_{k = 0}^{2\pi - \Delta} \big[1  +\mathrm{i}\Delta A_{\mathrm{elec}, c}(k)\big]\\
&=\lim_{L\rightarrow\infty}\prod_{k = 0}^{2\pi - \Delta} e^{\mathrm{i}\Delta A_{\mathrm{elec}, c}(k)}\\
&= e^{\mathrm{i}\int_{0}^{2\pi} \mathrm{d}k \:A_{\mathrm{elec}, c}(k)}.
\end{align}
As $L\rightarrow \infty$ therefore $(\lambda_R)^L\rightarrow e^{\mathrm{i}\gamma_c}$ where $\gamma_c$ is the Berry phase for the conduction band. It follows that, in the thermodynamic limit, the eigenvalues of the projected position operator $\hat{\mathcal{Z}}$ are, 
\begin{equation}
\lambda_R = e^{\mathrm{i}\Delta \left(\frac{\gamma_{c}}{2\pi} + R\right)},\label{eq:elecEigvalue}
\end{equation}
where $R \in \{0,1, \dots,  L-1\}$. This is because $(e^{\mathrm{i}\Delta R})^L = 1$ for all integers $R$. To get the periodic position operator $\hat Z_c$ we exponentiated the positions $R$ [Eq.~\eqref{eq:ApxZcElectronic}] so we expect to obtain the \emph{real space} Wannier centres by taking the logarithm of the eigenvalues,
\begin{equation}
-\frac{\mathrm{i}}{\Delta}  \log(\lambda_R) = \frac{\gamma_{c}}{2\pi} + R.
\end{equation}
We have therefore derived the result that the Berry phase gives the shift of the electronic Wannier states from the centre of the unit cell. For finite system size the eigenvalues are,
\begin{equation}
\lambda_R = (W_c)^{1/L} e^{\mathrm{i}\Delta R}.\label{eq:elecEigvalueFinite}
\end{equation}

Lastly, we show that the eigenstates are the Wannier states. We begin with Eq.~\eqref{eq:WannierStart} and plug in the eigenvalue expression in Eq.~\eqref{eq:elecEigvalueFinite},
\begin{align}
\mathcal{W}^R(q) &=\frac{1}{(\lambda_R)^{(q-p)/\Delta}} \left(\prod_{k = q-\Delta}^p t_{k}\right)\mathcal{W}^R(p)\\
&=\frac{1}{\left[(W_c)^{\frac{1}{L}} e^{\mathrm{i}\Delta R}\right]^{(q-p)/\Delta}} \left(\prod_{k = q-\Delta}^p t_{k}\right)\mathcal{W}^R(p)\\
&= e^{-\mathrm{i}(q-p)R} (W_c)^{-\mathrm{i}\frac{q - p}{2\pi}}
 \left(\prod_{k = q-\Delta}^p t_{k}\right)\mathcal{W}^R(p).
\end{align}
If we set $p = 0$ then we can construct the Wannier states using Eq.~\eqref{eq:eigenstateElec},
\begin{equation}
\ket{\mathcal{W}^R} = \mathcal{W}^R(0) \sum_{p} e^{-\mathrm{i} pR} (W_c)^{-\frac{p}{2\pi}} \prod_{k = 0}^{p-\Delta} t_{k}\ket{p}.
\end{equation}
We rewrite this as,
\begin{equation}
\ket{\mathcal{W}^R} = \mathcal{N} \sum_{p} e^{-\mathrm{i} pR} (W_c)^{-\frac{p}{2\pi}} \prod_{k = 0}^{p-\Delta} t_{k}\ket{p},\label{eq:ApxelectronicWannierStatesFinite}
\end{equation}
where we have rewritten $\mathcal{W}^R(0)$ as some normalisation $\mathcal{N}$. We can see that Eq.~\eqref{eq:ApxelectronicWannierStatesFinite} modifies the (arbitrary) gauge of $\ket{p}$ in order to maximally localise the Wannier functions. Firstly, the product of Bloch function overlaps ($t_k$) maximally aligns the phases of the Bloch functions ($\ket{u_{p, c}}$) at adjacent momenta. The Wilson loop term [$(W_c)^{-\frac{p}{2\pi}} \rightarrow e^{-\mathrm{i }\frac{\gamma_c p}{2\pi}}$] then equally distributes the total Berry phase across the BZ. This results in a smooth and continuous gauge across the whole BZ (known as the \emph{twisted parallel-transport gauge}), this gauge choice maximally localises the Wannier functions in 1D~\cite{MarzariVanderbiltSmoothGaugePRB97, Neupert_2018}. 

In the thermodynamic limit we can rewrite the Wannier states as,
\begin{align}
\ket{\mathcal{W}^R} &= \frac{1}{\sqrt{L}}\sum_{p} e^{-\mathrm{i} R p} e^{-\mathrm{i} \frac{\gamma_{c}}{2\pi}p} e^{\mathrm{i}\int_{0}^p \mathrm{d}q \: A_{\mathrm{elec}, c}(q)}  \ket{p}\label{eq:WannierStateElecThermLimit}.
\end{align}
where the electronic Berry phase is $\gamma_{c} = \int_{0}^p \mathrm{d} q \: A_{\mathrm{elec}, c}(q)$. This is exactly the form derived using the thermodynamic limit form of the position operator $\hat X_c\rightarrow \mathrm{i}\partial_p$ (see Ref.~\cite{Neupert_2018}). 

\subsection{Connection to $\hat P_{\mathrm{elec} }\hat X_c \hat P_\mathrm{elec}$}
We now discuss the fact the eigenstates of $\hat{\mathcal{Z}}$ are the same of those of the projected position operator $\hat P_{\mathrm{elec} }\hat X_c \hat P_\mathrm{elec}$ in the thermodynamic limit. We aim to give some intuition for why this is true. To do this we show that the diagonal elements of $\hat P_{\mathrm{elec} }\hat X_c \hat P_\mathrm{elec}$ projected into the basis of the eigenstates of $\hat{\mathcal{Z}}$ are the expected values in the thermodynamic limit. To fully demonstrate that the eigenstates of $\hat{\mathcal{Z}}$ are equal to those of $\hat P_{\mathrm{elec} }\hat X_c \hat P_\mathrm{elec}$ in the thermodynamic limit, it also necessary to show that the off-diagonal terms of $\hat P_{\mathrm{elec} }\hat X_c \hat P_\mathrm{elec}$ in this basis vanish. We refer the reader to the explicit calculations of the eigenstates of $\hat P_{\mathrm{elec} }\hat X_c \hat P_\mathrm{elec}$~\cite{Neupert_2018} in the thermodynamic limit in order to demonstrate this. 

We begin by noting that we cannot Taylor expand the operator $\hat{\mathcal{Z}}$ in terms of $R$ because we are summing over all $R$ from $0$ to $L-1$ so $R/L$ is not in general small. Instead we consider the operator $\hat{\mathcal{Z}}$ in the basis of its eigenvalues $\ket{\mathcal{W}^R}$,
\begin{equation}
\bra{\mathcal{W}^{R'}}\hat{\mathcal{Z}} \ket{\mathcal{W}^0} = \delta_{0, R'} e^{\mathrm{i}\Delta (\frac{\gamma_c}{2\pi})}\label{eq:matrixElementsWannierBasis}.
\end{equation}
We consider the diagonal terms of this matrix \emph{i.e.} $\bra{\mathcal{W}^{0}} \hat{\mathcal{Z}} \ket{\mathcal{W}^0}$. We note that the Wannier state $\ket{\mathcal{W}^0}$ is exponentially localised in 1D and so this means that terms in the sum $\bra{\mathcal{W}^{0}}\hat{\mathcal{Z}}\ket{\mathcal{W}^{0}} = \sum_{R, i} e^{\mathrm{i}\Delta R}\bra{\mathcal{W}^{0}} \hat n^c_{R, i}\ket{\mathcal{W}^{0}}$ with large $R$ vanish. It follows that, within the expectation value, we \emph{can} Taylor expand the operator in $R$,
\begin{align}
\bra{\mathcal{W}^0} \hat{\mathcal{Z}} \ket{\mathcal{W}^0} &= \bra{\mathcal{W}^0} \sum_{R, i} e^{\mathrm{i} \Delta R} \hat n^c_{R, i} \ket{\mathcal{W}^0}\\
&= \bra{\mathcal{W}^0} \sum_{R, i}\left\{ \left[1+\mathrm{i} \Delta R + (\mathrm{i}\Delta)^2 R^2 + \dots\right]  \hat n^c_{R, i}\right\} \ket{\mathcal{W}^0}\\
&= \bra{\mathcal{W}^0}\sum_{R, i} \hat n^c_{R, i}\ket{\mathcal{W}^0}  +\mathrm{i} \Delta \bra{\mathcal{W}^0}\sum_{R, i} R \: \hat n^c_{R, i}\ket{\mathcal{W}^0} + (\mathrm{i}\Delta)^2 \bra{\mathcal{W}^0}\sum_{R, i} R^2\: \hat n^c_{R, i} \ket{\mathcal{W}^0} + \dots\\
&= 1  +\mathrm{i} \Delta \bra{\mathcal{W}^0}\hat X_c \ket{\mathcal{W}^0} + (\mathrm{i}\Delta)^2 \bra{\mathcal{W}^0}\sum_{R, i} R^2\: \hat n^c_{R, i}\ket{\mathcal{W}^0} + \dots.
\end{align}
We can also Taylor expand the right hand side of Eq.~\eqref{eq:matrixElementsWannierBasis} to give,
\begin{equation}
1  +\mathrm{i} \Delta \bra{\mathcal{W}^0}\hat X_c \ket{\mathcal{W}^0} + (\mathrm{i}\Delta)^2 \bra{\mathcal{W}^0}\sum_{R, i} R^2\: \hat n^c_{R, i} \ket{\mathcal{W}^0} + \dots. = 1+ \mathrm{i}\Delta \frac{\gamma_c}{2\pi} + \frac{1}{2}(\mathrm{i}\Delta)^2 \left(\frac{\gamma_c}{2\pi}\right)^2.
\end{equation}
Matching the $\mathcal{O}(\Delta)$ terms on either side we find that $\bra{\mathcal{W}^0}\hat X_c \ket{\mathcal{W}^0} = \frac{\gamma_c}{2\pi}+\mathcal{O}(\Delta)$. Therefore in the thermodynamic limit $\bra{\mathcal{W}^0}\hat X_c \ket{\mathcal{W}^0} = \frac{\gamma_c}{2\pi}$. For the other diagonal elements we have $\bra{\mathcal{W}^R}\hat X_c \ket{\mathcal{W}^R} = \frac{\gamma_c}{2\pi} + R$. 

\newpage
\section{Projected position operator for excitons}
\label{Sec:ApxExcitonProjPosition}
\subsection{1-dimensional systems}
We now repeat the previous calculation for excitons, not single particle excitations. Let's consider a single exciton band with an exciton eigenstate at each total momentum $p$,
\begin{equation} 
\ket{p_{\mathrm{exc}}} = \sum_k \phi^{p}_{k} c^\dagger_{p+k, c} c_{k, v} \ket{\mathrm{GS}}.
\end{equation}
We have chosen $\alpha = 0$ from Eq.~\eqref{eq:ApxexcitonEigenstate}. We use $\ket{p_{\mathrm{exc}}}$ as a genuine many-body state (and do not construct some single-quantised wave function from it), therefore our final expressions will be independent of the exciton wave function convention used. We first define the projector onto the exciton band,
\begin{equation}
\hat P_{\mathrm{exc}} = \sum_p \ket{p_{\mathrm{exc}}}\bra{p_{\mathrm{exc}}}.
\end{equation}
As introduced in the main text, with excitons we can use either the position operator for the electrons in the conduction band or the holes in the valence band. We therefore consider projecting the number operator into the conduction and valence bands, first we rewrite the number operator in the band basis,
\begin{align}
\hat n_{R, i} &= c^\dagger_{R, i} c_{R, i}\\
&= \sum_{k, k', \alpha, \beta} \bar\psi^{k, \alpha}_{R, i} \psi^{k', \beta}_{R, i} c^\dagger_{k, \alpha} c_{k', \beta},
\end{align}
where $c^\dagger_{k, \alpha} = \sum_{R, i} \psi^{k, \alpha}_{R, i} c^\dagger_{R, i}$ for band $\alpha$. The band projection involves restricting the sum to only $\alpha = \beta = c$ or $\alpha = \beta = v$. This gives the operators,
\begin{align}
\hat n_{R, i}^c &= \sum_{k, k'} \bar\psi^{k, c}_{R, i} \psi^{k', c}_{R, i} c^\dagger_{k, c} c_{k', c}\\
\hat n_{R, i}^v &= \sum_{k, k'} \psi^{k, v}_{R, i} \bar \psi^{k', v}_{R, i} c_{k, v} c^\dagger_{k', v}.
\end{align}
where, for the valence band we are interested in the hole number operator and hence we band project $c_{R, i}c^\dagger_{R, i}$ instead of $\hat n_{R, i}$. 
Taking inspiration from the discussion in Sec.~\ref{Sec:ApxElectronicProjPosition}, we have two possible \emph{periodic} position operators. These are defined as,
\begin{equation}
\hat Z_{c/v} = \sum_{R, i} e^{\mathrm{i}\Delta R} \hat n_{R, i}^{c/v}.
\end{equation}
We then project these electron/hole position operators into the exciton band,
\begin{equation}
\hat{\mathcal{Z}}_{c/v} = \hat P_{\mathrm{exc}} \hat Z_{c/v} \hat P_{\mathrm{exc}}.
\end{equation}
We expect the eigenvectors of these operators to be the exciton Wannier states when we either maximally localise the electron or the hole components of the exciton. We do not expect these to be necessarily equal. We next calculate the matrix elements $[\mathcal{Z}_{c/v}]_{p, q}$,
\begin{align}
[\mathcal{Z}_{c/v}]_{p, q} &= \bra{p_{\mathrm{exc}}}\hat{\mathcal{Z}}_{c/v} \ket{q_{\mathrm{exc}}}\\
&=\bra{p_{\mathrm{exc}}}\sum_{R, i} e^{\mathrm{i}\Delta R} \hat n_{R, i}^{c/v}\ket{q_{\mathrm{exc}}}\\
&= \sum_{R, i, k, l} e^{\mathrm{i}\Delta R} \bar \phi^p_k \phi^q_{l} \bra{\mathrm{GS}} c^\dagger_{k, v} c_{p+k, c} \hat n_{R, i}^{c/v} c^\dagger_{q+l, c} c_{l, v}\ket{\mathrm{GS}}.
\end{align}
We consider the two operators separately. 

\subsubsection{Eigenspectrum of $\hat{\mathcal{Z}}_{c}$}
First we calculate the matrix elements for the conduction band projected position operator,
\begin{align}
[\mathcal{Z}_{c}]_{p, q} &= \sum_{R, i, k, l} e^{\mathrm{i}\Delta R} \bar \phi^p_k \phi^q_{l}  \bra{\mathrm{GS}} c^\dagger_{k, v} c_{p+k, c} \hat n_{R, i}^{c} c^\dagger_{q+l, c} c_{l, v}\ket{\mathrm{GS}}\\
&= \sum_{R, i, k, l, m, n} e^{\mathrm{i}\Delta R} \bar \phi^p_k \phi^q_{l}  \bar\psi^{m, c}_{R, i} \psi^{n, c}_{R, i}\bra{\mathrm{GS}} c^\dagger_{k, v} c_{p+k, c} c^\dagger_{m, c} c_{n, c} c^\dagger_{q+l, c} c_{l, v}\ket{\mathrm{GS}}\\
&= \sum_{R, i, k, l, m, n} e^{\mathrm{i}\Delta R} \bar \phi^p_k \phi^q_{l}  \bar\psi^{m, c}_{R, i} \psi^{n, c}_{R, i} (\delta_{k, l} \delta_{p+k, m} \delta_{n, q+l})\\
&= \sum_{R, i, k} e^{\mathrm{i}\Delta R} \bar \phi^p_k \phi^q_{k}  \bar\psi^{p+k, c}_{R, i} \psi^{q+k, c}_{R, i}\\
&= \frac{1}{L}\sum_{R, i, k} e^{\mathrm{i}R(\Delta +q-p)} \bar \phi^p_k \phi^q_{k}  \bar u^{p+k, c}_{i} u^{q+k, c}_{i}\\
&=  \delta_{p, q+\Delta} \sum_{i, k} \bar \phi^p_k \phi^q_{k}  \bar u^{p+k, c}_{i} u^{q+k, c}_{i}\\
&=  \delta_{p, q+\Delta} \sum_{k} \bar \phi^p_k \phi^q_{k}   \braket{u_{p+k, c}| u_{q+k, c}}\\
&=  \delta_{p, q+\Delta} \sum_{k} \bar \phi^{q+\Delta}_k \phi^q_{k}   \braket{u_{q+\Delta+k, c}| u_{q+k, c}}\\
&=  \delta_{p, q+\Delta} \sum_{k} \bar \phi^{q+\Delta}_{k - q} \phi^q_{k - q}   \braket{u_{\Delta+k, c}| u_{k, c}}.\label{eq:matrixelements1DZc}
\end{align}
Hence just like with the electronic projected position operator we can write, 
\begin{equation}
\hat{\mathcal{Z}}_{c} = \sum_{p} t^{\mathrm{exc}, c}_{p} \ket{(p+\Delta)_{\mathrm{exc}}} \bra{p_{\mathrm{exc}}},\label{eq:ZcOperatorForm}
\end{equation}
where
\begin{equation}
t^{\mathrm{exc}, c}_{p} = \sum_{k} \bar \phi^{p+\Delta}_{k - p} \phi^p_{k - p}   \braket{u_{k+\Delta, c}| u_{k, c}}.
\end{equation}

We can therefore find the eigenvalues and eigenstates of $\hat{\mathcal{Z}}_c$ following the method introduced in the previous section (Sec.~\ref{Sec:ApxElectronicProjPosition}). We define the eigenvalues of $\hat{\mathcal{Z}}_c$ as,
\begin{equation}
\ket{\mathcal{W}^R_{\mathrm{exc}, c}} = \:\sum_{p} \mathcal{W}^R_{\mathrm{exc}, c}(p)\: \ket{p}.\label{eq:eigenstate}
\end{equation}
with eigenvalue $\lambda_{R, c}$,
\begin{align}
\hat{\mathcal{Z}}_c \ket{\mathcal{W}^R_{\mathrm{exc}, c}}&= \lambda_{R, c} \ket{\mathcal{W}^R_{\mathrm{exc}, c}}\\
&=\sum_p   \lambda_{R, c}\: \mathcal{W}^R_{\mathrm{exc}, c}(p) \ket{p_{\mathrm{exc}}}.\label{eq:EXCEigValEquation}
\end{align}
From the form of $\hat{\mathcal{Z}}_{c}$ in Eq.~\eqref{eq:ZcOperatorForm} we can write,
\begin{align}
\hat{\mathcal{Z}}_c \ket{\mathcal{W}^R_{\mathrm{exc}, c}} &= \sum_p \mathcal{W}^R_{\mathrm{exc}, c}(p)\: t_p \ket{(p+\Delta)_{\mathrm{exc}}}\\
&= \sum_p \mathcal{W}^R_{\mathrm{exc}, c}(p-\Delta)\: t_{p-\Delta} \ket{p_{\mathrm{exc}}}.\label{eq:EXCEigValEquation2}
\end{align}
We solve for the components of the eigenvector ($\ket{\mathcal{W}^R_{\mathrm{exc, c}}}$) by equating  Eq.~\eqref{eq:EXCEigValEquation} and Eq.~\eqref{eq:EXCEigValEquation2}. We first find the relationship between $\mathcal{W}_{\mathrm{exc}, c}^R({p - \Delta})$ and $\mathcal{W}_{\mathrm{exc}, c}^R({p})$,
\begin{align}
\mathcal{W}_{\mathrm{exc}, c}^R({p - \Delta}) \:t^{\mathrm{exc}, c}_{p-\Delta} &= \mathcal{W}_{\mathrm{exc}, c}^R({p})  \lambda_{R, c},
\end{align}
therefore,
\begin{align}
\mathcal{W}_{\mathrm{exc}, c}^R({p+\Delta})  &=\frac{1}{\lambda_{R, c}} t^{\mathrm{exc}, c}_{p}\:\mathcal{W}_{\mathrm{exc}, c}^R(p).
\end{align}
Component $\mathcal{W}_{\mathrm{exc}, c}^R({p+2\Delta})$ is related to $\mathcal{W}_{\mathrm{exc}, c}^R(p)$ by,
\begin{align}
\mathcal{W}_{\mathrm{exc}, c}^R(p+2\Delta)  &=\frac{1}{(\lambda_{R, c})^2}\: t^{\mathrm{exc}, c}_{p+\Delta}\:\:t^{\mathrm{exc}, c}_{p}\:\mathcal{W}_{\mathrm{exc}, c}^R(p).
\end{align}
We can generalise this to give the relationship between arbitrary components $\mathcal{W}_{\mathrm{exc}, c}^R(q)$ and $\mathcal{W}_{\mathrm{exc}, c}^R(p)$,
\begin{align}
\mathcal{W}_{\mathrm{exc}, c}^R({p}) &=\frac{1}{(\lambda_{R, c})^{(p-q)/\Delta}} \left(\prod_{k = p-\Delta}^q t^{\mathrm{exc}, c}_{k}\right)\mathcal{W}_{\mathrm{exc}, c}^R({q}).\label{eq:EXCWannierStateCompRelation}
\end{align}
Finally, we find an expression for the eigenvalues ($\lambda_{R,c}$) by choosing $p = q+2\pi$, 
\begin{align}
\mathcal{W}_{\mathrm{exc}, c}^R({q+2\pi})  &= \frac{1}{(\lambda_{R, c})^{L}} \left(\prod_{k = q+2\pi - \Delta}^q t^{\mathrm{exc}, c}_{k}\right)\mathcal{W}_{\mathrm{exc}, c}^R(q),
\end{align}
and using $\mathcal{W}_{\mathrm{exc}, c}^R({q+2\pi}) = \mathcal{W}_{\mathrm{exc}, c}^R(q)$. This means that,
\begin{equation}
 \frac{1}{(\lambda_{R, c})^L}\prod_{k = q+2\pi - \Delta}^q t^{\mathrm{exc}, c}_{k} = 1.
\end{equation}
Therefore,
\begin{align}
(\lambda_{R, c})^L &= \prod_{k = q+2\pi - \Delta}^q t^{\mathrm{exc}, c}_{k}\\
&= \prod_{k = 0}^{2\pi - \Delta} t^{\mathrm{exc}, c}_{k}.
\end{align}
We note the similarity between this expression and the expression we derived for non-interacting electrons [Eq.~\eqref{eq:definitionofWilsongeneral}]. This indicates we should identify 
\begin{equation}
W_{\mathrm{exc},c} = \prod_{k = 0}^{2\pi - \Delta} t^{\mathrm{exc}, c}_{k}
\end{equation}
as an exciton Wilson loop. Just like the electronic Wilson loop [Eq.~\eqref{eq:definitionofWilsongeneral}], $W_{\mathrm{exc},c}$ has modulus 1 in the thermodynamic limit. This means it can be written as $e^{\mathrm{i}\gamma_{\mathrm{exc}, c}}$ where we identify $\gamma_{\mathrm{exc}, c}$ as an exciton Berry phase. Analogously to the electronic case described in Sec.~\ref{Sec:ApxElectronicProjPosition} we can express the eigenvalues and eigenvectors of $\hat{\mathcal{Z}}_c$ in terms of $W_{\mathrm{exc},c}$. The eigenvalues take the same form, now in terms of the \emph{exciton} Wilson loop $W_{\mathrm{exc}, c}$,
\begin{equation}
\lambda_{R, c} = (W_{\mathrm{exc}, c})^{1/L} e^{\mathrm{i}\Delta R},\label{eq:exciton_c_EigvalueFinite}
\end{equation}
where $R \in \{0,1, \dots,  L-1\}$. In the thermodynamic limit this tends to,
\begin{equation}
\lambda_{R, c} = e^{\mathrm{i}\Delta \left(\frac{\gamma_{\mathrm{exc}, c}}{2\pi}+R\right)}.\label{eq:exciton_c_EigvalueThermLimit}
\end{equation}
We saw that we could extract the real space centre of the \emph{electronic} Wannier centre by taking the logarithm. We can do exactly the same here to obtain the exciton Wannier centre,
\begin{equation}
-\frac{\mathrm{i}}{\Delta}  \log(\lambda_{R, c}) = \frac{\gamma_{\mathrm{exc}, c}}{2\pi} + R.\label{eq:EXCWannierCentres_c}
\end{equation}
What does this exciton Wannier centre actually represent? The exciton Wannier state $\ket{\mathcal{W}^R_{\mathrm{exc}, c}}$ was obtained by maximally localising the electron within the exciton Wannier state. Since $\hat{\mathcal{Z}}_c$ is the projected \emph{electron} position operator the centre is the average position of the \emph{electron} in $\ket{\mathcal{W}^R_{\mathrm{exc}, c}}$. We see that the shift of this position from the centre of the unit cells is given by some Berry phase $\gamma_{\mathrm{exc}, c}$ just as it was for the \emph{electronic} Wannier states. This means we were correct to identify $W_{\mathrm{exc}, c}$ as an exciton Wilson loop.

Using the eigenvalues $\lambda_{R, c}$ we can obtain expressions for the exciton Wannier states $\ket{\mathcal{W}^R_{\mathrm{exc},c}}$. We begin with Eq.~\eqref{eq:EXCWannierStateCompRelation} and plug in $\lambda_{R, c}$ from Eq.~\eqref{eq:exciton_c_EigvalueFinite},
\begin{align}
\mathcal{W}_{\mathrm{exc}, c}^R({q}) &=\frac{1}{(\lambda_{R, c})^{(q-p)/\Delta}} \left(\prod_{k = q-\Delta}^p t_{k}\right)\mathcal{W}_{\mathrm{exc}, c}^R({p})\\
&=\frac{1}{\left[(W_{\mathrm{exc}, c})^{1/L} e^{\mathrm{i}\Delta R}\right]^{(q-p)/\Delta}} \left(\prod_{k = q-\Delta}^p t_{k}\right)\mathcal{W}_{\mathrm{exc}, c}^R({p})\\
&= e^{-\mathrm{i}(q - p)R} (W_{\mathrm{exc}, c})^{-\frac{q-p}{2\pi}}  \left(\prod_{k = q-\Delta}^p t_{k}\right)\mathcal{W}_{\mathrm{exc}, c}^R({p}).
\end{align}
We choose $p = 0$ and write the exciton Wannier state $\ket{\mathcal{W}_{\mathrm{exc}, c}^R}$ as,
\begin{align}
\ket{\mathcal{W}^R_{\mathrm{exc}, c}} &= \mathcal{N}\sum_{q} e^{-\mathrm{i} R q} (W_{\mathrm{exc}, c})^{-\frac{q}{2\pi}} \left(\prod_{k = 0}^{q-\Delta} t^{\mathrm{exc}, c}_{k}\right)\ket{q_{\mathrm{exc}}}.
\end{align}
The maximally localised exciton Wannier states $\ket{\mathcal{W}^R_{\mathrm{exc}, c}}$ therefore take a very similar form to the maximally localised \emph{electronic} Wannier states [Eq.~\eqref{eq:ApxelectronicWannierStatesFinite}].

\subsubsection{Eigenspectrum of $\hat{\mathcal{Z}}_{v}$}
We now find the eigenvalues and eigenvectors of $\hat{\mathcal{Z}}_{v}$. The method is identical to that for $\hat{\mathcal{Z}}_{c}$. The matrix elements of $\hat{\mathcal{Z}}_{v}$ are,
\begin{align}
[\hat{\mathcal{Z}}_{v}]_{p, q} &= \sum_{R, i, k, l} e^{\mathrm{i}\Delta R} \bar \phi^p_k \phi^q_{l}  \bra{\mathrm{GS}} c^\dagger_{k, v} c_{p+k, c} \hat n_{R, i}^{v} c^\dagger_{q+l, c} c_{l, v}\ket{\mathrm{GS}}\\
&= \sum_{R, i, k, l, m, n} e^{\mathrm{i}\Delta R} \phi^p_k  \bar\phi^q_{l}  \psi^{m, v}_{R, i} \bar\psi^{n, v}_{R, i}\bra{\mathrm{GS}} c^\dagger_{k, v} c_{p+k, c} c_{m, v} c^\dagger_{n, v} c^\dagger_{q+l, c} c_{l, v}\ket{\mathrm{GS}}\\
&= \sum_{R, i, k, l, m, n} e^{\mathrm{i}\Delta R} \bar \phi^p_k \phi^q_{l}  \psi^{m, v}_{R, i} \bar\psi^{n, v}_{R, i} (\delta_{l, n} \delta_{p+k, q+l} \delta_{k, m})\\
&= \sum_{R, i, k} e^{\mathrm{i}\Delta R} \bar\phi^p_k \phi^q_{p+k-q} \psi^{k, v}_{R, i}  \bar\psi^{p+k-q, v}_{R, i}\\
&= \frac{1}{L}\sum_{R, i, k} e^{\mathrm{i}R(\Delta +q-p)} \bar\phi^p_k \phi^q_{p+k-q}  u^{k, v}_{i}  \bar u^{p+k-q, v}_{i}\\
&= \delta_{p, q+\Delta} \sum_{i, k} \bar\phi^p_k \phi^q_{p+k-q}  u^{k, v}_{i}  \bar u^{p+k-q, v}_{i}\\
&= \delta_{p, q+\Delta} \sum_{k} \bar\phi^p_k \phi^q_{p+k-q}  \braket{u_{p+k-q, v}|u_{k, v}}\\
&= \delta_{p, q+\Delta} \sum_{k} \bar\phi^{q+\Delta}_k \phi^q_{k+\Delta}  \braket{u_{k+\Delta, v}|u_{k, v}}\label{eq:matrixelements1DZv}.
\end{align}
We therefore define an exciton Wilson loop $W_{\mathrm{exc}, v}$ as,
\begin{align}
W_{\mathrm{exc}, v} &= \prod_{p = 0}^{2\pi - \Delta} t^{\mathrm{exc}, v}_{p},
\end{align}
where,
\begin{equation}
t^{\mathrm{exc}, v}_{p} = \sum_{k} \bar\phi^{p+\Delta}_k \phi^p_{k+\Delta}  \braket{u_{k+\Delta, v}|u_{k, v}}.
\end{equation}
The eigenvalues are,
\begin{equation}
\lambda_{R, v} = (W_{\mathrm{exc}, v})^{1/L} e^{\mathrm{i}\Delta R},\label{eq:exciton_v_EigvalueFinite}
\end{equation}
where $R \in \{0,1, \dots,  L-1\}$. Since the Wilson loop $W_{\mathrm{exc}, v}$ has modulus 1 in the thermodynamic limit we can write as $e^{\mathrm{i}\gamma_{\mathrm{exc}, v}}$ where $\gamma_{\mathrm{exc}, v}$ is an exciton Berry phase. In the thermodynamic limit the eigenvalues therefore tend to,
\begin{equation}
\lambda_{R, v} = e^{\mathrm{i}\Delta \left(\frac{\gamma_{\mathrm{exc}, v}}{2\pi}+R\right)}.\label{eq:exciton_v_EigvalueThermLimit}
\end{equation}
We obtain the exciton Wannier centres,
\begin{equation}
-\frac{\mathrm{i}}{\Delta}  \log(\lambda_{R, v}) = \frac{\gamma_{\mathrm{exc}, v}}{2\pi} + R.
\end{equation}
These exciton Wannier centres mean something different from those in Eq.~\eqref{eq:EXCWannierCentres_c}. The exciton Wannier state $\ket{\mathcal{W}^R_{\mathrm{exc}, v}}$ was obtained by maximally localising the hole within the exciton Wannier state. Since $\hat{\mathcal{Z}}_v$ is the projected \emph{hole} position operator the centre is the average position of the \emph{hole} in $\ket{\mathcal{W}^R_{\mathrm{exc}, v}}$. We see that the shift of this position from the centre of the unit cells is given by some Berry phase $\gamma_{\mathrm{exc}, v}$ just as it was for the \emph{electronic} Wannier states. 

Lastly, the corresponding exciton Wannier states take the form,
\begin{align}
\ket{\mathcal{W}^R_{\mathrm{exc}, v}} &= \frac{1}{\sqrt{L}}\sum_{q} e^{-\mathrm{i} R q} (W_{\mathrm{exc}, v})^{-\frac{q}{2\pi}} \left(\prod_{k = 0}^{q-\Delta} t^{\mathrm{exc}, v}_{k}\right)\ket{q_{\mathrm{exc}}}.
\end{align}

\subsection{Generalisation to higher dimensional systems}
We now generalise the previous discussion to give the Wilson loop in higher dimensions. This is performed in exactly the way described in Ref.~\onlinecite{Neupert_2018} for non-interacting electrons. In higher dimensions the projected position operators in different directions no longer have to commute, hence there can be an obstruction to generating exponentially localised Wannier functions. Instead we can construct hybrid Wannier states, these are maximally localised in one spatial dimension but are plane waves in the remaining. To construct these in a $d$-dimensional system, we consider a 1D incontractible loop in BZ which fixes $(d-1)$ components of the momentum (labelled $\boldsymbol{p}_\perp$). We then consider the projector onto all occupied band eigenstates along this path. The momenta along this path we label as $\boldsymbol{p}_\parallel$ such that the total momentum of all states along the path are $\boldsymbol{p} = \boldsymbol{p}_{\perp}+\boldsymbol{p}_\parallel$. This projector is then labelled by $\boldsymbol{p}_{\perp}$,
\begin{equation}
\hat P_{\mathrm{exc}}(\boldsymbol{p}_{\perp}) = \sum_{\boldsymbol{p}_\parallel} \ket{(\boldsymbol{p}_{\perp}+\boldsymbol{p}_\parallel)_{\mathrm{exc}}}\bra{(\boldsymbol{p}_{\perp}+\boldsymbol{p}_\parallel)_{\mathrm{exc}}}.
\end{equation}
We define $\boldsymbol{\mathcal{G}}_\parallel$ as the reciprocal lattice vector parallel pointing along the loop direction. We use the normal vector in this direction $\hat{\boldsymbol{\mathcal{G}}}_\parallel = \boldsymbol{\mathcal{G}}_\parallel/|\boldsymbol{\mathcal{G}}_\parallel|$.  Then $R_{\parallel} = \boldsymbol{R}\cdot\hat{\boldsymbol{\mathcal{G}}}_\parallel $ is the distance along the spatial dimension conjugate to the momentum $\boldsymbol{p}_\parallel$ (\emph{i.e.} this would be $x$ if we are considering loops along $p_x$). Without loss of generality we choose to have $L$ unit cells along each lattice vector direction. We define the \emph{periodic} position operator in this direction as,
\begin{equation}
\hat Z_{c/v, \parallel} = \sum_{\boldsymbol{R}, i} e^{\mathrm{i}\Delta R_\parallel} \:\hat n_{\boldsymbol{R}, i}^{c/v}.
\end{equation}
The hybrid Wannier states at $\boldsymbol{p}_{\perp}$ are then the eigenvectors of the projected position operator,
\begin{equation}
\hat{\mathcal{Z}}_{c/v}(\boldsymbol{p}_{\perp}) = \hat P_{\mathrm{exc}}(\boldsymbol{p}_{\perp}) \hat Z_{c/v, \parallel} \hat P_{\mathrm{exc}}(\boldsymbol{p}_{\perp}).
\end{equation}
We now repeat the calculation as performed in the previous section. We begin by finding the matrix elements of $\hat{\mathcal{Z}}_{c}(\boldsymbol{p}_{\perp})$,
\begin{align}
[\mathcal{Z}_{c}(\boldsymbol{p}_{\perp})]_{\boldsymbol{p}_\parallel, \boldsymbol{q}_\parallel} &= \bra{(\boldsymbol{p}_{\perp}+\boldsymbol{p}_\parallel)_{\mathrm{exc}}}\hat{\mathcal{Z}}_{c} (\boldsymbol{p}_{\perp})\ket{(\boldsymbol{p}_{\perp}+\boldsymbol{q}_\parallel)_{\mathrm{exc}}}\\
&=\bra{(\boldsymbol{p}_{\perp}+\boldsymbol{p}_\parallel)_{\mathrm{exc}}}\sum_{\boldsymbol{R}, i} e^{\mathrm{i}\Delta R_\parallel} \:\hat n_{\boldsymbol{R}, i}^{c}\ket{(\boldsymbol{p}_{\perp}+\boldsymbol{q}_\parallel)_{\mathrm{exc}}}\\
&= \sum_{\boldsymbol{R}, i, \boldsymbol{k}, \boldsymbol{l}} e^{\mathrm{i}\Delta R_{\parallel}} \bar \phi^{\boldsymbol{p}_{\perp}+\boldsymbol{p}_\parallel}_{\boldsymbol{k}} \phi^{\boldsymbol{p}_{\perp}+\boldsymbol{q}_\parallel}_{\boldsymbol{l}} \bra{\mathrm{GS}} c^\dagger_{\boldsymbol{k}, v} c_{\boldsymbol{p}_{\perp}+\boldsymbol{p}_\parallel+\boldsymbol{k}, c} \hat n_{\boldsymbol{R}, i}^{c} c^\dagger_{\boldsymbol{p}_{\perp}+\boldsymbol{q}_\parallel+\boldsymbol{l}, c} c_{\boldsymbol{l}, v}\ket{\mathrm{GS}}.
\end{align}
We write the volume of the BZ as $\mathcal{V} = L^d$. We now explicitly find the expressions for these matrix elements,
\begin{align}
[\mathcal{Z}_{c}(\boldsymbol{p}_{\perp})]_{\boldsymbol{p}_\parallel, \boldsymbol{q}_\parallel} &= \sum_{\boldsymbol{R}, i, \boldsymbol{k}, \boldsymbol{l}} e^{\mathrm{i}\Delta R_{\parallel}} \bar \phi^{\boldsymbol{p}_{\perp}+\boldsymbol{p}_\parallel}_{\boldsymbol{k}} \phi^{\boldsymbol{p}_{\perp}+\boldsymbol{q}_\parallel}_{\boldsymbol{l}} \bra{\mathrm{GS}} c^\dagger_{\boldsymbol{k}, v} c_{\boldsymbol{p}_{\perp}+\boldsymbol{p}_\parallel+\boldsymbol{k}, c} \hat n_{\boldsymbol{R}, i}^{c} c^\dagger_{\boldsymbol{p}_{\perp}+\boldsymbol{q}_\parallel+\boldsymbol{l}, c} c_{\boldsymbol{l}, v}\ket{\mathrm{GS}}\\
&= \sum_{\boldsymbol{R}, i, \boldsymbol{k}, \boldsymbol{l}, \boldsymbol{m}, \boldsymbol{n}} e^{\mathrm{i}\Delta R_{\parallel}} \bar \phi^{\boldsymbol{p}_{\perp}+\boldsymbol{p}_\parallel}_{\boldsymbol{k}} \phi^{\boldsymbol{p}_{\perp}+\boldsymbol{q}_\parallel}_{\boldsymbol{l}}\bar\psi^{\boldsymbol{m}, c}_{\boldsymbol{R}, i} \psi^{\boldsymbol{n}, c}_{\boldsymbol{R}, i} \bra{\mathrm{GS}} c^\dagger_{\boldsymbol{k}, v} c_{\boldsymbol{p}_{\perp}+\boldsymbol{p}_\parallel+\boldsymbol{k}, c} c^\dagger_{\boldsymbol{m}, c} c_{\boldsymbol{n}, c} c^\dagger_{\boldsymbol{p}_{\perp}+\boldsymbol{q}_\parallel+\boldsymbol{l}, c} c_{\boldsymbol{l}, v}\ket{\mathrm{GS}}\\
&= \sum_{\boldsymbol{R}, i, \boldsymbol{k}, \boldsymbol{l}, \boldsymbol{m}, \boldsymbol{n}} e^{\mathrm{i}\Delta R_{\parallel}} \bar \phi^{\boldsymbol{p}_{\perp}+\boldsymbol{p}_\parallel}_{\boldsymbol{k}} \phi^{\boldsymbol{p}_{\perp}+\boldsymbol{q}_\parallel}_{\boldsymbol{l}}\bar\psi^{\boldsymbol{m}, c}_{\boldsymbol{R}, i} \psi^{\boldsymbol{n}, c}_{\boldsymbol{R}, i} (\delta_{\boldsymbol{k}, \boldsymbol{l}} \delta_{\boldsymbol{p}_\perp + \boldsymbol{p}_\parallel +\boldsymbol{k}, \boldsymbol{m}} \delta_{\boldsymbol{p}_\perp + \boldsymbol{q}_\parallel +\boldsymbol{l}, \boldsymbol{n}})\\
&= \sum_{\boldsymbol{R}, i, \boldsymbol{k}} e^{\mathrm{i}\Delta R_{\parallel}} \bar \phi^{\boldsymbol{p}_{\perp}+\boldsymbol{p}_\parallel}_{\boldsymbol{k}} \phi^{\boldsymbol{p}_{\perp}+\boldsymbol{q}_\parallel}_{\boldsymbol{k}}\bar\psi^{\boldsymbol{p}_\perp + \boldsymbol{p}_\parallel +\boldsymbol{k}, c}_{\boldsymbol{R}, i} \psi^{\boldsymbol{p}_\perp + \boldsymbol{q}_\parallel +\boldsymbol{k}, c}_{\boldsymbol{R}, i}\\
&= \frac{1}{\mathcal{V}}\sum_{\boldsymbol{R}, i, \boldsymbol{k}} e^{\mathrm{i}\Delta R_{\parallel}}  e^{-\mathrm{i} \boldsymbol{R}\cdot \left(\boldsymbol{p}_\perp + \boldsymbol{p}_\parallel +\boldsymbol{k}\right)} e^{\mathrm{i} \boldsymbol{R}\cdot \left(\boldsymbol{p}_\perp + \boldsymbol{q}_\parallel +\boldsymbol{k}\right)}\bar \phi^{\boldsymbol{p}_{\perp}+\boldsymbol{p}_\parallel}_{\boldsymbol{k}} \phi^{\boldsymbol{p}_{\perp}+\boldsymbol{q}_\parallel}_{\boldsymbol{k}}\bar u^{\boldsymbol{p}_\perp + \boldsymbol{p}_\parallel +\boldsymbol{k}, c}_{i} u^{\boldsymbol{p}_\perp + \boldsymbol{q}_\parallel +\boldsymbol{k}, c}_{i}\\
&= \frac{1}{\mathcal{V}}\sum_{\boldsymbol{R}, i, \boldsymbol{k}} e^{\mathrm{i}R_{\parallel} (\Delta + |\boldsymbol{q}_\parallel| - |\boldsymbol{p}_\parallel|)} \bar \phi^{\boldsymbol{p}_{\perp}+\boldsymbol{p}_\parallel}_{\boldsymbol{k}} \phi^{\boldsymbol{p}_{\perp}+\boldsymbol{q}_\parallel}_{\boldsymbol{k}}\bar u^{\boldsymbol{p}_\perp + \boldsymbol{p}_\parallel +\boldsymbol{k}, c}_{i} u^{\boldsymbol{p}_\perp + \boldsymbol{q}_\parallel +\boldsymbol{k}, c}_{i}\\
&= \delta_{|\boldsymbol{p}_\parallel|, \Delta + |\boldsymbol{q}_\parallel|}
\sum_{i, \boldsymbol{k}} \bar \phi^{\boldsymbol{p}_{\perp}+\boldsymbol{p}_\parallel}_{\boldsymbol{k}} \phi^{\boldsymbol{p}_{\perp}+\boldsymbol{q}_\parallel}_{\boldsymbol{k}}\bar u^{\boldsymbol{p}_\perp + \boldsymbol{p}_\parallel +\boldsymbol{k}, c}_{i} u^{\boldsymbol{p}_\perp + \boldsymbol{q}_\parallel +\boldsymbol{k}, c}_{i}\\
&= \delta_{|\boldsymbol{p}_\parallel|, \Delta + |\boldsymbol{q}_\parallel|}
\sum_{\boldsymbol{k}} \bar \phi^{\boldsymbol{p}_{\perp}+\boldsymbol{p}_\parallel}_{\boldsymbol{k}} \phi^{\boldsymbol{p}_{\perp}+\boldsymbol{q}_\parallel}_{\boldsymbol{k}} \braket{u_{\boldsymbol{p}_\perp + \boldsymbol{p}_\parallel +\boldsymbol{k}, c}| u_{\boldsymbol{p}_\perp + \boldsymbol{q}_\parallel +\boldsymbol{k}, c}}.\\
\end{align}
We further simplify this using $|\boldsymbol{p}_\parallel|= \Delta + |\boldsymbol{q}_\parallel|$ this means that $\boldsymbol{p}_\parallel = \Delta \,\hat{\boldsymbol{\mathcal{G}}}_\parallel + \boldsymbol{q}_\parallel$ since $\hat{\boldsymbol{\mathcal{G}}}_\parallel$ is the unit vector along the reciprocal lattice vector direction (\emph{i.e.} pointing along the loop direction). Using this we rewrite the matrix elements as,
\begin{align}
[\mathcal{Z}_{c}(\boldsymbol{p}_{\perp})]_{\boldsymbol{p}_\parallel, \boldsymbol{q}_\parallel}&=\delta_{|\boldsymbol{p}_\parallel|, \Delta + |\boldsymbol{q}_\parallel|}
\sum_{\boldsymbol{k}} \bar \phi^{\boldsymbol{p}_{\perp}+\boldsymbol{q}_\parallel + \Delta \hat{\boldsymbol{\mathcal{G}}}_\parallel}_{\boldsymbol{k}} \phi^{\boldsymbol{p}_{\perp}+\boldsymbol{q}_\parallel}_{\boldsymbol{k}} \braket{u_{\boldsymbol{p}_\perp  +\boldsymbol{q}_\parallel + \Delta \hat{\boldsymbol{\mathcal{G}}}_\parallel +\boldsymbol{k}, c}| u_{\boldsymbol{p}_\perp + \boldsymbol{q}_\parallel +\boldsymbol{k}, c}}\\
&=\delta_{|\boldsymbol{p}_\parallel|, \Delta + |\boldsymbol{q}_\parallel|}
\sum_{\boldsymbol{k}} \bar \phi^{\boldsymbol{p}_{\perp}+\boldsymbol{q}_\parallel + \Delta \hat{\boldsymbol{\mathcal{G}}}_\parallel}_{\boldsymbol{k} - \boldsymbol{p}_\perp  -\boldsymbol{q}_\parallel}
\phi^{\boldsymbol{p}_{\perp}+\boldsymbol{q}_\parallel}_{\boldsymbol{k}- \boldsymbol{p}_\perp  -\boldsymbol{q}_\parallel} 
\braket{u_{\Delta\hat{\boldsymbol{\mathcal{G}}}_\parallel+\boldsymbol{k}, c}|u_{\boldsymbol{k}, c}}.
\end{align}
This expression is seen to be very similar in form to the matrix elements in 1D given in equation Eq.~\eqref{eq:matrixelements1DZv}. Similarly the $\mathcal{Z}_{v}(\boldsymbol{p}_{\perp})$ matrix elements are given by,
\begin{equation}
[\mathcal{Z}_{v}(\boldsymbol{p}_{\perp})]_{\boldsymbol{p}_\parallel, \boldsymbol{q}_\parallel}=\delta_{|\boldsymbol{p}_\parallel|, \Delta + |\boldsymbol{q}_\parallel|}
\sum_{\boldsymbol{k}} \bar \phi^{\boldsymbol{p}_{\perp}+\boldsymbol{q}_\parallel + \Delta \hat{\boldsymbol{\mathcal{G}}}_\parallel}_{\boldsymbol{k}}
\phi^{\boldsymbol{p}_{\perp}+\boldsymbol{q}_\parallel}_{\boldsymbol{k}+\Delta\hat{\boldsymbol{\mathcal{G}}}_\parallel} 
\braket{u_{\Delta\hat{\boldsymbol{\mathcal{G}}}_\parallel+\boldsymbol{k}, v}|u_{\boldsymbol{k}, v}}.
\end{equation}
Hence we can write the two operators as,
\begin{equation}
\hat{\mathcal{Z}}_{c/v}(\boldsymbol{p}_{\perp}) = \sum_{\boldsymbol{p}_{\parallel}} t^{\mathrm{exc}, c/v}_{\boldsymbol{p}_{\parallel}}(\boldsymbol{p}_{\perp}) \ket{(\boldsymbol{p}_{\perp}+\boldsymbol{p}_{\parallel}+\Delta\hat{\boldsymbol{\mathcal{G}}}_\parallel)_{\mathrm{exc}}} \bra{(\boldsymbol{p}_{\perp}+\boldsymbol{p}_{\parallel})_{\mathrm{exc}}},
\end{equation}
where,
\begin{align}
t^{\mathrm{exc}, c}_{\boldsymbol{p}_{\parallel}}(\boldsymbol{p}_{\perp})  &= \sum_{\boldsymbol{k}} \bar \phi^{\boldsymbol{p}_{\perp}+\boldsymbol{p}_\parallel + \Delta \hat{\boldsymbol{\mathcal{G}}}_\parallel}_{\boldsymbol{k} - \boldsymbol{p}_\perp  -\boldsymbol{p}_\parallel}
\phi^{\boldsymbol{p}_{\perp}+\boldsymbol{p}_\parallel}_{\boldsymbol{k}- \boldsymbol{p}_\perp  -\boldsymbol{p}_\parallel} 
\braket{u_{\Delta\hat{\boldsymbol{\mathcal{G}}}_\parallel+\boldsymbol{k}, c}|u_{\boldsymbol{k}, c}},\\
t^{\mathrm{exc}, v}_{\boldsymbol{p}_{\parallel}}(\boldsymbol{p}_{\perp})  &= \sum_{\boldsymbol{k}} \bar \phi^{\boldsymbol{p}_{\perp}+\boldsymbol{p}_\parallel + \Delta \hat{\boldsymbol{\mathcal{G}}}_\parallel}_{\boldsymbol{k}}
\phi^{\boldsymbol{p}_{\perp}+\boldsymbol{p}_\parallel}_{\boldsymbol{k}+\Delta\hat{\boldsymbol{\mathcal{G}}}_\parallel} 
\braket{u_{\Delta\hat{\boldsymbol{\mathcal{G}}}_\parallel+\boldsymbol{k}, v}|u_{\boldsymbol{k}, v}}.
\end{align}
The Wilson loops are constructed as in the 1D derivation above. We obtain,
\begin{equation}
W_{\mathrm{exc}, c/v}(\boldsymbol{p}_{\perp}) = \prod_{\boldsymbol{p}_\parallel} t^{\mathrm{exc}, c/v}_{\boldsymbol{p}_{\parallel}}(\boldsymbol{p}_{\perp}),
\end{equation}
where the product is over all $\boldsymbol{p}_\parallel$ in the 1D loop at fixed $\boldsymbol{p}_\perp$. Hence we obtain hybrid Wannier centres at fixed $\boldsymbol{p}_\perp$,
\begin{equation}
\lambda_{R, c/v}(\boldsymbol{p}_\perp) = \left[W_{\mathrm{exc}, c/v}(\boldsymbol{p}_{\perp})\right]^{1/L} e^{\mathrm{i}\Delta R},
\end{equation}
where $R \in \{0,1, \dots,  L-1\}$.
The hybrid Wannier states are of the form,
\begin{align}
\ket{\mathcal{W}^R_{\mathrm{exc}, c/v}(\boldsymbol{p}_{\perp})} = \frac{1}{\sqrt{L}}\sum_{\boldsymbol{p_\parallel}} e^{-\mathrm{i} R_{\parallel} |\boldsymbol{p}_{\parallel}|} \left[W_{\mathrm{exc}, c/v}(\boldsymbol{p}_{\perp})\right]^{-\frac{|\boldsymbol{p}_\parallel|}{2\pi}} 
\left(\prod_{\boldsymbol{k}_{\parallel} = 0}^{\boldsymbol{p}_\parallel-\Delta \hat{\boldsymbol{\mathcal{G}}}_{\parallel}} t^{\mathrm{exc}, c/v}_{\boldsymbol{k}_{\parallel}}(\boldsymbol{p}_{\perp})\right)\ket{(\boldsymbol{p}_{\perp}+\boldsymbol{p}_{\parallel})_{\mathrm{exc}}}.
\end{align}
The results in 1D therefore readily generalise to higher dimensions.

\newpage
\section{Exciton Berry connections}
\label{sec:ApxExcitonBerryConnections}
\subsection{Exciton Berry connection in 1D}
In the previous section we derived two exciton Wilson loops,
\begin{align}
W_{\mathrm{exc}, c} &= \prod_p \big[\sum_k \bar \phi^{p+\Delta}_{k- p} \phi^{p}_{k- p} \braket{u_{k+\Delta, c}| u_{k, c}}]\label{eq:conductionExcWilsonLoop}\\
W_{\mathrm{exc}, v} &= \prod_p \big[\sum_k \bar \phi^{p+\Delta}_{k} \phi^{p}_{k+\Delta} \braket{u_{k+\Delta, v}| u_{k, v}}]\label{eq:valenceExcWilsonLoop}.
\end{align}
These have modulus 1 in the thermodynamic limit ($L\rightarrow\infty$ and $\Delta\rightarrow0$) and so they can be written as $e^{\mathrm{i}\gamma_{\mathrm{exc}, c/v}}$ where $\gamma_{\mathrm{exc}, c/v}$ are exciton Berry phases. We find expressions for $\gamma_{\mathrm{exc}, c/v}$ by taking the thermodynamic limit of Eq.~\eqref{eq:conductionExcWilsonLoop} and Eq.~\eqref{eq:valenceExcWilsonLoop}.

We begin with $W_{\mathrm{exc}, c}$. We assume we have a smooth gauge for the exciton wave function $\phi^p_k$ and the periodic Bloch functions $\ket{u_{k, \alpha}}$. We can perform the Taylor expansions,
\begin{align}
W_{\mathrm{exc}, c} &= \prod_p \big[\sum_k \bar \phi^{p+\Delta}_{k- p} \phi^{p}_{k- p} \braket{u_{k+\Delta, c}| u_{k, c}}]\\
&= \prod_p \big[\sum_k \bar \phi^{p+\Delta}_{k} \phi^{p}_{k} \braket{u_{p+k+\Delta, c}| u_{p+k, c}}]\\
&=\prod_p \left\{\int_0^{2\pi} \frac{\mathrm{d}k}{\Delta} (\bar\phi^{p}_{k}+ \Delta \partial_p \bar\phi^{p}_{k}) \phi^{p}_{k} \left[\bra{u_{p+k, c}} + \Delta \partial_k\left(\bra{u_{p+k, c}}\right)\right] \ket{u_{p+k, c}}+\mathcal{O}(\Delta^2)\right\}\\
&=\prod_p \left\{\int_0^{2\pi} \frac{\mathrm{d}k}{\Delta}   (\bar\phi^{p}_{k} \phi^{p}_{k} + \Delta \phi^{p}_{k}\partial_p \bar\phi^{p}_{k} )  \left[1 + \Delta \partial_k\left(\bra{u_{p+k, c}}\right)\ket{u_{p+k, c}}\right]+\mathcal{O}(\Delta^2) \right\}\\
&=\prod_p \int_0^{2\pi} \frac{\mathrm{d}k}{\Delta}   \left[|\phi^{p}_{k}|^2 + \Delta \phi^{p}_{k}\partial_p \bar\phi^{p}_{k} +\Delta|\phi^{p}_{k}|^2 \partial_k\left(\bra{u_{p+k, c}}\right)\ket{u_{p+k, c}}+\mathcal{O}(\Delta^2)\right] \\
&=\prod_p \left\{1+ \Delta \int_0^{2\pi} \frac{\mathrm{d}k}{\Delta}\left[ \phi^{p}_{k}\partial_p \bar\phi^{p}_{k} +|\phi^{p}_{k}|^2 \partial_k\left(\bra{u_{p+k, c}}\right)\ket{u_{p+k, c}}\right]+\mathcal{O}(\Delta^2)\right\}
\end{align}
We use
\begin{equation}
\bra{u_{p+k, c}}\partial_k\ket{u_{p+k, c}} = -\partial_k\left(\bra{u_{p+k, c}}\right)\ket{u_{p+k, c}}
\end{equation}
and 
\begin{equation}
\int_0^{2\pi} \frac{\mathrm{d}k}{\Delta} \phi^{p}_{k}\partial_p \bar\phi^{p}_{k} = -\int_0^{2\pi} \frac{\mathrm{d}k}{\Delta}  \bar\phi^{p}_{k}\partial_p\phi^{p}_{k}.
\end{equation}
We drop the $\mathcal{O}(\Delta^2)$ terms in the thermodynamic limit,
\begin{align}
W_{\mathrm{exc}, c} &=\lim_{L\rightarrow\infty}\prod_p \left\{1- \Delta \int_0^{2\pi} \frac{\mathrm{d}k}{\Delta}\left[ \bar\phi^{p}_{k}\partial_p \phi^{p}_{k} +|\phi^{p}_{k}|^2 \bra{u_{p+k, c}}\partial_k\ket{u_{p+k, c}}\right]\right\}\\
&=\lim_{L\rightarrow\infty}\prod_p \left\{1+ \mathrm{i}\Delta \int_0^{2\pi} \frac{\mathrm{d}k}{\Delta}\left[ \bar\phi^{p}_{k}\mathrm{i}\partial_p \phi^{p}_{k} +|\phi^{p}_{k}|^2 A_{\mathrm{elec}, c}(p+k)\right]\right\}\\
&= \lim_{L\rightarrow\infty}\prod_p \exp\left\{\mathrm{i}\Delta \int_0^{2\pi} \frac{\mathrm{d}k}{\Delta}\left[ \bar\phi^{p}_{k}\mathrm{i}\partial_p \phi^{p}_{k} +|\phi^{p}_{k}|^2 A_{\mathrm{elec}, c}(p+k)\right]\right\}\\
&=\exp\left\{\mathrm{i}\int_{0}^{2\pi} \mathrm{d} p\int_0^{2\pi} \frac{\mathrm{d}k}{\Delta}\left[ \bar\phi^{p}_{k}\mathrm{i}\partial_p \phi^{p}_{k} +|\phi^{p}_{k}|^2 A_{\mathrm{elec}, c}(p+k)\right]\right\}.
\end{align}
We ignore terms of $\mathcal{O}(\Delta^2)$ in the thermodynamic limit. In the thermodynamic limit we can therefore write the Wilson loop $W_{\mathrm{exc}, c}$ as,
\begin{align}
W_{\mathrm{exc}, c} = \exp\left[\mathrm{i}\int_{0}^{2\pi} \mathrm{d}p \:A_{\mathrm{exc},c }(p)\right],
\end{align}
where the exciton Berry connection $A_{\mathrm{exc},c }(p)$ is
\begin{equation}
A_{\mathrm{exc},c }(p) = \int_0^{2\pi} \frac{\mathrm{d}k}{\Delta}\left[ \bar\phi^{p}_{k}\mathrm{i}\partial_p \phi^{p}_{k} +|\phi^{p}_{k}|^2 A_{\mathrm{elec}, c}(p+k)\right].
\end{equation}
We see that $A_{\mathrm{exc},c }(p) = A^{\alpha=0}_{\mathrm{exc}}(p)$.

We repeat this for the second exciton Wilson loop $W_{\mathrm{exc}, v}$.
\begin{align}
W_{\mathrm{exc}, v} &= \lim_{L\rightarrow\infty}\prod_p \big[\sum_k \bar \phi^{p+\Delta}_{k} \phi^{p}_{k+\Delta} \braket{u_{k+\Delta, v}| u_{k, v}}]\\
&= \lim_{L\rightarrow\infty}\prod_p \left\{\int_0^{2\pi} \frac{\mathrm{d}k}{\Delta} \left(\bar \phi^{p}_{k} + \Delta \partial_p\bar \phi^{p}_{k}\right) \left(\phi^{p}_{k}+\Delta\partial_k \phi^{p}_{k}\right) \left[\bra{u_{k, v}}+ \Delta\partial_k\left(\bra{u_{k, v}}\right)\right]\ket{u_{k, v}}\right\}\\
&= \lim_{L\rightarrow\infty}\prod_p \left[\int_0^{2\pi} \frac{\mathrm{d}k}{\Delta} |\phi^{p}_{k}|^2 + \Delta (\partial_p\bar \phi^{p}_{k}) \phi^{p}_{k} +\Delta\bar \phi^{p}_{k}\partial_k \phi^p_k +  |\phi^{p}_{k}|^2 \Delta\partial_k\left(\bra{u_{k, v}}\right)\ket{u_{k, v}}\right]\\
&= \lim_{L\rightarrow\infty}\prod_p \left[1+\Delta \int_0^{2\pi} \frac{\mathrm{d}k}{\Delta}(\partial_p\bar \phi^{p}_{k}) \phi^{p}_{k} +\bar \phi^{p}_{k}\partial_k \phi^p_k +|\phi^{p}_{k}|^2 \Delta\partial_k\left(\bra{u_{k, v}}\right)\ket{u_{k, v}}\right]\\
&= \lim_{L\rightarrow\infty}\prod_p \left[1+\mathrm{i}\Delta \int_0^{2\pi} \frac{\mathrm{d}k}{\Delta}\left(\bar\phi^{p}_{k}\mathrm{i}\partial_p \phi^{p}_{k} -\bar \phi^{p}_{k}\mathrm{i}\partial_k \phi^p_k  + |\phi^{p}_{k}|^2 A_{\mathrm{elec}, v}(k)\right)\right]\\
&= \lim_{L\rightarrow\infty}\prod_p \exp\left[\mathrm{i}\Delta \int_0^{2\pi} \frac{\mathrm{d}k}{\Delta}\left(\bar\phi^{p}_{k}\mathrm{i}\partial_p \phi^{p}_{k}  -\bar \phi^{p}_{k}\mathrm{i}\partial_k \phi^p_k+ |\phi^{p}_{k}|^2 A_{\mathrm{elec}, v}(k)\right)\right]\\
&= \exp\left[\mathrm{i}\int_{0}^{2\pi}\mathrm{d}p\: \int_0^{2\pi} \frac{\mathrm{d}k}{\Delta}\left(\bar\phi^{p}_{k}\mathrm{i}\partial_p \phi^{p}_{k} -\bar \phi^{p}_{k}\mathrm{i}\partial_k \phi^p_k + |\phi^{p}_{k}|^2 A_{\mathrm{elec}, v}(k)\right)\right]
\end{align}
We see that,
\begin{align}
W_{\mathrm{exc}, v} = \exp\left[\mathrm{i}\int_{0}^{2\pi} \mathrm{d}p \:A_{\mathrm{exc},v }(p)\right],
\end{align}
where the exciton Berry connection $A_{\mathrm{exc},v }(p)$ is
\begin{align}
A_{\mathrm{exc},v }(p) &= \int_0^{2\pi} \frac{\mathrm{d}k}{\Delta}\left(\bar\phi^{p}_{k}\mathrm{i}\partial_p \phi^{p}_{k} -\bar \phi^{p}_{k}\mathrm{i}\partial_k \phi^p_k + |\phi^{p}_{k}|^2 A_{\mathrm{elec}, v}(k)\right)\\
&= \int_0^{2\pi} \frac{\mathrm{d}k}{\Delta}\left(\bar\phi^{p}_{k-p}\mathrm{i}\partial_p \phi^{p}_{k-p} + |\phi^{p}_{k-p}|^2 A_{\mathrm{elec}, v}(k-p)\right)
\end{align}
Therefore, $A_{\mathrm{exc},v }(p) = A^{\alpha=1}_{\mathrm{exc}}(p)$.
\subsection{Exciton Berry connection in higher dimensions}
In higher dimensions, the exciton Wilson loops can be written in terms of the Berry connection in a similar way,
\begin{align}
W_{\mathrm{exc}, c/v}(\boldsymbol{p}_{\perp}) &= \exp \left(\mathrm{i} \int_l \boldsymbol{A}_{\mathrm{exc}, c/v}(\boldsymbol{p}_{\perp}+\boldsymbol{p}_{\parallel})\cdot \mathrm{d}\boldsymbol{p}_{\parallel}\right)\\
&= \exp \left(\mathrm{i} \int_{0}^{2\pi} \left[\boldsymbol{A}_{\mathrm{exc}, c/v}(\boldsymbol{p}_{\perp}+\boldsymbol{p}_{\parallel})\right]_{\parallel} \mathrm{d}p_{\parallel}\right)
\end{align}
where $l$ is the BZ loop at $\boldsymbol{p}_{\perp}$ and $\left[\boldsymbol{A}_{\mathrm{exc}, c/v}(\boldsymbol{p}_{\perp}+\boldsymbol{p}_{\parallel})\right]_{\parallel}$ is the component of the vector Berry connection along the direction of the path $l$ in the BZ. We define the unit vector tangent to the path $l$ as $\hat{\boldsymbol{n}}_\parallel$ so $\left[\boldsymbol{A}_{\mathrm{exc}, c/v}(\boldsymbol{p}_{\perp}+\boldsymbol{p}_{\parallel})\right]_{\parallel} = \boldsymbol{A}_{\mathrm{exc}, c/v}(\boldsymbol{p}_{\perp}+\boldsymbol{p}_{\parallel})\cdot \hat{\boldsymbol{n}}_\parallel$  We find that,
\begin{align}
\left[\boldsymbol{A}_{\mathrm{exc},c}(\boldsymbol{p}_{\perp}+\boldsymbol{p}_{\parallel})\right]_{\parallel} &= \int_{BZ} \frac{\mathrm{d}^d\boldsymbol{k}}{\Delta^d}\left[ \bar\phi^{\boldsymbol{p}_{\perp}+\boldsymbol{p}_{\parallel}}_{\boldsymbol{k}}\mathrm{i}\:\boldsymbol{\hat{n}_\parallel}\cdot \boldsymbol{\nabla}_{\boldsymbol{p}_{\parallel}} \phi^{\boldsymbol{p}_{\perp}+\boldsymbol{p}_{\parallel}}_{\boldsymbol{k}} +|\phi^{\boldsymbol{p}_{\perp}+\boldsymbol{p}_{\parallel}}_{\boldsymbol{k}}|^2 \boldsymbol{\hat{n}_\parallel} \cdot \boldsymbol{A}_{\mathrm{elec}, c}(\boldsymbol{p}_{\perp}+\boldsymbol{p}_{\parallel}+\boldsymbol{k})\right],\\
\left[\boldsymbol{A}_{\mathrm{exc},v}(\boldsymbol{p}_{\perp}+\boldsymbol{p}_{\parallel})\right]_{\parallel} &= \int_{BZ}  \frac{\mathrm{d}^d\boldsymbol{k}}{\Delta^d}\left[ \bar\phi^{\boldsymbol{p}_{\perp}+\boldsymbol{p}_{\parallel}}_{\boldsymbol{k}-\boldsymbol{p}_{\perp}-\boldsymbol{p}_{\parallel}}\mathrm{i}\:\boldsymbol{\hat{n}_\parallel}\cdot \boldsymbol{\nabla}_{\boldsymbol{p}_{\parallel}} \phi^{\boldsymbol{p}_{\perp}+\boldsymbol{p}_{\parallel}}_{\boldsymbol{k}-\boldsymbol{p}_{\perp}-\boldsymbol{p}_{\parallel}} +|\phi^{\boldsymbol{p}_{\perp}+\boldsymbol{p}_{\parallel}}_{\boldsymbol{k}-\boldsymbol{p}_{\perp}-\boldsymbol{p}_{\parallel}}|^2 \boldsymbol{\hat{n}_\parallel} \cdot \boldsymbol{A}_{\mathrm{elec}, v}(\boldsymbol{k}-\boldsymbol{p}_{\perp}-\boldsymbol{p}_{\parallel})\right],
\end{align}
using $\boldsymbol{A}_{\mathrm{elec}, c/v}(\boldsymbol{k}) = \bra{u_{\boldsymbol{k}, c/v}} \mathrm{i}\boldsymbol{\nabla}_{\boldsymbol{k}} \ket{u_{\boldsymbol{k}, c/v}}$. It follows that the vector exciton Berry connections are
\begin{align}
\boldsymbol{A}_{\mathrm{exc},c}(\boldsymbol{p}) &= \int_{BZ}  \frac{\mathrm{d}^d\boldsymbol{k}}{\Delta^d}\left[ \bar\phi^{\boldsymbol{p}}_{\boldsymbol{k}}\mathrm{i}\boldsymbol{\nabla}_{\boldsymbol{p}} \phi^{\boldsymbol{p}}_{\boldsymbol{k}} +|\phi^{\boldsymbol{p}}_{\boldsymbol{k}}|^2\boldsymbol{A}_{\mathrm{elec}, c}(\boldsymbol{p}+\boldsymbol{k})\right]\\
\boldsymbol{A}_{\mathrm{exc},v}(\boldsymbol{p}) &= \int_{BZ}  \frac{\mathrm{d}^d\boldsymbol{k}}{\Delta^d}\left[ \bar\phi^{\boldsymbol{p}}_{\boldsymbol{k}-\boldsymbol{p}}\mathrm{i} \boldsymbol{\nabla}_{\boldsymbol{p}} \phi^{\boldsymbol{p}}_{\boldsymbol{k}-\boldsymbol{p}} +|\phi^{\boldsymbol{p}}_{\boldsymbol{k}-\boldsymbol{p}}|^2 \boldsymbol{A}_{\mathrm{elec}, v}(\boldsymbol{k}-\boldsymbol{p})\right].
\end{align}

\subsection{Exciton Berry connections for all $\alpha$}
Here we demonstrate the relationship between the different possible projected position operators and the infinite family of Berry connections introduced in Sec.~\ref{sec:ApxFirstExcitonBerryPhaseSec}. If we want to study the location of the weighted centre of mass of the exciton,
\begin{equation}
R_\alpha = \alpha r_{v}+(1-\alpha) r_{c}.
\end{equation}
This weighted mixture is parametrised just by a single parameter ($\alpha$) since we want the prefactors ($1-\alpha$ and $\alpha$) to add up to 1. 

In the previous section we saw that $\frac{\gamma_{\mathrm{exc}, c}}{2\pi}$ is the shift of the electron from the centre of unit cell when we maximally localise the electron within the exciton. Similarly, $\frac{\gamma_{\mathrm{exc}, v}}{2\pi}$ is the shift of the hole when we instead maximally localise the hole within the exciton. The electron/hole centres within the exciton ($r_{c/v}$) are therefore given by the exciton Berry phases,
\begin{equation}
r_{c/v} = \frac{1}{2\pi}\int_0^{2\pi}  A_{\mathrm{elec}, c/v}(p) \:\mathrm{d}p,
\end{equation}
and so the weighted position $R_\alpha$ is,
\begin{equation}
R_{\alpha} = \frac{1}{2\pi}\left[  \alpha\int \mathrm{d}p\:A_{\mathrm{exc}, v}(p) + (1-\alpha)\int \mathrm{d}p\: A_{\mathrm{exc}, c}(p)\right].
\end{equation}
So the appropriate Berry connection to calculate the weighted mixture $R_\alpha$ is identical to the exciton Berry connection [$A^{\alpha}_{\mathrm{exc}}(p)$] introduced in Eq.~\eqref{eq:ApxnaiveBerryConnection},
\begin{equation}
A^\alpha_{\mathrm{exc}}(p) = \alpha A_{\mathrm{exc}, v}(p) + (1-\alpha) A_{\mathrm{exc}, c}(p)
\end{equation}
The exciton Berry connections introduced in Sec.~\ref{sec:ApxFirstExcitonBerryPhaseSec} are therefore related to the  location of the different weighted mixtures of electron and hole positions ($R_\alpha$).

Similarly in higher dimensions we have 
\begin{equation}
\boldsymbol{A}^\alpha_{\mathrm{exc}}(p) = \alpha \boldsymbol{A}_{\mathrm{exc}, v}(p) + (1-\alpha) \boldsymbol{A}_{\mathrm{exc}, c}(p)
\end{equation}
where,
\begin{equation}
\begin{aligned}
\boldsymbol{A}^\alpha_{\mathrm{exc}}(\boldsymbol{p}) =  \int_{BZ}  \frac{\mathrm{d}^d\boldsymbol{k}}{\Delta^d}  \bar \phi^{\boldsymbol{p}}_{\boldsymbol{k} - \alpha \boldsymbol{p}} \mathrm{i} \boldsymbol{\nabla}_{\boldsymbol{p}} \phi^{\boldsymbol{p}}_{\boldsymbol{k} - \alpha \boldsymbol{p}} +|\phi^{\boldsymbol{p}}_{\boldsymbol{k} - \alpha \boldsymbol{p}}|^2 \big\{ (1-\alpha) \boldsymbol{A}_{\mathrm{elec}, c}\big[(1-\alpha) \boldsymbol{p}+\boldsymbol{k}\big] + \alpha \boldsymbol{A}_{\mathrm{elec}, v}\big[-\alpha \boldsymbol{p}+\boldsymbol{k}\big]\big\}.
\end{aligned}
\end{equation}

\newpage
\section{The electric polarisation of excitons}
\label{sec:ApxElectronicPolarisationExcitons}
In this section, we show that the difference between the two exciton Berry connections [$A_{\mathrm{exc},c/v }(p)$] gives the electric polarisation of the exciton at each momentum $p$. The difference between the exciton Berry connections is,
\begin{align}
&\mathcal{F}(p) = A_{\mathrm{exc}, c}(p) - A_{\mathrm{exc}, v}(p)~\label{eq:ApxFofp}\\
&= \int_0^{2\pi} \frac{\mathrm{d}k}{\Delta}\:\left\{\bar\phi^p_k \mathrm{i} \partial_k  \phi^p_k + |\phi^p_k|^2 \left[A_{\mathrm{elec},c}(p+k)  - A_{\mathrm{elec},v}(k)\right]\right\}.
\end{align}
This can be shown to be gauge invariant under both gauge transformations of the electronic bands and the total exciton wave function. We now show that it is equal to the electric polarisation of the exciton at momentum $p$. The electric polarisation is equal to the expectation value of the electron-hole distance in the exciton (in units where the charge of the electron is $e = 1$). To calculate this, we begin by rewriting the exciton wave function ($\phi^p_k$) at total momentum $p$ in real space. We transform the exciton eigenstate into the basis of the maximally localised \emph{electronic} Wannier functions. The maximally localised \emph{electronic} Wannier functions are defined as,
\begin{equation}
c^\dagger_{R, c/v} = \frac{1}{\sqrt{L}} \sum_k e^{-\mathrm{i} kR} c^\dagger_{k, c/v},
\end{equation}
where we assume that the $c^\dagger_{k, c/v}$ are in the gauge where the  electronic Wannier function are maximally localised. In this basis we can write the exciton eigenstate as,
\begin{align}
\ket{p_{\mathrm{exc}}} &= \sum_k \phi^p_k c^\dagger_{p+k, c} c_{k, v} \ket{\mathrm{GS}}\\
&= \frac{1}{L}\sum_{k, R, r} \phi^p_k e^{\mathrm{i} (p+k)(R+r)} e^{-\mathrm{i} kR}c^\dagger_{R+r, c} c_{R, v} \ket{\mathrm{GS}}\\
&= \frac{1}{L}\sum_{k, R, r} \phi^p_k e^{\mathrm{i} p(R+r)} e^{\mathrm{i} kr}c^\dagger_{R+r, c} c_{R, v} \ket{\mathrm{GS}}\\
&= \frac{1}{L}\sum_{R, r} \left(\sum_k \phi^p_k e^{\mathrm{i} p(R+r)} e^{\mathrm{i} kr}\right)c^\dagger_{R+r, c} c_{R, v} \ket{\mathrm{GS}}\\
&= \sum_{R, r} \phi^p_{R, r} c^\dagger_{R+r, c} c_{R, v} \ket{\mathrm{GS}},
\end{align}
where we define,
\begin{equation}
\phi^p_{R, r} = \frac{1}{L}\sum_k \phi^p_k e^{\mathrm{i} p(R+r)} e^{\mathrm{i} kr}.
\end{equation}
We are interested in the electron-hole distance $r$. Since we have translational symmetry over $R$ then we can study the distribution over $r$ for $R = 0$ without loss of generality. Therefore we define $\phi^p_{r}\equiv \sqrt{L} \:\phi^p_{R=0, r}$ \emph{i.e.}
\begin{equation}
\phi^p_{r} = \frac{e^{\mathrm{i}p r}}{\sqrt{L} }\sum_{k}\phi^p_k e^{\mathrm{i} kr},
\end{equation}
this is normalised such that $\sum_{r} |\phi^p_r|^2 = 1$. We now use $\phi^p_r$ to calculate the expectation value of the electron-hole separation within the exciton. This consists of two contributions: the first is the relative coordinate, $r$, which represents the electron-hole separation in \emph{unit cell} coordinates. In addition there is a contribution from the positions of the electronic Wannier centres within the unit cell. The electric polarisation of the excitons is therefore equal to the expectation value of $\xi = r + x_c - x_v$ where $x_{c/v}$ are the electronic Wannier centres. Even if the unit cell separation $r = 0$ then there can still be a contribution to the electron-hole distance $\xi$ from the (possibly differing) electronic Wannier centres for the electron and the hole. We would like to evaluate the expectation value,
\begin{equation}
\langle \xi\rangle_p = \sum_{r} (r+x_c - x_v)|\phi^p_{r}|^2.
\end{equation}
Instead of calculating this directly though, we instead relate it to another expectation value ($\langle e^{\mathrm{i} \xi \Delta}\rangle_p$) which is easier to calculate,
\begin{align}
\langle e^{\mathrm{i} \xi \Delta}\rangle_p &= \langle 1 + \mathrm{i} \xi \Delta + \mathcal{O}\left(\xi^2 \Delta^2\right)\rangle_p\\
&= \langle 1 \rangle_p + \langle \mathrm{i} \xi \Delta \rangle_p+ \langle\mathcal{O}\left(\xi^2 \Delta^2\right)\rangle_p\\
&= 1 + \mathrm{i} \Delta \langle  \xi \rangle_p+ \mathcal{O}\left( \Delta^2\right)\langle\xi^2\rangle_p.
\end{align}
We can Taylor expand $e^{\mathrm{i} \xi \Delta}$ in powers of $\xi$ (or equivalently $r$ since $\xi \propto r$) because the excitons are bound so $\phi^p_r$ decays exponentially in $r$. Therefore only terms with $\xi \ll L$ (or equivalently $\xi\Delta \ll 1$) contribute. Therefore $\langle \xi \rangle_p$ can be calculated using $\langle e^{\mathrm{i} \xi \Delta}\rangle_p$, 
\begin{equation}
\langle  \xi \rangle_p = \lim_{\Delta\rightarrow0}\frac{\langle e^{\mathrm{i} \xi \Delta}\rangle_p - 1}{\mathrm{i}\Delta}.\label{eq:expXiCalc}
\end{equation}
\\
We begin by calculating the expectation value,
\begin{align}
\langle e^{\mathrm{i} \xi \Delta}\rangle_p = \sum_{r} e^{\mathrm{i}\Delta (r+x_c - x_v)}|\phi^p_{r}|^2.\label{eq:ExpectationValuePolarisation}
\end{align}
This evaluates to,
\begin{align}
\langle e^{\mathrm{i} \xi \Delta}\rangle_p &= \frac{1}{L} \sum_{r} e^{\mathrm{i} (r+x_c - x_v)\Delta } \left|\sum_k \phi^p_k e^{\mathrm{i}kr} \right|^2\\
&= \frac{1}{L} \sum_{r, k, k'} e^{\mathrm{i} (r+x_c - x_v)\Delta }  \bar \phi^p_{k'}\phi^p_k  e^{\mathrm{i}(k - k')r} \\
&= \frac{1}{L} \sum_{r, k, k'} e^{\mathrm{i}(k - k' + \Delta)r} e^{\mathrm{i} (x_c - x_v)\Delta }  \bar \phi^p_{k'}\phi^p_k   \\
&=  \sum_{k, k'} \delta_{k', k+\Delta} e^{\mathrm{i} (x_c - x_v)\Delta }  \bar \phi^p_{k'}\phi^p_k   \\
&=e^{\mathrm{i} (x_c - x_v)\Delta } \sum_{k} \bar \phi^p_{k}\phi^p_{k - \Delta}.
\end{align}
We assume $L$ is large (or equivalently $\Delta$ small) and Taylor expand,
\begin{align}
\langle e^{\mathrm{i} \xi \Delta}\rangle_p &= \left[1+\mathrm{i} (x_c - x_v)\Delta \right] \int_0^{2\pi}\frac{\mathrm{d}k}{\Delta} \bar \phi^p_{k}\left(\phi^p_{k}  -\Delta \partial_k \phi^p_k\right) + \mathcal{O}(\Delta^2)\\
&= 1 + \mathrm{i}\Delta \left\{\left[ \int_0^{2\pi}\frac{\mathrm{d}k}{\Delta}\: \left(\bar \phi^p_k\mathrm{i}\partial_k \phi^p_k\right)\right] + (x_c -x_v)\right\} + \mathcal{O}(\Delta^2)\\
&= 1 + \mathrm{i}\Delta \left\{\left[ \int_0^{2\pi}\frac{\mathrm{d}k}{\Delta}\: \left(\bar \phi^p_k \mathrm{i}\partial_k \phi^p_k\right)\right] + \frac{\gamma_c}{2\pi}  -\frac{\gamma_v}{2\pi} \right\} + \mathcal{O}(\Delta^2).
\end{align}
Here we expressed the electronic Wannier centres ($x_{c,v}$) in terms of the electronic Berry phases $\gamma_{c/v}$. This expression is written in the gauge which maximally localises the \emph{electronic} Wannier states. We wish to manipulate this expression into something which is gauge invariant so we can evaluate it in any (smooth) gauge. Firstly we can turn the last two terms into an integral over $k$,
\begin{align}
\lim_{L\rightarrow \infty}\langle e^{\mathrm{i} \xi \Delta}\rangle_p &= 1 + \mathrm{i}\Delta \left\{\left[ \int_0^{2\pi}\frac{\mathrm{d}k}{\Delta}\: \left(\bar \phi^p_k \mathrm{i}\partial_k \phi^p_k\right)\right] + \frac{\gamma_c}{2\pi}  -\frac{\gamma_v}{2\pi} \right\} + \mathcal{O}(\Delta^2)\\
&= 1 + \mathrm{i}\Delta \left\{\int_0^{2\pi}\frac{\mathrm{d}k}{\Delta}\: \left[\bar \phi^p_k \mathrm{i}\partial_k \phi^p_k +  |\phi^p_k|^2 \left(\frac{\gamma_c}{2\pi}  -\frac{\gamma_v}{2\pi}\right)\right]\right\} + \mathcal{O}(\Delta^2),
\end{align}
using the normalisation of the wave function $\int_0^{2\pi}\frac{\mathrm{d}k}{\Delta}\: |\phi^p_k|^2 = 1$. In the gauge which maximally localises the electronic Wannier states, the electronic Berry connection is equal to $\frac{\gamma_{c/v}}{2\pi}$~\cite{MarzariVanderbiltSmoothGaugePRB97}. Therefore we replace these terms with the general expressions for the electronic Berry connections $\bra{u_{k, c/v}} \mathrm{i} \partial_k \ket{u_{k, c/v}}$. There is a unique way to do this to ensure the final expression is gauge invariant,
\begin{align}
\langle e^{\mathrm{i} \xi \Delta}\rangle_p 
&= 1 + \mathrm{i}\Delta \left\{\int_0^{2\pi}\frac{\mathrm{d}k}{\Delta}\: \left(\bar\phi^p_k \mathrm{i} \partial_k  \phi^p_k + |\phi^p_k|^2 \bra{u_{p+k, c}} \mathrm{i} \partial_k \ket{u_{p+k, c}}  - |\phi^p_k|^2 \bra{u_{k, v}} \mathrm{i} \partial_k \ket{u_{k, v}}\right) \right\} + \mathcal{O}(\Delta^2)\\
&= 1 + \mathrm{i}\Delta \:\mathcal{F}(p) + \mathcal{O}(\Delta^2).
\end{align}
We have identified that the $\mathcal{O}(\Delta)$ term can be expressed in terms of the $\mathcal{F}(p)$ introduced earlier. Finally we use Eq.~\eqref{eq:expXiCalc} to find $\langle  \xi \rangle_p$,
\begin{equation}
\langle  \xi \rangle_p = \mathcal{F}(p).
\end{equation}
To summarise, the difference between the two exciton Berry connections is a gauge invariant quantity which is equal to the electric polarisation of the exciton. It is desirable to have a formulation of $\mathcal{F}(p)$ that does not require a smooth gauge for the electronic Bloch functions. This will allow it to be easily calculated using standard electronic structure methods. We find this expression in the next section.

\subsection{$\mathcal{F}(p)$ without a smooth gauge}
In this section we obtain an expression for $\mathcal{F}(p)$ [Eq.~\eqref{eq:ApxFofp}] which does not require a smooth gauge for the Bloch functions. Since the expression only involves the exciton wave function $\phi^p_k$ for one total momentum $p$, we already have an expression that does not need a smooth gauge for the exciton wave function over $p$. We begin with the expression for $\mathcal{F}(p)$ in the \emph{electronic} gauge which maximally localises the electronic Wannier functions (\emph{i.e.} the twisted parallel transport gauge~\cite{MarzariVanderbiltSmoothGaugePRB97}). We denote the exciton wave function in this gauge choice as $\tilde \phi^p_k$ with creation operators for the bands (in the twisted parallel gauge) denoted as $\tilde c^\dagger_{k, c/v}$. Therefore the exciton state can be written as,
\begin{equation}
\ket{p_{\mathrm{exc}}} = \sum_k \tilde \phi^p_k \tilde c^\dagger_{p+k, c} \tilde c_{k, v} \ket{\mathrm{GS}}.
\end{equation}
The expression for $\langle \xi\rangle$ in this gauge is,
\begin{align}
\langle \xi \rangle &= - \lim_{\Delta \rightarrow 0}\left[\frac{\mathrm{i}}{\Delta}\log \left(e^{\mathrm{i} (x_c - x_v)\Delta } \sum_{k}  (\tilde \phi^p_{k})^* \tilde \phi^p_{k - \Delta}\right)\right]\\
&= - \lim_{\Delta \rightarrow 0}\left[\frac{\mathrm{i}}{\Delta}\log \left((W_c)^{\frac{1}{L}} (W_v)^{-\frac{1}{L}}  \sum_{k}  (\tilde \phi^p_{k})^* \tilde \phi^p_{k - \Delta}\right)\right].\label{eq:ApxxiExpression}
\end{align}
We write the exciton wave function in some general (and not necessarily smooth) electronic gauge as $\phi^p_k$. The corresponding creation operators for the bands we write as $c^\dagger_{p, c/v}$. We want to express $\tilde \phi^p_k$ in terms of $\phi^p_k$. If we make gauge transformations of the electron bands ($ c^\dagger_{p, v} =  e^{\mathrm{i}\eta(p)}\tilde c^\dagger_{p, v}$ and $ c^\dagger_{p, c}= e^{\mathrm{i}\xi(p)}\tilde c^\dagger_{p, c}$) then this should still leave the exciton state $\ket{p_{\mathrm{exc}}}$ invariant. Hence the exciton wave function must transform oppositely to the electronic bands, 
\begin{equation}
\phi^p_{k} = e^{\mathrm{i}\eta ( k)} e^{-\mathrm{i} \xi(p + k)} \tilde \phi^{p}_{k}.
\end{equation} 
We calculate the phase factors which transform the electronic bands from some arbitrary non-smooth gauge to the twisted parallel transport gauge. In Sec.~\ref{Sec:ApxElectronicProjPosition} (and described in Ref.~\onlinecite{MarzariVanderbiltSmoothGaugePRB97}) we saw that these were,
\begin{align}
e^{-\mathrm{i}\xi(p)} &= (W_c)^{-\frac{p}{2\pi}} \prod_{q = 0}^{p-\Delta} \braket{u_{q+\Delta, c}|  u_{q, c}},\\
e^{-\mathrm{i}\eta(p)} &= (W_v)^{-\frac{p}{2\pi}} \prod_{q = 0}^{p-\Delta} \braket{u_{q+\Delta, v}|  u_{q, v}}.
\end{align}
Therefore we can relate $\phi^p_k$ and $\tilde \phi^p_k$ via,
\begin{equation}
\tilde \phi^p_k = \left[(W_c)^{\frac{p+k}{2\pi}} (W_v)^{-\frac{k}{2\pi}} \prod_{q = 0}^{p+k - \Delta} \braket{u_{q, c}|  u_{q+\Delta, c}} \prod_{l = 0}^{k- \Delta} \braket{u_{l+\Delta, v}|  u_{l, v}}\right]  \phi^p_k.
\end{equation}
We rewrite the $\tilde \phi^p_k$ dependent part of  Eq.~\eqref{eq:ApxxiExpression} in terms of the arbitrary gauge exciton wave function $ \phi^p_k$,
\begin{align}
\sum_{k} (\tilde \phi^p_k)^* \tilde \phi^p_{k-\Delta} &= \sum_k \left[(W_c)^{-\frac{p+k}{2\pi}} (W_v)^{\frac{k}{2\pi}}  \prod_{q = 0}^{p+k- \Delta} (\braket{u_{q, c}|  u_{q+\Delta, c}})^*\prod_{l = 0}^{k- \Delta} (\braket{u_{l+\Delta, v}|  u_{l, v}})^*\right] \nonumber \\&\qquad\cdot \left[(W_c)^{\frac{p+k-\Delta}{2\pi}} (W_v)^{-\frac{k-\Delta}{2\pi}}  \prod_{q = 0}^{p+k - 2\Delta} \braket{u_{q, c}|  u_{q+\Delta, c}} \prod_{l = 0}^{k- 2\Delta} \braket{u_{l+\Delta, v}|  u_{l, v}} \right]\left(\phi^p_k\right)^* \phi^p_{k-\Delta}\\
&= \sum_k \left(\phi^p_k\right)^* \phi^p_{k-\Delta}(W_c)^{-\frac{\Delta}{2\pi}} (W_v)^{\frac{\Delta}{2\pi}} \braket{u_{p+k,c}| u_{p+k-\Delta, c}} \braket{u_{k - \Delta,v}| u_{k, v}}\\
&= (W_c)^{-\frac{1}{L}} (W_v)^{\frac{1}{L}} \sum_k \left(\phi^p_k\right)^*  \phi^p_{k-\Delta}\braket{u_{p+k,c}| u_{p+k-\Delta, c}} \braket{u_{k - \Delta,v}| u_{k, v}}
\end{align}
Therefore we can find an expression for $\langle \xi \rangle$,
\begin{align}
\langle \xi \rangle &= - \lim_{\Delta \rightarrow 0}\left[\frac{\mathrm{i}}{\Delta}\log \left((W_c)^{\frac{1}{L}} (W_v)^{-\frac{1}{L}}  \sum_{k}  (\tilde \phi^p_{k}) \tilde\phi^p_{k - \Delta}\right)\right]\\ &=- \lim_{\Delta \rightarrow 0}\left[\frac{\mathrm{i}}{\Delta}\log \left( \sum_k \left(\phi^p_k\right)^* \phi^p_{k-\Delta}\braket{u_{p+k,c}| u_{p+k-\Delta, c}} \braket{u_{k - \Delta,v}| u_{k, v}}\right)\right].
\end{align}
We have obtained a \emph{gauge invariant} expression for $\langle \xi \rangle$ which does not require a smooth gauge for the exciton wave function and the electronic Bloch functions,
\begin{equation}
\langle \xi \rangle \approx -\mathrm{i}\frac{L}{2 \pi}\log \left( \sum_k \bar \phi^p_k \phi^p_{k-\Delta}\braket{u_{p+k,c}| u_{p+k-\Delta, c}} \braket{u_{k - \Delta,v}| u_{k, v}}\right),
\end{equation}
where the expression becomes an equality in the limit $\Delta \rightarrow 0$ (or equivalently $L\rightarrow \infty$). Or equivalently,
\begin{equation}
e^{\mathrm{i}\Delta \mathcal{F}(p)} =\lim_{L\rightarrow\infty}\sum_k \bar \phi^p_k \phi^p_{k-\Delta}\braket{u_{p+k,c}| u_{p+k-\Delta, c}} \braket{u_{k - \Delta,v}| u_{k, v}}.
\end{equation}
\subsection{Generalisation to higher dimensions}\label{apx:PolarisationHigherDim}
The difference between the (vector) exciton Berry connections is,
\begin{align}
&\boldsymbol{\mathcal{F}}(\boldsymbol{p}) = \boldsymbol{A}_{\mathrm{exc}, c}(\boldsymbol{p}) - \boldsymbol{A}_{\mathrm{exc}, v}(\boldsymbol{p})\\
&= \int_{BZ}  \frac{\mathrm{d}^d\boldsymbol{k}}{\Delta^d}\:\left\{\bar\phi^{\boldsymbol{p}}_{\boldsymbol{k}} \:\mathrm{i} \boldsymbol{\nabla}_{\boldsymbol{k}} \phi^{\boldsymbol{p}}_{\boldsymbol{k}} + |\phi^{\boldsymbol{p}}_{\boldsymbol{k}}|^2 \left[\boldsymbol{A}_{\mathrm{elec},c}(\boldsymbol{p}+\boldsymbol{k})  - \boldsymbol{A}_{\mathrm{elec},v}(\boldsymbol{k})\right]\right\}.
\end{align}
It also follows directly from the 1D result that $\boldsymbol{\mathcal{F}}(\boldsymbol{p})$ is the polarisation of the exciton (which is now also a vector in higher dimensions). This is because we can calculate the expectation value in Eq.~\eqref{eq:ExpectationValuePolarisation} for different spatial directions and obtain the components of $\boldsymbol{\mathcal{F}}(\boldsymbol{p})$. We can again construct a gauge invariant expression for the components of the polarisation. If we have an orthonormal basis with basis vectors $\hat{\boldsymbol{n}}_i$, then the component of the polarisation along $\hat{\boldsymbol{n}}_i$ is,
\begin{equation}
e^{\mathrm{i}\Delta \boldsymbol{\mathcal{F}}(\boldsymbol{p})\cdot \hat{\boldsymbol{n}}_i } =\lim_{L\rightarrow\infty}\sum_{\boldsymbol{k}} \bar \phi^{\boldsymbol{p}}_{\boldsymbol{k}} \phi^{\boldsymbol{p}}_{\boldsymbol{k}- \hat{\boldsymbol{n}}_i \Delta}\braket{u_{\boldsymbol{p}+\boldsymbol{k},c}| u_{\boldsymbol{p}+\boldsymbol{k}- \hat{\boldsymbol{n}}_i \Delta , c}} \braket{u_{\boldsymbol{k} -  \hat{\boldsymbol{n}}_i\Delta  ,v}| u_{\boldsymbol{k}, v}}.
\end{equation}

\newpage
\section{Exciton Wannier states}
\label{sec:ApxExcitonWannierStates}
\subsection{Relation between the two sets of exciton Wannier states}
In this section we derive the general condition for the two sets of exciton Wannier functions ($\ket{\mathcal{W}^R_{c/v}}$) (derived in Sec.~\ref{Sec:ApxExcitonProjPosition}) to be equal. 

    Recall that the two sets of exciton Wannier states are,
\begin{align}
\ket{\mathcal{W}^R_{\mathrm{exc}, c/v}} &= \mathcal{N}\sum_{q} e^{-\mathrm{i}  qR} (W_{\mathrm{exc}, c/v})^{-\frac{q}{2\pi}} \left(\prod_{k = 0}^{q-\Delta} t^{\mathrm{exc}, c/v}_{k}\right)\ket{q_{\mathrm{exc}}}.
\end{align}
In the thermodynamic limit we can write (the normalised) Wannier states as,
\begin{align}
\ket{\mathcal{W}^R_{\mathrm{exc}, c/v}} &= \frac{1}{\sqrt{2\pi}}\int_{0}^{2\pi} \mathrm{d}q\: e^{-\mathrm{i}  qR} e^{-\mathrm{i}\frac{\gamma_{\mathrm{exc}, c/v}}{2\pi}q} e^{\mathrm{i}\int_{0}^{q} \mathrm{d}p \:A_{\mathrm{exc, c/v}}(p) }\ket{q_{\mathrm{exc}}}.
\end{align}
We want to find the condition for these two sets of exciton Wannier states (\emph{i.e.} those that maximally localise the electron or the hole) to be equal. We consider the overlap,
\begin{align}
\braket{\mathcal{W}^{R = 0}_{v}| \mathcal{W}^{R = 0}_{c}} &= \frac{1}{2\pi} \int_{0}^{2\pi} \int_{0}^{2\pi}\mathrm{d}q\:\mathrm{d}k\: e^{\mathrm{i} \frac{\gamma_{\mathrm{exc}, v} k}{2\pi}}e^{-\mathrm{i} \frac{\gamma_{\mathrm{exc}, c} q}{2\pi}} e^{-\mathrm{i}\int_{0}^q \mathrm{d}p \:A_{\mathrm{exc}, v}(p)}  e^{\mathrm{i}\int_{0}^k \mathrm{d}p' \:A_{\mathrm{exc}, c}(p')} \braket{q_{\mathrm{exc}}|k_{\mathrm{exc}}}\\
&= \frac{1}{2\pi} \int_{0}^{2\pi} \int_{0}^{2\pi}\mathrm{d}q\:\mathrm{d}k\:  e^{\mathrm{i} \frac{\gamma_{\mathrm{exc}, v} k}{2\pi}}e^{-\mathrm{i} \frac{\gamma_{\mathrm{exc}, c} q}{2\pi}} e^{-\mathrm{i}\int_{0}^q \mathrm{d}p \:A_{\mathrm{exc}, v}(p)}  e^{\mathrm{i}\int_{0}^k \mathrm{d}p' \:A_{\mathrm{exc}, c}(p')} \delta(q- k)\\
&= \frac{1}{2\pi} \int_{0}^{2\pi}\mathrm{d}q\:  e^{\mathrm{i} \frac{(\gamma_{\mathrm{exc}, v}-\gamma_{\mathrm{exc}, c}) q}{2\pi}}   e^{\mathrm{i}\int_{0}^q \mathrm{d}p \:[A_{\mathrm{exc}, c}(p) -  \:A_{\mathrm{exc}, v}(p)]}.
\end{align}
For the two sets of Wannier states to be equal we require $\braket{\mathcal{W}^{R = 0}_{v}| \mathcal{W}^{R = 0}_{c}} = 1$. We recall the definition of $\mathcal{F}(p)$,
\begin{align}
\mathcal{F}(p) &= A_{\mathrm{exc}, c}(p) -  \:A_{\mathrm{exc}, v}(p)\\ &=\int_{0}^{2\pi}\frac{\mathrm{d}k}{\Delta}\:\left(\bar\phi^p_k \mathrm{i} \partial_k  \phi^p_k + |\phi^p_k|^2 \bra{u_{p+k, c}} \mathrm{i} \partial_k \ket{u_{p+k, c}}  - |\phi^p_k|^2 \bra{u_{k, v}} \mathrm{i} \partial_k \ket{u_{k, v}}\right)
\end{align}
which in Sec.~\ref{sec:ApxElectronicPolarisationExcitons} we identified as the electric polarisation of the exciton \emph{i.e.} the expected separation between the electron and the hole in the exciton. It is gauge invariant under both gauge transformations of the exciton band ($\ket{\phi^p}\rightarrow e^{\mathrm{i}\theta(p)} \ket{\phi^p}$) and gauge transformations of the underlying electronic bands. We can rewrite the overlap in terms of $\mathcal{F}(p)$,
\begin{align}
\braket{\mathcal{W}^{R = 0}_{v}| \mathcal{W}^{R = 0}_{c}} &= \frac{1}{2\pi} \int_{0}^{2\pi}\mathrm{d}q\:  e^{\mathrm{i} \frac{(\gamma_{\mathrm{exc}, v}-\gamma_{\mathrm{exc}, c}) q}{2\pi}}   e^{\mathrm{i}\int_{0}^q \mathrm{d}p \:\mathcal{F}(p)} .
\end{align}
This is an integral of a quantity that is modulus 1 at all $q$. For this overlap to equal 1 (\emph{i.e.} for the two Wannier states to be equal), the integrand must therefore equal 1 at all momenta $q$.
We therefore require that,
\begin{equation}
e^{\mathrm{i}\frac{(\gamma_{\mathrm{exc}, v} - \gamma_{\mathrm{exc}, c}) q}{2\pi}}e^{\mathrm{i}\int_{0}^q \mathrm{d}p \:\mathcal{F}(p)} = 1.
\end{equation}
This is true if and only if $e^{\mathrm{i}\int_{0}^q \mathrm{d}p \:\mathcal{F}(p)} = e^{\mathrm{i} \frac{(\gamma_{\mathrm{exc}, c} - \gamma_{\mathrm{exc}, v}) q}{2\pi}}$. Therefore for the two sets of exciton Wannier states to be equal $\mathcal{F}(p)$ must be independent of momentum $p$ [$\mathcal{F}(p) = \mathcal{F}$]. This condition has a clear physical interpretation. Physically, if the electron-hole separation is the same at all momenta $p$, then maximally localising the hole is clearly equivalent to maximally localising the electron. As a result, the two sets of exciton Wannier states must be exactly equal. But the condition that the electron-hole separation is the same at all momenta $p$ is equivalent to $\mathcal{F}(p)$ being constant. This is the condition we derived above.

\subsection{Exciton Wannier state plots}
For the plots of the exciton Wannier states in the main text, we calculate components in the basis of the maximally localised \emph{electronic} Wannier states,
\begin{align}
\ket{\mathcal{W}^{R'}_{c/v}} &= \sum_{p} \mathcal{W}^{R'}_{c/v}(p) \ket{p_{\mathrm{exc}}}\\
 &= \sum_{p, k} \mathcal{W}^{R'}_{c/v}(p) \phi^p_k c^\dagger_{p+k, c} c_{k, v} \ket{\mathrm{GS}}\\
 &= \frac{1}{L}\sum_{p, k, R, r} \mathcal{W}^{R'}_{c/v}(p) \phi^p_k e^{-\mathrm{i}(p+k) (R+r)} e^{\mathrm{i}k R}c^\dagger_{R+r, c} c_{R, v} \ket{\mathrm{GS}}\\
 &= \frac{1}{L}\sum_{p, k,  R, r} \mathcal{W}^{R'}_{c/v}(p) \phi^p_k e^{-\mathrm{i}p (R+r)} e^{-\mathrm{i}k \Delta} c^\dagger_{R+r, c} c_{R, v} \ket{\mathrm{GS}}\\
  &= \sum_{R, r} \mathcal{W}^{R'}_{c/v}(R, r) \:c^\dagger_{R+r, c} c_{R, v} \ket{\mathrm{GS}}
\end{align}
where the components of the exciton Wannier state in this basis are,
\begin{equation}
\mathcal{W}^{R'}_{c/v}(R, r) = \frac{1}{L}\sum_{p, k} \mathcal{W}^{R'}_{c/v}(p) \phi^p_k e^{-\mathrm{i}p (R+r)} e^{-\mathrm{i}k r}.
\end{equation}
\newpage
\section{Exciton calculations}
In this section we describe the method we use to numerically calculate the exciton band structures and exciton wave functions shown in the paper. The general approach is to project the many-body Hamiltonian into the variational basis,
\begin{equation}
\ket{p, k} = c^\dagger_{p+k, c} c_{k, v} \ket{\mathrm{GS}}.
\end{equation}
The total momentum of $\ket{p, k}$ is $p$ whereas $k$ labels the relative momentum. Since the Hamiltonian is translationally invariant $p$ is conserved by the Hamiltonian, that is to say we can diagonalise the Hamiltonian,
\begin{equation}
\bra{p, k} \hat H \ket{p, k'},
\end{equation}
at each fixed total momentum $p$. 

The total Hamiltonian consists of a kinetic part $\hat H_0$ and an interaction part $\hat H_{\mathrm{int}}$. We project the total Hamiltonian $\hat H = \hat H_{0} + \hat H_{\mathrm{int}}$ into the variational basis shown above and calculate the matrix elements,
\begin{equation}
\bra{p, k} \hat H \ket{p, k'}.
\end{equation}
First we evaluate
\begin{equation}
\bra{p, k} \hat H_{0} \ket{p, k'}.
\end{equation}
This can be seen to be equal to
\begin{equation}
\bra{p, k} \hat H_{0} \ket{p, k'} = \delta_{k,k'} (E_{\mathrm{GS}} - \epsilon_{k', v} + \epsilon_{p+k', c}),
\end{equation}
where $E_{\mathrm{GS}} = \sum_k \epsilon_{k,v}$ is the energy of the ground state.

We project the interaction term into this basis. We begin with the general form of the interaction Hamiltonian for density-density interactions,
\begin{equation}
\hat H_{\mathrm{int}} = \sum_{R,R',i,j} U_{R-R', i, j} \hat n_{R,i} \hat n_{R',j} ,
\label{eq:Hint_orig} 
\end{equation}
where $\hat n_{R,i}$ is the number operator at unit cell $R$ and $i$ labels the sublattice. We write this in the momentum basis. First note that
\begin{align}
c^\dagger_{k,i} &= \frac{1}{\sqrt{L}} \sum_R e^{ikR} c^\dagger_{R,i},\\
c^\dagger_{R,i} &= \frac{1}{\sqrt{L}} \sum_R e^{-ikR} c^\dagger_{k,i}.
\end{align}
Therefore we can rewrite the number operators as
\begin{align}
\hat n_{R,i} &= c^\dagger_{R,i} c_{R,i}\\
 &= \frac{1}{L} \sum_{p,q} e^{-i(p-q)R} c^\dagger_{p,i} c_{q,i}.
\end{align}
We can rewrite Eq.~\eqref{eq:Hint_orig} as,
\begin{align}
\hat H_{\mathrm{int}} &= \sum_{R,R', i, j} \sum_{p,q,p',q'} \frac{1}{L^2} U_{R-R', ij} e^{-i(p-q)R - i(p'-q')R'} c^\dagger_{pi} c_{qi} c^\dagger_{p'j} c_{q'j}\\
 &= \sum_{R,R', i, j} \sum_{p,q,p',q'} \frac{1}{L^2} U_{R, ij} e^{-i(p-q)R - i(p'-q'+p'-q')R'} c^\dagger_{pi} c_{qi} c^\dagger_{p'j} c_{q'j},
\end{align}
where the final line sets $R-R' \rightarrow R$. This can be further simplified by performing the sum over $R'$ and $q'$ to give
\begin{equation}
\hat H_{\mathrm{int}} = \frac{1}{L} \sum_{R,i,j} \sum_{pqp'} U_{R,ij} e^{-i(p-q)R} c^\dagger_{p,i} c_{q,i} c^\dagger_{p',j} c_{p-q+p',j}.
\end{equation}
Defining $U_{p,ij} = \frac{1}{L} \sum_R U_{R,ij} e^{-ipR}$, we have
\begin{equation}
\hat H_{\mathrm{int}} = \sum_{pqq'ij} U_{p,ij} c^\dagger_{p+q,i} c_{q,i} c^\dagger_{q',j} c_{p+q',j}.
\label{eq:ham_int_sublattice_basis}
\end{equation}
We wish to calculate the matrix elements
\begin{equation}
\bra{p, k} \hat H_{\mathrm{int}} \ket{p, k'},
\label{eq:int_matrix_element}
\end{equation}
so we first express the interaction Hamiltonian $\hat H_{\mathrm{int}}$ in terms of the creation operators for the bands. These are related to the $c^\dagger_{k,i}$ operators via the periodic Bloch functions $u^{k,\alpha}_{i}$. For example
\begin{align}
c^\dagger_{k,\alpha} &= \sum_i u^{k,\alpha}_i c^\dagger_{k,i}\\
c_{k,\alpha} &= \sum_i \bar u^{k,\alpha}_i c_{k,i}
\end{align}
where $\alpha \in \{c, v\}$ labels the two bands. We use the notation $\bar u^{k, \alpha}_i = (u^{k, \alpha}_i)^*$.
Plugging this relation into Eq.~\eqref{eq:ham_int_sublattice_basis} gives
\begin{align}
\hat H_{\mathrm{int}} &= \sum_{pqq'ij} U_{p,ij} c^\dagger_{p+q,i} c_{q,i} c^\dagger_{q',j} c_{p+q',j}\\
 &= \sum_{pqq'ij\alpha\alpha'\beta\beta'} U_{p,ij} \bar{u}^{p+q, \alpha}_i u^{q\alpha'}_{i} \bar{u}^{q', \beta}_j u^{p+q'\beta'}_{j} c^\dagger_{p+q,\alpha}   c_{q,\alpha'} c^\dagger_{q',\beta} c_{p+q',\beta'}.
\label{eq:ham_int_band_basis}
\end{align}
Defining $U^{pqq'}_{\alpha\alpha'\beta\beta'} = \sum_{ij} (U_{p,ij} \bar{u}^{p+q, \alpha}_i u^{q\alpha'}_{i} \bar{u}^{q', \beta}_j u^{p+q'\beta'}_{j})$, then we can write
\begin{equation}
\hat H_{\mathrm{int}} = \sum_{pqq'\alpha\alpha'\beta\beta'} U^{pqq'}_{\alpha\alpha'\beta\beta'} c^\dagger_{p+q,\alpha}   c_{q,\alpha'} c^\dagger_{q',\beta} c_{p+q',\beta'}.
\end{equation}
The matrix elements can then be expressed as,
\begin{equation}
\bra{p, k} \hat H_{\mathrm{int}} \ket{p, k'} = \sum_{lqq'\alpha\alpha'\beta\beta'} U^{lqq'}_{\alpha\alpha'\beta\beta'} \bra{\mathrm{GS}} c^\dagger_{k,v} c_{p+k,c} c^\dagger_{l+q,\alpha}   c_{q,\alpha'} c^\dagger_{q',\beta} c_{l+q',\beta'} c^\dagger_{p+k',c} c_{k',v} \ket{\mathrm{GS}}.
\label{eq:final_h_int}
\end{equation}

The following expectation value can be simplified using Wick's theorem,
\begin{equation}
\bra{\mathrm{GS}} c^\dagger_{k,v} c_{p+k,c} c^\dagger_{l+q,\alpha}   c_{q,\alpha'} c^\dagger_{q',\beta} c_{l+q',\beta'} c^\dagger_{p+k',c} c_{k',v} \ket{\mathrm{GS}},
\end{equation}
for readability we will simplify the notation to
\begin{align}
\mu^\dagger &= c^\dagger_{k,v}\\
\nu &= c_{p+k,c}\\
\zeta^\dagger &= c^\dagger_{l+q, \alpha}\\
\sigma &= c_{q,\alpha'}\\
\chi^\dagger &= c^\dagger_{q',\beta}\\
\gamma &= c_{l+q', \beta'}\\
\lambda^\dagger &= c^\dagger_{p+k',c}\\
\xi &= c_{k',v}.
\end{align}
The expectation value becomes
\begin{equation}
\langle \mu^\dagger \nu \zeta^\dagger \sigma \chi^\dagger \gamma \lambda^\dagger \xi \rangle.
\end{equation}
This can be expanded using Wick's theorem as
\begin{multline}
\langle \mu^\dagger \nu \zeta^\dagger \sigma \chi^\dagger \gamma \lambda^\dagger \xi \rangle = \langle \mu^\dagger \sigma \rangle (\langle \nu \zeta^\dagger \rangle \langle \chi^\dagger \xi \rangle \langle \gamma \lambda^\dagger \rangle - \langle \nu \chi^\dagger \rangle \langle \zeta^\dagger \xi \rangle \langle \gamma \lambda^\dagger \rangle - \langle \nu \lambda^\dagger \rangle (\langle \zeta^\dagger \xi \rangle \langle \chi^\dagger \gamma \rangle - \langle \zeta^\dagger \gamma \rangle \langle \chi^\dagger \xi \rangle)) \\
+ \langle \mu^\dagger \gamma \rangle (- \langle\nu \zeta^\dagger \rangle \langle \sigma \lambda^\dagger\rangle\langle \chi^\dagger \zeta \rangle + \langle \nu \chi^\dagger \rangle \langle \zeta^\dagger \xi \rangle \langle \sigma \lambda^\dagger \rangle - \langle v \lambda^\dagger \rangle (\langle \zeta^\dagger \sigma \rangle \langle \chi^\dagger \xi \rangle + \langle \zeta^\dagger \xi \rangle \langle \sigma \chi^\dagger \rangle))\\
+ \langle \mu^\dagger \xi\rangle ( \langle \nu \zeta^\dagger \rangle (\langle \sigma \chi^\dagger \rangle \langle \gamma \lambda^\dagger \rangle + \langle \sigma \lambda^\dagger \rangle \langle \chi^\dagger \gamma \rangle ) + \langle \nu \chi^\dagger \rangle ( \langle \zeta^\dagger \sigma \rangle \langle \gamma \lambda^\dagger \rangle - \langle \zeta^\dagger \gamma \rangle \langle \sigma \lambda^\dagger \rangle ) + \\ \langle \nu \lambda^\dagger \rangle ( \langle \zeta^\dagger \sigma \rangle \langle \chi^\dagger \gamma \rangle + \langle \zeta^\dagger \gamma \rangle \langle \sigma \chi^\dagger \rangle)).
\end{multline}
We calculate the two particle correlation functions to find the exciton matrix elements. After performing the sum in Eq.~\eqref{eq:final_h_int} we obtain,
\begin{multline}
\bra{p, k} \hat H_{\mathrm{int}} \ket{p, k'} = U^{p,k,k'}_{cvvc} + U^{-p,p+k',p+k}_{vccv} - U^{k'-k,k,p+k}_{vvcc} - U^{k-k',p+k',k'}_{ccvv}\\ + \delta_{kk'} \sum_q ( U^{q,k,k}_{vvvv} - U^{0,k,q}_{vvvv} - U^{0,q,k}_{vvvv} -U^{q,k-q,k-q}_{vccv} + U^{p+k-q,q,q}_{cccc} + U^{0,p+k,q}_{ccvv} + U^{0,q,p+k}_{vvcc} - U^{q,p+k,p+k}_{vccv})\\ + \delta_{kk'} \sum_{q,q'} (U^{0, q, q'}_{vvvv} + U^{q, q', q'}_{vccv})\label{eq:InteractingHam}.
\end{multline}
Finally we note that the final term is equal to the shift of the ground state energy due to the interaction \emph{i.e.}
\begin{equation}
\bra{\mathrm{GS}} \hat H_{\mathrm{int}}\ket{\mathrm{GS}} = \sum_{q,q'} (U^{0, q, q'}_{vvvv} + U^{q, q', q'}_{vccv}).
\end{equation}
The exciton energy above the modified ground state energy is therefore obtained by subtracting this term from the exciton Hamiltonian ($\bra{p, k} \hat H_{\mathrm{int}} \ket{p, k'}$). 

\newpage
\section{Crystalline symmetries}
\label{sec:ApxSymmetries}
\subsection{Inversion symmetry}
We first consider inversion ($\mathcal{I}$) symmetry. We derived two different Wilson loops in Sec.~\ref{Sec:ApxExcitonProjPosition},
\begin{align}
W_{\mathrm{exc}, c} &= \prod_p \big[\sum_k\bar \phi^{p+\Delta}_{k- p} \phi^{p}_{k- p} \braket{u_{k+\Delta, c}| u_{k, c}}]\\
W_{\mathrm{exc}, v} &= \prod_p \big[\sum_k\bar \phi^{p+\Delta}_{k} \phi^{p}_{k+\Delta} \braket{u_{k+\Delta, v}| u_{k, v}}].
\end{align}
Both of these Wilson loops can be re-expressed as products of projectors,
\begin{equation}
W_{\mathrm{exc}, c/v} = \bra{\mathcal{U}^{\mathrm{exc}, c/v}_{p_1}} \lim_{R\rightarrow\infty} \prod^2_{i = R} \ket{\mathcal{U}^{\mathrm{exc}, c/v}_{p_i}} \bra{\mathcal{U}^{\mathrm{exc}, c/v}_{p_i}}\ket{\mathcal{U}^{\mathrm{exc}, c/v}_{p_1}}
\end{equation}
using,
\begin{align}
\ket{\mathcal{U}^{\mathrm{exc}, c}_{p}} &= \sum_{k} \phi^p_k \ket{k} \otimes \ket{u_{p+k, c}} \otimes \ket{\bar u_{k, v}}\label{eq:Wavefunctions4WilsonLoops1}
\\
\ket{\mathcal{U}^{\mathrm{exc}, v}_{p}} &= \sum_{k} \phi^p_k \ket{p+k} \otimes \ket{u_{p+k, c}} \otimes \ket{\bar u_{k, v}},
\label{eq:Wavefunctions4WilsonLoops2}
\end{align}
respectively. The first quantised Bloch vectors depend on the first quantised momentum eigenstates $\ket{k}$ or $\ket{p+k}$. If your real space basis is a set of unit cells $R$ each with orbitals $\alpha$ then we could write the basis as $\ket{R, \alpha} = \ket{R} \otimes \ket{\alpha}$. Then we define $\ket{k} = \frac{1}{\sqrt{L}} \sum_{R = 0}^{L-1} e^{\mathrm{i} R k} \ket{R}$.

We can show that these two Wilson loops are equal when the system is inversion symmetric. We do this in two ways below, firstly using the Wilson loop and secondly in the thermodynamic limit from the Berry phase. 

In inversion symmetric systems, the Bloch functions for band $\alpha$ and momentum $p$ obey $\hat U_I \ket{u_{p, \alpha}} = e^{\mathrm{i}\xi_{\alpha}(p)} \ket{u_{-p, \alpha}}$. The unitary matrix $\hat U_I$ represents $\mathcal{I}$ symmetry in the single-electron Hilbert space. The phase factor $e^{\mathrm{i}\xi_{\alpha}(p)}$ reflects the arbitrary gauge of the Bloch functions. At the level of the creation operators, for band $\alpha$ and momentum $p$, inversion symmetry implies that $\hat I c^\dagger_{p, \alpha} \hat I = e^{\mathrm{i}\xi_{\alpha}(p)}c^\dagger_{-p,\alpha}$ where $\hat I$ represents $\mathcal{I}$ symmetry in the full many-body Hilbert space. 

If we have $\mathcal{I}$ symmetry then the exciton Wave function must transform as follows under inversion, 
\begin{align}
\hat I\ket{p_{\mathrm{exc}}} &= e^{\mathrm{i} \beta(p)}\ket{-p_{\mathrm{exc}}}\\
\hat I \sum_k \phi^p_k c^\dagger_{p+k, c} c_{k, v}\ket{\mathrm{GS}} &= e^{\mathrm{i} \beta(p)} \sum_k \phi^{-p}_k c^\dagger_{-p+k, c} c_{k, v}\ket{\mathrm{GS}}.
\end{align}
Since $-p = p$ when $p = 0, \pi$, the phase factor ($e^{\mathrm{i} \beta(p)}$) at these momenta are eigenvalues of the inversion operator. We label these as $\lambda^I_{0}$ and $\lambda^I_{\pi}$ respectively. Since inversion squares to 1 ($\hat I^2 = 1 $) the inversion eigenvalues have values~$\pm 1$. 

The representation of $\mathcal{I}$ acting on the first quantised exciton wave functions $\ket{\mathcal{U}^{\mathrm{exc}, c/v}_{p}}$ is the unitary operator,
\begin{align}
\hat \pi_I = \sum_k \ket{-k} \bra{k} \otimes \hat U_I\otimes \hat U_I.
\end{align}
It can be shown that,
\begin{align}
\hat \pi_I \ket{\mathcal{U}^{\mathrm{exc}, c}_{p}} &= e^{\mathrm{i} \beta(p)} \ket{\mathcal{U}^{\mathrm{exc}, c}_{-p}}\\
\hat \pi_I \ket{\mathcal{U}^{\mathrm{exc}, v}_{p}} &= e^{\mathrm{i} \beta(p)} \ket{\mathcal{U}^{\mathrm{exc}, v}_{-p}}.
\end{align}
From this alone it follows that both of these Wilson loops are equal to the product of the inversion eigenvalues of the exciton wave function at $p = 0$ and $p=\pi$ since,
\begin{align}
W_{\mathrm{exc}, c/v} &= \bra{\mathcal{U}^{\mathrm{exc}, c/v}_{0}}\ket{\mathcal{U}^{\mathrm{exc}, c/v}_{\Delta}} 
       \bra{\mathcal{U}^{\mathrm{exc}, c/v}_{\Delta}}\ket{\mathcal{U}^{\mathrm{exc}, c/v}_{2\Delta}} 
       \bra{\mathcal{U}^{\mathrm{exc}, c/v}_{2\Delta}} \dots \ket{\mathcal{U}^{\mathrm{exc}, c/v}_{\pi}} \\
       &\quad\quad \cdot\bra{\mathcal{U}^{\mathrm{exc}, c/v}_{\pi}} \dots \ket{\mathcal{U}^{\mathrm{exc}, c/v}_{2\pi-2\Delta}} 
       \bra{\mathcal{U}^{\mathrm{exc}, c/v}_{2\pi-2\Delta}}\ket{\mathcal{U}^{\mathrm{exc}, c/v}_{2\pi-\Delta}} 
       \bra{\mathcal{U}^{\mathrm{exc}, c/v}_{2\pi-\Delta}}\ket{\mathcal{U}^{\mathrm{exc}, c/v}_{0}}\nonumber \\[10pt]
&= \bra{\mathcal{U}^{\mathrm{exc}, c/v}_{0}}\hat \pi_I \hat \pi_I \ket{\mathcal{U}^{\mathrm{exc}, c/v}_{\Delta}} 
   \bra{\mathcal{U}^{\mathrm{exc}, c/v}_{\Delta}}\hat \pi_I \hat \pi_I\ket{\mathcal{U}^{\mathrm{exc}, c/v}_{2\Delta}} 
   \bra{\mathcal{U}^{\mathrm{exc}, c/v}_{2\Delta}}\hat \pi_I \dots \hat \pi_I\ket{\mathcal{U}^{\mathrm{exc}, c/v}_{\pi}} \notag \\
&\quad\quad \cdot \bra{\mathcal{U}^{\mathrm{exc}, c/v}_{\pi}} \dots \ket{\mathcal{U}^{\mathrm{exc}, c/v}_{2\pi-2\Delta}} 
   \bra{\mathcal{U}^{\mathrm{exc}, c/v}_{2\pi-2\Delta}}\ket{\mathcal{U}^{\mathrm{exc}, c/v}_{2\pi-\Delta}} 
   \bra{\mathcal{U}^{\mathrm{exc}, c/v}_{2\pi-\Delta}}\ket{\mathcal{U}^{\mathrm{exc}, c/v}_{0}} \\[10pt]
&= \bra{\mathcal{U}^{\mathrm{exc}, c/v}_{0}} e^{-\mathrm{i} \beta(0)} e^{\mathrm{i} \beta(\Delta)} 
   \ket{\mathcal{U}^{\mathrm{exc}, c/v}_{2\pi-\Delta}} 
   \bra{\mathcal{U}^{\mathrm{exc}, c/v}_{2\pi-\Delta}} e^{-\mathrm{i} \beta(\Delta)} 
   e^{\mathrm{i} \beta(2\Delta)}\ket{\mathcal{U}^{\mathrm{exc}, c/v}_{2\pi-2\Delta}} \notag \\
&\quad\quad \cdot \bra{\mathcal{U}^{\mathrm{exc}, c/v}_{2\pi-2\Delta}} e^{-\mathrm{i} \beta(2\Delta)} 
   \dots e^{\mathrm{i} \beta(\pi)}\ket{\mathcal{U}^{\mathrm{exc}, c/v}_{\pi}} 
   \bra{\mathcal{U}^{\mathrm{exc}, c/v}_{\pi}} \dots \ket{\mathcal{U}^{\mathrm{exc}, c/v}_{2\pi-2\Delta}} 
   \bra{\mathcal{U}^{\mathrm{exc}, c/v}_{2\pi-2\Delta}}\ket{\mathcal{U}^{\mathrm{exc}, c/v}_{2\pi-\Delta}} 
   \bra{\mathcal{U}^{\mathrm{exc}, c/v}_{2\pi-\Delta}}\ket{\mathcal{U}^{\mathrm{exc}, c/v}_{0}} \\[10pt]
&= e^{-\mathrm{i} \beta(0)} e^{\mathrm{i} \beta(\pi)} 
   \bra{\mathcal{U}^{\mathrm{exc}, c/v}_{0}} \ket{\mathcal{U}^{\mathrm{exc}, c/v}_{2\pi-\Delta}} 
   \bra{\mathcal{U}^{\mathrm{exc}, c/v}_{2\pi-\Delta}} \ket{\mathcal{U}^{\mathrm{exc}, c/v}_{2\pi-2\Delta}} 
   \bra{\mathcal{U}^{\mathrm{exc}, c/v}_{2\pi-2\Delta}} \dots \ket{\mathcal{U}^{\mathrm{exc}, c/v}_{\pi}} \notag \\
&\quad\quad \cdot \bra{\mathcal{U}^{\mathrm{exc}, c/v}_{\pi}} \dots \ket{\mathcal{U}^{\mathrm{exc}, c/v}_{2\pi-2\Delta}} 
   \bra{\mathcal{U}^{\mathrm{exc}, c/v}_{2\pi-2\Delta}} \ket{\mathcal{U}^{\mathrm{exc}, c/v}_{2\pi-\Delta}} 
   \bra{\mathcal{U}^{\mathrm{exc}, c/v}_{2\pi-\Delta}} \ket{\mathcal{U}^{\mathrm{exc}, c/v}_{0}} \\[10pt]
&= e^{-\mathrm{i} \beta(0)} e^{\mathrm{i} \beta(\pi)} 
   \left| \bra{\mathcal{U}^{\mathrm{exc}, c/v}_{0}} \ket{\mathcal{U}^{\mathrm{exc}, c/v}_{2\pi-\Delta}} 
   \bra{\mathcal{U}^{\mathrm{exc}, c/v}_{2\pi-\Delta}} \ket{\mathcal{U}^{\mathrm{exc}, c/v}_{2\pi-2\Delta}} 
   \bra{\mathcal{U}^{\mathrm{exc}, c/v}_{2\pi-2\Delta}} \dots \ket{\mathcal{U}^{\mathrm{exc}, c/v}_{\pi}} \right|^2\label{eq:InversionWilsonLoop} \\[10pt]
&= \lambda^I_0 \lambda^I_\pi.
\end{align}
Therefore, in the thermodynamic limit, the Wilson loops ($W_{\mathrm{exc}, c/v}$) are both equal to the product of the inversion eigenvalues of the many-body exciton eigenstates at the high-symmetry points. We have also therefore shown that these two Wilson loops ($W_{\mathrm{exc}, c/v}$) are equal in the presence of inversion symmetry. 

We can also show this using the exciton Berry connections [$A_{\mathrm{exc}, c/v}(p)$]. The difference between the exciton Berry connections is $\mathcal{F}(p)$. This represents the electron-hole separation. Under inversion an exciton with momentum $p$ is mapped to one with $-p$. In addition the electron-hole separation also must flip sign because under inversion, if the electron was to the right of the hole then it must be flipped to being on the left. It follows that $\mathcal{F}(p) = -\mathcal{F}(-p)$. Therefore, 
\begin{align}
\int_0^{2\pi}\mathrm{d}p\: \big[A_{\mathrm{exc}, c}(p) - A_{\mathrm{exc}, v}(p) \big] &= \int_0^{2\pi}\mathrm{d}p\: \mathcal{F}(p)\\
&= \int_0^{\pi}\mathrm{d}p\: \mathcal{F}(p) + \int_{-\pi}^{0}\mathrm{d}p\: \mathcal{F}(p)\\
&= \int_0^{\pi}\mathrm{d}p\: \mathcal{F}(p) - \int_{-\pi}^{0}\mathrm{d}p\: \mathcal{F}(-p)\\
&= \int_0^{\pi}\mathrm{d}p\: \mathcal{F}(p) - \int_{0}^{\pi}\mathrm{d}p\: \mathcal{F}(p)\\
&= 0
\end{align}
Therefore the two Berry phases are equal $\int_0^{2\pi}\mathrm{d}p\: A_{\mathrm{exc}, c}(p) = \int_0^{2\pi}\mathrm{d}p\: A_{\mathrm{exc}, v}(p)$. As a result the Wilson loops, which are related to the Berry phases ($W_{\mathrm{exc}, c/v} = e^{\mathrm{i}\int_0^{2\pi}\mathrm{d}p\: A_{\mathrm{exc}, c/v}(p)}$), are also equal in the thermodynamic limit.

To conclude, $\mathcal{I}$ symmetry means that the two Wilson loops $W_{\mathrm{exc}, c/v}$ are equal in the thermodynamic limit. Furthermore, the Wilson loops are equal to the product of the inversion eigenvalues of the many-body exciton eigenstates at the high-symmetry points. However, the two sets of exciton Wannier states $\ket{\mathcal{W}^R_{\mathrm{exc}, c/v}}$ are not necessarily equal in $\mathcal{I}$ symmetric systems. This is because inversion symmetry requires that $\mathcal{F}(p) = -\mathcal{F}(-p)$ whereas $\mathcal{F}(p)$ must be constant/independent of $p$ in order for $\ket{\mathcal{W}^R_{\mathrm{exc}, c}} = \ket{\mathcal{W}^R_{\mathrm{exc}, v}}$. However, since the two Wilson loops are equal, the exciton Wannier centres of these two sets of (possibly differing) exciton Wannier states do have to be equal. 

\subsection{$C_2 \mathcal{T}$ symmetry}
We now consider the anti-unitary symmetry $\hat C_2 \hat{\mathcal{T}}$. This also quantises the Berry phase. We first demonstrate that this is true for non-interacting electrons. 

We use $\hat C_2 \hat{\mathcal{T}}$ to represent the operator in the many-body Hilbert space. In the single particle Hilbert space (and for spinless models) we can represent time-reversal symmetry as complex conjugation $\mathcal{K}$. The $C_2$ symmetry is denoted by $U_{C_2}$ in the single-particle Hilbert space. 

Unlike for $\mathcal{I}$ symmetry, we can't use symmetry indicators to diagnose the Berry phase in $C_2 \mathcal{T}$ symmetric systems. This is because $C_2 \mathcal{T}$ is anti-unitary and so doesn't have eigenvalues. Consider the representation of  $C_2 \mathcal{T}$ in the single particle Hilbert space ($U_{C_2} \mathcal{K}$). In $C_2\mathcal{T}$ symmetric systems, if we apply $U_{C_2} \mathcal{K}$ to one of the periodic Bloch functions $\ket{u_{p, \alpha}}$ (for band $\alpha$ at momentum $p$) we get the same Bloch function back up to a phase,
\begin{align}
U_{C_2} \mathcal{K} \ket{u_{p, \alpha}} = \lambda \ket{u_{p, \alpha}}.
\end{align}
However $\lambda$ is not an eigenvalue of $U_{C_2} \mathcal{K}$ since we can change $\lambda$ by changing the gauge of the Bloch functions $\ket{u_{p, \alpha}}\rightarrow e^{\mathrm{i}\phi}\ket{u_{p, \alpha}}$,
\begin{align}
U_{C_2} \mathcal{K} e^{\mathrm{i} \phi}\ket{u_{p, \alpha}} &= e^{-\mathrm{i} \phi} U_{C_2} \mathcal{K} \ket{ u_{p, \alpha}} \\
&= e^{-\mathrm{i} \phi} \lambda \ket{ u_{p, \alpha}}\\
&= e^{-2\mathrm{i} \phi} \lambda (e^{\mathrm{i} \phi}\ket{u_{p, \alpha}}).
\end{align}
Hence the eigenvalue changes ($\lambda \rightarrow e^{-2\mathrm{i}\phi} \lambda$) under the gauge change $\ket{u_{p, \alpha}}\rightarrow e^{\mathrm{i}\phi}\ket{u_{p, \alpha}}$. Therefore, the eigenvalues are not well-defined and so cannot be used as indicators for the topology. Instead we must rely on the Berry phase to classify the bands. We write $\lambda = e^{\mathrm{i}\beta(p)}$ and show that, for electronic systems, the Berry phase is still quantised. The \emph{electronic} Berry phase for a band $\alpha$ is $\gamma_{\alpha}$,
\begin{align}
\gamma_\alpha &= \int_{0}^{2\pi}\mathrm{d}p\: \bra{u_{p, \alpha}} \mathrm{i}\partial_p \ket{u_{p, \alpha}}\label{eq:C2TElectronicBerryPhase}\\
&= \int_{0}^{2\pi}\mathrm{d}p\: \bra{u_{p, \alpha}}(U_{C_2}\mathcal{K})^2 \mathrm{i}\partial_p (U_{C_2}\mathcal{K})^2\ket{u_{p, \alpha}}\\
&= \int_{0}^{2\pi}\mathrm{d}p\: \bra{u_{p, \alpha}}e^{-\mathrm{i}\beta(p)}(U_{C_2}\mathcal{K}) \mathrm{i}\partial_p (U_{C_2}\mathcal{K}) e^{\mathrm{i}\beta(p)}\ket{u_{p, \alpha}}\\
&= -\int_{0}^{2\pi}\mathrm{d}p\: \bra{u_{p, \alpha}} e^{-\mathrm{i}\beta(p)}\mathrm{i}\partial_p (U_{C_2}\mathcal{K})^2 e^{\mathrm{i}\beta(p)}\ket{u_{p, \alpha}}\\
&= -\int_{0}^{2\pi}\mathrm{d}p\: \bra{u_{p, \alpha}} e^{-\mathrm{i}\beta(p)}\mathrm{i}\partial_p e^{\mathrm{i}\beta(p)}\ket{u_{p, \alpha}}\\
&= -\gamma_\alpha +[\beta(2\pi) - \beta(0)]\\
&= -\gamma_\alpha +2\pi X,
\end{align}
where $X\in\mathbb{Z}$. This means that $\gamma_\alpha$ is quantised to $0, \pi$ ($\mathrm{mod}\: 2\pi$). The exciton Berry phases $\gamma_{\mathrm{exc}, c/v}$ are also quantised by $C_2\mathcal{T}$. This follows from the fact that we can can write the exciton Berry phases in the same way as the electronic in terms of first quantised exciton wave functions [\emph{e.g.} see Eq.~\eqref{eq:FirstQuantisedExcitonBerryConnection}].

We now show that in addition to the two exciton Berry phases $\gamma_{\mathrm{exc}, c/v}$ both being quantised, they also have to be quantised to the same value (as with $\mathcal{I}$ symmetry). The difference between the two exciton Berry phases is,
\begin{equation}
\gamma_{\mathrm{exc}, c} - \gamma_{\mathrm{exc}, v} = \int_{0}^{2\pi}\mathrm{d}p \: \mathcal{F}(p).
\end{equation}
If a system is $C_2\mathcal{T}$ symmetric then we can show that $\mathcal{F}(p) = 0$ for all momenta. This has a very physical origin in terms of how the electron-hole separation is transformed under $C_2 \mathcal{T}$ symmetry (see main text). Here we instead demonstrate that $\mathcal{F}(p) =0$ mathematically. We begin by noting that $\mathcal{F}(p)$ can be rewritten in a particularly useful form,
\begin{align}
\mathcal{F}(p) &= \int_0^{2\pi} \frac{\mathrm{d}k}{\Delta}\:\left\{\bar\phi^p_k \mathrm{i} \partial_k  \phi^p_k + |\phi^p_k|^2 \left[\bra{u_{c, p+k}}\mathrm{i}\partial_k\ket{u_{c, p+k}}  - \bra{u_{v, k}}\mathrm{i}\partial_k\ket{u_{v, k}} \right]\right\}\\
 &= \int_0^{2\pi} \frac{\mathrm{d}k}{\Delta}\:\left\{\bar\phi^p_k \mathrm{i} \partial_k  \phi^p_k + |\phi^p_k|^2 \left[\bra{u_{c, p+k}}\mathrm{i}\partial_k\ket{u_{c, p+k}}  + \bra{\bar u_{v, k}}\mathrm{i}\partial_k\ket{\bar u_{v, k}} \right]\right\}\\
  &= \int_0^{2\pi} \frac{\mathrm{d}k}{\Delta}\:\left[\left(\bar\phi^p_k \bra{u_{c, p+k}} \otimes  \bra{\bar u_{v, k}}\right) \mathrm{i} \partial_k  \left(\phi^p_k \ket{u_{c, p+k}} \otimes  \ket{\bar u_{v, k}}\right) \right]\label{eq:FofpExpressionC2T}.
\end{align}
The $C_2\mathcal{T}$ symmetry means,
\begin{equation}
\hat C_2\hat{\mathcal{T}} \ket{p_{\mathrm{exc}}} = e^{\mathrm{i}\beta(p)}\ket{p_{\mathrm{exc}}}.
\end{equation}
Consider first-quantised exciton wave functions of the form,
\begin{equation}
\ket{\phi^p_{\mathrm{exc}}} = \sum_k \left(\phi^p_k \ket{u_{c, p+k}} \otimes  \ket{\bar u_{v, k}}\right).
\end{equation}
The representation of $C_2 \mathcal{T}$ symmetry acting on $\ket{\phi^p_{\mathrm{exc}}}$ is $\hat{\mathcal{Q}} = (U_{C_2}\otimes U_{C_2})\mathcal{K}$. If we apply this to $\ket{\phi^p_{\mathrm{exc}}}$ we get,
\begin{equation}
\hat{\mathcal{Q}}\ket{\phi^p_{\mathrm{exc}}} = e^{\mathrm{i}\beta(p)}\ket{\phi^p_{\mathrm{exc}}}.
\end{equation}
We return to the expression for $\mathcal{F}(p)$ in Eq.~\eqref{eq:FofpExpressionC2T}.
\begin{align}
\mathcal{F}(p) &= \int_0^{2\pi} \frac{\mathrm{d}k}{\Delta}\:\left[\left(\bar\phi^p_k \bra{u_{c, p+k}} \otimes  \bra{\bar u_{v, k}}\right) \mathrm{i} \partial_k  \left(\phi^p_k \ket{u_{c, p+k}} \otimes  \ket{\bar u_{v, k}}\right) \right]\\
&= \int_0^{2\pi} \frac{\mathrm{d}k}{\Delta}\:\left[\left(\bar\phi^p_k \bra{u_{c, p+k}} \otimes  \bra{\bar u_{v, k}}\right) \hat{\mathcal{Q}}^2\:\mathrm{i} \partial_k \hat{\mathcal{Q}}^2 \left(\phi^p_k \ket{u_{c, p+k}} \otimes  \ket{\bar u_{v, k}}\right) \right]\\
&= \int_0^{2\pi} \frac{\mathrm{d}k}{\Delta}\:\left[\left(\bar\phi^p_k \bra{u_{c, p+k}} \otimes  \bra{\bar u_{v, k}}\right)e^{-\mathrm{i}\beta(p)} \hat{\mathcal{Q}}\:\mathrm{i} \partial_k \hat{\mathcal{Q}} e^{\mathrm{i}\beta(p)} \left(\phi^p_k \ket{u_{c, p+k}} \otimes  \ket{\bar u_{v, k}}\right) \right]\\
&= -\int_0^{2\pi} \frac{\mathrm{d}k}{\Delta}\:\left[\left(\bar\phi^p_k \bra{u_{c, p+k}} \otimes  \bra{\bar u_{v, k}}\right)e^{-\mathrm{i}\beta(p)} \:\mathrm{i} \partial_k \hat{\mathcal{Q}}^2 e^{\mathrm{i}\beta(p)} \left(\phi^p_k \ket{u_{c, p+k}} \otimes  \ket{\bar u_{v, k}}\right) \right]\\
&= -\int_0^{2\pi} \frac{\mathrm{d}k}{\Delta}\:\left[\left(\bar\phi^p_k \bra{u_{c, p+k}} \otimes  \bra{\bar u_{v, k}}\right) \:\mathrm{i} \partial_k  \left(\phi^p_k \ket{u_{c, p+k}} \otimes  \ket{\bar u_{v, k}}\right) \right]\\
&=-\mathcal{F}(p).
\end{align}
We have therefore shown that $\mathcal{F}(p) = 0$ for all momenta $p$. 

To conclude, $C_2\mathcal{T}$ symmetry means that the two Wilson loops $W_{\mathrm{exc}, c/v}$ are equal in the thermodynamic limit and quantised to $\pm1$. The two sets of exciton Wannier states $\ket{\mathcal{W}^R_{\mathrm{exc}, c/v}}$ are equal in $C_2\mathcal{T}$ symmetric systems. This is because $C_2\mathcal{T}$ symmetry requires that $\mathcal{F}(p) = 0$ for all $p$. Therefore $\mathcal{F}(p)$ is constant/independent of $p$ and this was the required condition for $\ket{\mathcal{W}^R_{\mathrm{exc}, c}} = \ket{\mathcal{W}^R_{\mathrm{exc}, v}}$. The exciton Wannier centres must also be equal since $\gamma_{\mathrm{exc}, c} = \gamma_{\mathrm{exc}, v}$. In the main text we show a simple model which demonstrates these results. In particular we explore a regime where the electronic bands are trivial (\emph{i.e.} the electronic Wannier centres are at the centre of the unit cell) whilst the exciton band is non-trivial (\emph{i.e.} the exciton Wannier centres are at the edge of the unit cell).

\newpage
\section{No crystalline symmetries}
\label{sec:ApxNoSymmetries}
We state the model studied in Sec.~\ref{sec:NoSymmetries} in the main text. The model consists of two stacked SSH models, with different on-site potential ($V$) to separate their band structures in energy. We also add density-density interactions which exponentially decay with distance,
\begin{align}
\hat H &= v\sum_{R} \big( \,c^\dagger_{R,A} c^{\vphantom{\dagger}}_{R, B} + \, c^\dagger_{R,B} c^{\vphantom{\dagger}}_{R, A}\big) +w\sum_{R}\big( c^\dagger_{R, B} c^{\vphantom{\dagger}}_{R+1, A}+c^\dagger_{R+1, A} c^{\vphantom{\dagger}}_{R, B} \big)\\&\quad+v\sum_{R} \big( \,c^\dagger_{R,C} c^{\vphantom{\dagger}}_{R, D} + \, c^\dagger_{R,C} c^{\vphantom{\dagger}}_{R, D}\big) +w\sum_{R}\big( c^\dagger_{R, D} c^{\vphantom{\dagger}}_{R+1, C}+c^\dagger_{R+1, C} c^{\vphantom{\dagger}}_{R, D} \big)  \nonumber\\&\quad+ V\sum_{R}\left(c^\dagger_{R, C}c^{\vphantom{\dagger}}_{R, C} +c^\dagger_{R, D}c^{\vphantom{\dagger}}_{R, D} \right)  \nonumber\\&\quad+ \:\sum_{\substack{R, R', \\i\in\{A, B\}, j\in\{C, D\}}} U(R - R') n_{R, i} n_{R', j} \nonumber .
\end{align}
Here $R$ labels the unit cells and $\{A, B, C, D\}$ labels the four atomic orbitals within the unit cell (see Fig.~\ref{fig:NoSymmetries}a in the main text). 
We break inversion symmetry - we do this at the level of the interactions by choosing an interaction that exponentially decays in $|R|$ but with different rates for $R>0$ compared to $R< 0$,
\begin{equation}
U(R) = 
\begin{cases}
  U_0 e^{-\alpha |R|} & \text{if $R\ge 0$} \\
   U_0 e^{-\beta |R|} & \text{if $R< 0$}.
\end{cases}
\end{equation}
The results shown in the main text (Fig.~\ref{fig:NoSymmetries}) correspond to $U_0 = 0.5$, $\alpha = 0.3$ and $\beta = 2.0$. The remaining parameters used are $v = 1.0$, $w = 0.2$ and $V = 4.5$. We study the model at half-filling (\emph{i.e.} the two lowest energy electronic bands are occupied in the ground state). 

\end{document}